\documentclass[11pt,a4paper,english]{article}
\usepackage{amssymb}
\usepackage{amsmath}
\usepackage{amscd}
\usepackage{latexsym}
\usepackage{graphics}
\usepackage{color}
\usepackage{babel,cite}
\usepackage{epsfig}
\usepackage{amsthm}

\makeatletter

\catcode`@=11
\renewcommand{\section}
{\@startsection{section}{1}{0pt}{\medskipamount}{\medskipamount}{\large\bf}}
\makeatletter\renewcommand{\subsection}
{\@startsection{subsection}{2}{\z@}{-3.25ex plus -1ex minus -.2ex}
{1.5ex plus .2ex}{\it }}

\numberwithin{equation}{section}
\catcode`@=12

\newcommand{\be}{\begin{equation}}
\newcommand{\ee}{\end{equation}}
\newcommand{\eq}[1]{(\ref{#1})}
\def\beq{\be}
\def\eeq{\ee}

\def\nn{\nonumber}
\def\bea{\begin{eqnarray}}
\def\eea{\end{eqnarray}}
\newcommand{\barr}{\begin{array}}
\newcommand{\earr}{\end{array}}
\def\obar{\overline}
\def\one{\mbox{1 \kern-.59em {\rm l}}}

\newcommand{\tr}[1]{\:{\rm tr}\,#1}
\newcommand{\Tr}[1]{\:{\rm Tr}\,#1}
\def\ii{{\,{\rm i}\,}}
\def\ddd{{\rm d}}
\def\eee{{\,\rm e}\,}
\def\vol{{\rm vol}}

\def\a{\alpha}          
\def\b{\beta}           
    
\def\d{\delta}  \def\D{\Delta}  
\def\e{\epsilon}                
        
\def\g{\gamma}

 \def\la{\lambda}

\def\cN{\mathcal{N}}

\def\one{\mbox{1 \kern-.59em {\rm l}}}

  \def\cC{{\cal C}}

\def\cJ{{\cal J}}  \def\cL{{\cal L}}
\def\cM{{\cal M}} \def\cN{{\cal N}} \def\cO{{\cal O}}
 \def\cQ{{\cal Q}} 
  
 \def\cW{{\cal W}}

\def\mg{\mathfrak{g}}
\def\mh{\mathfrak{h}}
\def\mr{\mathfrak{r}}
\def\mt{\mathfrak{t}}
\def\mS{\mathfrak{S}}

\def\ms{\mathfrak{s}}
\def\mun{\mathfrak{u}}

\def\bit{\begin{itemize}}
\def\eit{\end{itemize}}

\def\sK{{\sf K}}

\def\diag{\mbox{diag}}

\def\tens{\otimes}

\def\pfaff{\mbox{pfaff}}

\def\a{\alpha}
\def\b{\beta}
\def\g{\gamma}

\def\e{\epsilon}

\def\beq{\begin{equation}}
\def\eeq{\end{equation}}
\def\bea{\begin{eqnarray}}
\def\eea{\end{eqnarray}}

\renewcommand{\e}{\,\mathrm{e}\,}

\newcommand{\R}{{\mathbb{R}}}
\newcommand{\N}{{\mathbb{N}}}
\newcommand{\C}{{\mathbb{C}}}
\newcommand{\Z}{{\mathbb{Z}}}

\newcommand{\ad}{{\rm ad}}

\def\>{\rangle}
\def\<{\langle}
\def\+{\dagger}
\def\={\ =\ }

\begin{document}

\begin{titlepage}
\setcounter{page}{0}
\begin{flushright}
UWThPh--2007--01\\
HWM--06--47\\
EMPG--06--12\\
\end{flushright}

\vskip 1.8cm

\begin{center}

{\Large\bf Localization for Yang-Mills Theory on the Fuzzy Sphere}

\vspace{15mm}

{\large Harold Steinacker}
\\[2mm]
\noindent{\em Institut f\"ur Theoretische Physik\\
Universit\"at Wien \\ Boltzmanngasse 5, A-1090 Wien, Austria}\\
{Email: {\tt harold.steinacker@univie.ac.at}}
\\[10mm]
{\large Richard J. Szabo}
\\[2mm]
\noindent{\em Department of Mathematics}\\ and \\{\em Maxwell
  Institute for Mathematical Sciences\\ Heriot-Watt University\\
Colin Maclaurin Building, Riccarton, Edinburgh EH14 4AS, U.K.}\\
{Email: {\tt R.J.Szabo@ma.hw.ac.uk}}

\vspace{15mm}

\begin{abstract}
\noindent

We present a new model for Yang-Mills theory on the fuzzy
sphere in which the configuration space of gauge fields is given by a
coadjoint orbit. In the classical limit it reduces to ordinary
Yang-Mills theory on the sphere. We find all classical solutions of
the gauge theory and use nonabelian localization techniques to write
the partition function entirely as a sum over local contributions from
critical points of the action, 
which are evaluated explicitly. The partition function of ordinary
Yang-Mills theory on the sphere is recovered in the classical limit as
a sum over instantons. We also apply abelian localization techniques
and the geometry of symmetric spaces to derive an explicit 
combinatorial expression for the partition function, and compare the
two approaches. These extend the standard techniques for solving
gauge theory on the sphere to the fuzzy case in a rigorous framework.

\end{abstract}

\end{center}
\end{titlepage}

\tableofcontents

\newpage

\section{Introduction and summary \label{Intro}}

Gauge theory on the fuzzy sphere has been of interest for many years
as the simplest example of a noncommutative gauge theory with finitely
many degrees of freedom which retains all of the classical symmetries
of the corresponding undeformed field theory
(see for
instance~\cite{Madore:1991bw,Grosse:1995ar,Klimcik:1997mg,Carow-Watamura:1998jn,Baez:1998he,Grosse:2001qt,Grosse:2001ss,Presnajder:2003ak,matrixsphere,
Castro-Villarreal:2004vh,Ydri:2006xw,Aschieri:2006uw}
and references therein). It can be formulated as an $N\times N$ matrix
model, which provides a natural regularization preserving all
symmetries of quantum gauge theory on the classical sphere which is
recovered in the large $N$ limit. At the classical level one finds  
non-trivial gauge field configurations such as monopoles which can be
naturally described in terms of the noncommutative topology of
projective modules.
Besides Yang-Mills gauge theory which is the focus of this paper, 
certain other gauge theories on the fuzzy sphere naturally emerge 
in string theory upon 
quantizing the worldvolume dynamics of spherical
D2-branes~\cite{Alekseev:2000fd}, obtained for instance as expansions
about vacua of matrix models with a Chern-Simons
term~\cite{Iso:2001mg,Azuma:2004zq} describing superstrings in pp-wave
backgrounds~\cite{Berenstein:2002jq}. These models contain additional 
scalar degrees of freedom and are not considered here.

The formulation of Yang-Mills theory  as an $N\times N$ matrix
model allows a nonperturbative quantization in terms of a 
finite-dimensional path integral~\cite{matrixsphere}. This can then be
evaluated in terms of an $N$-dimensional integral, and 
the classical result as a sum over two-dimensional
instantons~\cite{witten1,Minahan:1993tp,Gross:1994mr}
is recovered in the commutative limit $N \to \infty$. 
A different approach to evaluate the path integral was 
given in~\cite{Ydri:2006xw}, which is also restricted to the large
$N$ limit. This indicates in particular that
the model is void of the usual perturbative ambiguities which plague
noncommutative gauge theories in higher dimensions, such as UV/IR mixing 
(see~\cite{Douglas:2001ba,Szabo:2001kg} for reviews).

In this paper we will formulate a new model for quantum Yang-Mills
theory on the fuzzy sphere, and solve it exactly. 
The model reduces to pure
Yang-Mills theory on the classical sphere when $N\to\infty$ without
any spurious auxilliary scalar fields. The classical theory admits 
topologically non-trivial solutions as in previous matrix model
formulations~\cite{matrixsphere},
including some purely noncommutative ones. 
Its main virtue is that the finite-dimensional
configuration space of gauge fields can be described as a compact
coadjoint orbit, which is naturally a symplectic manifold with a
hamiltonian action of a nonabelian Lie symmetry group. The Yang-Mills
action is the square of the corresponding moment map, and therefore
our model can be solved exactly using nonabelian localization
techniques~\cite{witten1,JK1,Paradan1,JKKW1,Woodward:2004xz,witten} to
cast the partition function as a sum over local
contributions from the classical solutions of the gauge
theory.  It can also be solved by abelian localization
techniques which exploit the usual Duistermaat-Heckman theorem
(see~\cite{Blau:1995rs,szaboloc} for extensive treatments) and which
provide an interesting alternative to the semiclassical expansion.
Although the model described in this paper is fundamentally
different from the fuzzy gauge theories that naturally emerge in
string theory, which contain a Chern-Simons term in their action,
nonabelian localization bears certain remarkable similarities to the
nonabelian localization of Chern-Simons theory on Seifert homology
spheres~\cite{witten}.

There are two main motivations behind the present work. Firstly, in
the commutative case, two-dimensional gauge theories are exactly
solvable and can be solved explicitly, either at strong coupling by exploiting
the Migdal formula~\cite{Migdal:1975zg,Rusakov:1990rs} which expresses
it in terms of a sum over irreducible representations of its gauge
group, or at weak coupling by using Poisson resummation techniques to
cast it as a sum over two-dimensional
instantons~\cite{witten1,Minahan:1993tp,Gross:1994mr}. 
One would therefore
like to have a similar picture in the noncommutative case.
The instanton
expansion can be readily generalized to provide the exact solution for
gauge theory on a two-dimensional noncommutative
torus~\cite{szabo,szaborev1}. However, in previous formulations  of
gauge theory on the fuzzy sphere this is not possible,
either because
extra scalar degrees of freedom not normally present in commutative
Yang-Mills theory destroy the topological nature of the gauge theory
and hence its exact solvability, or else because the exact solution
does not decompose neatly into isolated contributions from classical
solutions.
Our model fills this gap, providing a gauge theory on the
fuzzy sphere whose exact solution is on a unified footing with 
that of gauge theory on the noncommutative torus, in the same way that all
two-dimensional gauge theories admit universal solutions. This is even
apparent from the strong coupling expansions of the two noncommutative
gauge theories~\cite{szaborev1,Paniak:2003gn}, which exhibit the same
degrees of complexity.

However, the precise implementation of the
nonabelian localization principle is rather different in the two
cases. In the case of the torus, one starts from a rational
noncommutative gauge theory and exploits Morita equivalence with
commutative gauge theory to extract the exact instanton expansion, and
then uses continuity arguments to extend the expansion to generic
values of the noncommutativity parameter. On the fuzzy sphere, Morita
equivalence is not available in this manner, and we will have to
evaluate the quantum fluctuation integrals required in the
semiclassical expansion explicitly. This entails a significantly
larger amount of analysis and work than in the case of the torus.

Secondly, our formulation of gauge theory on the fuzzy sphere provides
a new finite-dimensional model which can be solved explicitly by nonabelian
localization techniques. In particular, we draw heavily on techniques
developed recently in~\cite{witten} to analyse higher critical points
in ordinary two-dimensional Yang-Mills theory. In our case, the
analysis is intrinsically finite-dimensional and in accord with rigorous
results established in~\cite{Paradan1,Woodward:2004xz}. The techniques
we exploit in this paper involve a beautiful mix of methods from
random matrix theory and (both abelian and nonabelian)
localization. In particular, we will throughout compare with some
analogous results obtained directly from random matrix theory
in~\cite{matrixsphere}. Our approach thereby extends the toolkit of
methods which can be generally used to treat gauge theories on fuzzy
spaces.

The outline of this paper is as follows. In Section~\ref{SymplModel}
we introduce our new symplectic model for gauge theory on the fuzzy
sphere, showing that it reduces to pure Yang-Mills theory on the
classical sphere in the large $N$ limit. We also describe in detail
the standard construction of the symplectic structure on the coadjoint
orbit space of gauge fields. In Section~\ref{ClassSols} we classify all
classical solutions of the gauge theory, finding fuzzy versions of the
usual instantons and monopoles as well hosts of purely noncommutative
solutions such as fluxons~\cite{Gross:2000ss}. We then give a detailed
description of the local geometry of the configuration space near each
Yang-Mills critical point. In Section~\ref{NonabLoc} we review some general
aspects of nonabelian localization, and apply it to compute precisely
the contributions to the path integral from the vacuum and also higher
unstable critical points, showing in each case that the standard
instanton contributions on the sphere are recovered at
$N\to\infty$. In Sections~\ref{Abelianization}, \ref{IZ-loc}, and
\ref{Abel-loc} we give an alternative
description of the exact path integral in terms of abelian
localization, which exploits the fact that the configuration space is
a hermitian symmetric space to express the gauge field degrees of
freedom in a suitable system of coordinates~\cite{helgason1}. These
coordinates have been previously used to evaluate integrals arising in
random matrix theory in~\cite{casmag1,szabo1}. Finally, in
Section~\ref{AbLoc} we compare the abelian and nonabelian localization
approaches, indicating how to map between the Yang-Mills critical
points and those of the abelianized localization. This is similar to
the abelianized localization at higher critical points of ordinary
Yang-Mills theory studied in~\cite{Blau:1995rs}, although in the fuzzy
case the mapping is not one-to-one and is thus far more intricate.

\bigskip

\section{Symplectic model for Yang-Mills theory on the fuzzy
  sphere\label{SymplModel}}

In this section we will introduce our new symplectic model for gauge
theory on the fuzzy sphere. A similar formulation was given for gauge
theory on fuzzy $\C P^2$ in~\cite{CP2paper}. This formulation will be
particularly suitable for the approach that we take later on to
computing the path integral using localization techniques.

\subsection{The fuzzy sphere\label{FuzzySphere}}

Let $N\in\N$, and let $\xi_i$, $i=1,2,3$ be the $N\times N$ hermitian
coordinate generators of the fuzzy sphere $S^2_N\cong{\rm Mat}_N$
which satisfy the relations
\beq \label{FS-l}
\epsilon^{ij}{}_{k}\, \xi_i \,\xi_j = \ii \xi_k \qquad
 \mbox{and}\qquad \xi_i\, \xi^i  = \mbox{$\frac14$}\,\left(N^2-1
\right)~\one_N
\eeq
where throughout repeated upper and lower indices are implicitly
summed over. The deformation parameter is $\frac1N$ and $S_N^2$
becomes the algebra of functions on the classical unit sphere $S^2$
in the limit $N\to\infty$.
The quantum space $S_N^2$ preserves the
classical invariance under global rotations as follows.     
The $\xi_i$ generate an $N$-dimensional representation of
the global $SU(2)$ isometry group. Under the adjoint action of
$SU(2)$, this representation decomposes covariantly into
$p$-dimensional irreducible representations $(p)$ of $SU(2)$ as
\beq
{\rm Mat}_N\cong(1)\oplus(3)\oplus\cdots\oplus(2N-1) \ ,
\label{MatNSU2decomp}\eeq
which are interpreted as fuzzy spherical harmonics.
This decomposition defines a natural 
map from $S^2_N$ to the space of functions 
on the commutative sphere.
The integral of a function
$f\in S_N^2$ over the fuzzy sphere is given by the trace of $f$,
which  coincides with the usual integral on $S^2$
\beq
\Tr(f) = \frac N{4\pi}\,\int_{S^2}\,\ddd\Omega~f \ 
\label{fuzzyint}\eeq
where the above map is understood.
Rotational invariance of the integral 
then corresponds to invariance of the matrix trace under the
adjoint action of $SU(2)$.

Following~\cite{matrixsphere}, let us combine the generators $\xi_i$
into a larger hermitian $\cN\times\cN$ matrix
\be
\Xi = \mbox{$\frac 12$}\, \one_N\otimes\sigma^0 + \xi_i \otimes\sigma^i
\label{Xi-collective}
\ee 
where $\cN =  2N$, $\sigma^0 = \one_2$, while
\beq
\sigma^1=\begin{pmatrix}0&1\\1&0\end{pmatrix} \ , \quad
\sigma^2=\begin{pmatrix}0&\ii\\-\ii&0\end{pmatrix} \quad \mbox{and}
\quad \sigma^3=\begin{pmatrix}1&0\\0&-1\end{pmatrix}
\label{Pauli}\eeq
are the Pauli spin matrices obeying
\beq
\Tr\big(\sigma^i\big)=0 \qquad \mbox{and} \qquad
\sigma^i\,\sigma^j=\delta^{ij}~\one_2+\ii\epsilon^{ij}{}_k\,\sigma^k \ .
\label{Pauliids}\eeq
One easily finds from (\ref{FS-l}) and (\ref{Pauliids}) the identities
\be
\Xi^2 = \mbox{$\frac{N^2}4$}~\one_{\cN} \qquad \mbox{and} \qquad
\Tr(\Xi) = N \ .
\ee
Since $\xi_i\otimes\sigma^i$ is an intertwiner of the Clebsch-Gordan
decomposition $(N)\otimes(2)=(N-1)\oplus(N+1)$, this implies that
$\Xi$ has eigenvalues $\pm \,\frac N2$ with respective multiplicities
$N_\pm=N\pm1$.

\subsection{Configuration space of gauge fields\label{ConfSpace}}

We will now describe the gauge field degrees of freedom in our
formulation. To elucidate the construction in as transparent a way as
possible, we begin with the abelian case of $U(1)$ gauge theory. To
introduce $\mun(N)$ gauge fields $A_i$ on $S_N^2$, consider the
covariant coordinates~\cite{Madore:2000en}
\be
C_i = \xi_i + A_i \qquad \mbox{and} \qquad C_0 = 
\mbox{$\frac 12$}~\one_N + A_0
\label{covcoordsdef}\ee
which transform under the gauge group $G = U(N)$ as $C_\mu \mapsto
U^{-1}\, C_\mu\, U$ for $\mu=0,1,2,3$ and $U \in U(N)$. We can again
assemble them into a larger $\cN\times\cN$ matrix
\be
C = C_\mu \otimes\sigma^\mu \ .
\label{C-collective}
\ee
Generically, these would consist of four independent fields, and we
have to somehow reduce them to two tangential fields on $S_N^2$. There
are several ways to do this. For example, one can impose the
constraints $A_0 =0$ and $C_i\, C^i = \frac{N^2-1}4~\one_\cN$ as
in~\cite{matrixsphere}, leading to a constrained hermitian multi-matrix
model describing quantum gauge theory on the fuzzy sphere which
recovers Yang-Mills theory on the classical sphere in the large $N$
limit.

Here we will use a different approach and impose the constraints
\be
C^2 = \mbox{$\frac{N^2}4$}~\one_{\cN} \qquad \mbox{and} \qquad \Tr(C)
=N
\label{constraint}
\ee
which is equivalent to requiring that $C$ has eigenvalues $\pm\,\frac
N2$ with multiplicities $N_\pm=N\pm1$. In terms of the components of
(\ref{C-collective}), this amounts to the constraints
\be
C_i \,C^i + C_0^2=\mbox{$\frac{N^2}{4}$}~\one_{\cN} \qquad \mbox{and}
\qquad \ii\epsilon_{i}{}^{jk}\,C_j \,C_k+ \{C_0,C_i\}= 0 \ .
\label{C2}
\ee
We checked in Section~\ref{FuzzySphere} above that this is satisfied
for $A_\mu=0$, wherein $C=\Xi$. We can then consider the action of the
unitary group $U(2N)$ given by
\be
C ~\longmapsto~ U^{-1}\, C \, U
\ee
which generates a coadjoint orbit of $U(2N)$ and preserves
the constraint \eq{constraint}. The gauge fields $A_\mu$ are in this
way interpreted as fluctuations about the coordinates of the quantum
space $S_N^2$. The constraint (\ref{constraint}) ensures that the
covariant coordinates (\ref{C-collective}) describe a dynamical fuzzy
sphere. The gauge group $G=U(N)$ and the global isometry group $SU(2)$
of the sphere are subgroups of the larger symmetry group $U(2N)$. 
In particular, the generators of the gauge group are given by 
elements of
the form $\phi =\phi_0 \otimes\sigma^0 \, \in\, \mg := \mun(N)\subset\mun(\cN\,)$, which defines the gange 
algebra $\mg$.

We thus claim that a possible {configuration space of gauge
  fields} is given by the {\em single} coadjoint orbit
\be
\cO := \cO(\Xi) = \big\{ C = U^{-1}\, \Xi\, U~\big|~  U \in U(\cN\,) 
\big\}
\label{orbit-2}
\ee
where $\Xi\in \mun(2N)$ is given by (\ref{Xi-collective}). Explicitly,
dividing by the stabilizer of $\Xi$ gives a representation of the
orbit (\ref{orbit-2}) as the symmetric space $\cO\cong
U(2N)/U(N+1)\times U(N-1)$ of dimension $\dim(\cO)=2(N^2-1)$. A
similar construction was given in~\cite{CP2paper} for the case of $\C P^2$, and
applied to $S^2_N$ in a different way in~\cite{Ydri:2006xw}. To justify
this claim, we must check that the orbit $\cO$ captures the correct
number of degrees of freedom at least in the commutative limit
$N\to\infty$, i.e. that the gauge fields $A_i$ are essentially
tangent vector fields on $S_N^2$.

The tangent space to $\cO(\Xi)$ at a point $C$ is
isomorphic to $T_C\cO\cong\mun(\cN\,)/\mr$, where
$\mr=\mun(N_+)\times\mun(N_-)$ is the stabilizer subalgebra of
$\Xi$. This identification is equivariant with respect to the natural
adjoint action of the Lie group $U(\cN\,)$. Explicitly, tangent
vectors to $\cO(\Xi)$ at $C$ have the form\footnote{To streamline
  notation, we will not write explicitly the local dependences of
  fields and operators defined at points $C\in\cO$.}
\be
V_\phi = \ii[C,\phi]
\label{tangentvectors}
\ee
for any hermitian element $\phi \in \mun(\cN\,)/\mr$,\footnote{With
  our conventions, the vector fields \eq{tangentvectors} are real.}
which are just the generators of the unitary group $U(\cN\,)$ acting
on $\cO(\Xi)$ by the adjoint action. These actually describe vector
fields on the entire orbit space $\cO(\Xi)$. Here and in the following
we use the symbol $C$ to denote both elements of $\cO(\Xi)$, as well
as the matrix of overcomplete coordinate functions on $\cO(\Xi)$
defined using the embeddings
$\cO(\Xi)\hookrightarrow\mun(\cN\,)\hookrightarrow\C^{\cN^2}$.

\subsubsection*{\it The map $\cJ$}

Following~\cite{CP2paper}, we can make the description of the tangent
space to $\cO$, spanned by the vectors $V_\phi$, more
explicit as follows. Consider for $C \in\cO$ the map
\be
\cJ\,:\, \mun(\cN\,)~ \longrightarrow~ \ms\mun(\cN\,)
\ee
defined by
\beq
\cJ(\phi) = \mbox{$\frac1N$}\,V_\phi=
\mbox{$\frac{\ii}N$}\, \big[C\,,\,\phi\big] \ .
\label{calJmapdef}\eeq
Using \eq{constraint} one finds that it satisfies
\be
\cJ^3 = -\cJ
\label{cJprojprop}\ee
and hence amounts to suitable projectors.
 Moreover, the map $\cJ$ is
an antihermitian operator with respect to the invariant Cartan-Killing
inner product $\Tr(\phi \, \psi)$ on $\mun(\cN\,)$, since
\be
\Tr\big(\phi\, \cJ( \psi)\big) =  \mbox{$\frac{\ii}{N}$}\,
\Tr\big(\phi \,[C,\psi]\big)
  = - \mbox{$\frac{\ii}{N}$}\,\Tr\big([C,\phi]\,\psi\big) = -
\Tr\big(\cJ(\phi)\, \psi\big) \ .
\label{cJantiherm}\ee
The map $\cJ$ will play an instrumental role in this paper and its
geometrical properties will be studied in more detail in the next
section.

Here we simply note the meaning of $\cJ$ in the commutative limit
$N\to\infty$. In component form with $\phi = \phi_\mu
\otimes\sigma^\mu$, it acts as\footnote{Throughout,
  the notation $\approx$ will always mean an equality which is valid
  in the large $N$ commutative limit.}
\bea
 \cJ(\phi)&\approx& -\mbox{$\frac{\ii}N$}\,
\big[\phi_\mu \otimes\sigma^\mu\,,\,C_j\otimes\sigma^j\big]
\nonumber\\[4pt] &\approx&
-\mbox{$\frac{\ii}N$}\, \big[\phi_\mu \,,\,C_j\big]
\otimes\sigma^\mu\,
\sigma^j + \mbox{$\frac{\ii}N$}\,\phi_\mu\,
 C_j\otimes\big[\sigma^\mu\,,\,\sigma^j\big]
\label{calJcomponent}\eea
where we have set $C_0 \approx \frac 12~\one_N$ in the large $N$ limit
as will be justified below. Thus at large $N$ this reduces to
\beq
 \cJ(\phi)\approx O\big(\mbox{$\frac1N$}\big) -
\epsilon^{ij}{}_{k}\,\phi_i\,x_j \otimes \sigma^k
\eeq
for ``almost'' commutative functions describing the gauge field
fluctuations $A_\mu$. Here $\xi_i\approx\frac N2\,x_i$ define homogeneous
coordinates $x_i$ on the sphere. This result means that if we interpret
$\phi_i$ as a three-component vector field on the fuzzy sphere,
including radial components, then the operator $\cJ$ vanishes
on the normal component and essentially coincides with the complex
structure for tangential fields on the K\"ahler manifold $S^2$. In
particular, the image of $\cJ$, i.e. the space of tangent vectors  
\eq{tangentvectors} to $\cO(\Xi)$ or 
small variations of the gauge field, indeed admits two independent
field degrees of freedom. This implies that the orbit \eq{orbit-2}
describes two tangent vector fields on $S^2_N$. Hence the tangent
space to $\cO$ can be interpreted precisely the space of tangent
vector fields on the fuzzy sphere. 
This nicely reflects the affine nature of the space of gauge fields.

\subsubsection*{\it Nonabelian gauge theory}

The generalization to nonabelian $U(n)$ gauge theory is very
simple. One now takes 
\be
\cN = 2 n \,N
\label{cN2nN}\ee
and enlarges the matrix \eq{Xi-collective} to $\Xi\otimes\one_n$
(which we continue to denote as $\Xi$ for ease of notation). The
configuration space is given by the $U(\cN\,)$ orbit (\ref{orbit-2})
with $C^2 = \frac{N^2}4~\one_\cN$ and
\be
\Tr(C) = n\,N \ .
\label{UnTrconstr}\ee
Then $C$ has eigenvalues $\pm \,\frac N2$ of respective multiplicities
$n\,(N\pm1)$. The configuration space
\beq
\cO = U(2n\,N)/U(n\,N_+)\times U(n\,N_-)
\label{nonaborbit}\eeq
describes $\mun(n)$ -- valued gauge fields on $S^2_N$. Its dimension is
given by
\be
\dim(\cO) = 2 n^2\,\left(N^2-1\right) \ .
\label{dim-nonabelian}
\ee
 The gauge group
is now given by $G = U(nN)$, and acts on the covariant coordinates 
 $C_i = \xi_i\otimes \one_n + A_i, 
\,\, C_0 = \frac 12\, \one_{nN} + A_0$
as $C_\mu \to U^{-1} C_\mu U$. This leads to the expected
transformation law for the $\mun(n)$ -- valued gauge fields $A_i$.
The corresponding gauge algebra is now 
$\mg := \mun(n N)\subset\mun(\cN\,)$,
consisting of elements of
the form $\phi =\phi_0 \otimes\sigma^0 \, \in\, \mg$.

\subsection{The Yang-Mills action\label{YMAction}}

Consider the action
\be
S=S(C) := \mbox{$\frac{N}g$}\,\Tr\big(C_0-\mbox{$\frac
  12$}~\one_{n\,N}\big)^2 
\label{YM-action}
\ee
for $C \in \cO$, which is invariant under the group of gauge 
transformations $G$ as well as global $SU(2)$ rotations. 
We claim that it reduces in the 
commutative limit $N \to \infty$ to the
usual Yang-Mills action on the sphere $S^2$, and
can therefore be taken
as a definition of the Yang-Mills action on the fuzzy sphere
$S^2_N$. We establish this explicitly below in the abelian case
$n=1$, the extension to general $n$ being obvious.

Consider the three-component field strength~\cite{matrixsphere}
\bea
F_i &:=& \ii\epsilon_{i}{}^{jk}\,C_j\, C_k + C_i  \nn\\[4pt]
 &=& \ii\epsilon_{i}{}^{jk}\,[\xi_j, A_k] + 
\ii\epsilon_{i}{}^{jk}A_j\, A_k + A_i \ 
\label{fieldstrength-3}
\eea
where $C_i = \xi_i + A_i$ as in \eq{covcoordsdef}.
To understand its significance, consider the ``north pole'' of $S_N^2$
where $\xi_3 \approx \frac N2 \,x_3 = \frac N2~\one_N$ (with unit radius),
and one can replace the operators
\be
\ii\,\ad_{\xi_i} \;\;\longrightarrow \;\;
-\varepsilon_{i}{}^{j}\,\partial_j:=
-\varepsilon_{ij}\,\mbox{$\frac{\partial }{\partial x_{j}}$}
\ee
in the commutative limit for $i,j = 1,2$. Hence upon identifying the
commutative gauge fields $A^{\rm cl}_i$ through
\be
A^{\rm cl}_i = -\varepsilon_{i}{}^{j}\, A_j \ ,
\ee
the ``radial'' component $F_3$ of the field strength
\eq{fieldstrength-3} reduces in the commutative limit to the standard
expression
\be
F_3 \approx \partial_1 A_{2}^{\rm cl}-\partial_2 A_{1}^{\rm cl} 
+ \ii \big[A_1^{\rm cl}\,,\,A_2^{\rm cl}\big] \ .
\ee
The constraint \eq{C2} now implies
\bea
F_i  + \big\{C_0-\mbox{$\frac 12$}~\one_N\,,\,C_i\big\} =
F_i  + \big\{A_0\,,\,C_i\big\} &=& 0 \ , \nn\\
\{\xi_i, A^i\} + A_0 + A_i\, A^i + A_0\, A_0 &=& 0 \ .
\label{constraint-F}
\eea
Since only configurations with $A_0 = O(\frac 1N)$
have finite action \eq{YM-action} and $\xi_3$ is of order $N$, this
implies that  $A_3$, $F_1$ and $F_2$ are of order $\frac 1N$ at the north
pole, while  $A_{1}$ and $A_2$ can be finite of order $1$.  In
particular, only the radial component $F_3$ survives the $N \to
\infty$ limit, with 
\be
F_3 = -\{A_0,C_3\} \approx -N\, A_0 \ .
\label{rho-F}
\ee
This analysis can be made global by considering the ``radial'' field
strength  $F_r = x^i\,  F_i$, which reduces to the usual field
strength scalar on $S^2$. The action \eq{YM-action} thus indeed 
reduces to the usual Yang-Mills action in the commutative limit
with dimensionless gauge coupling $g$, giving
\be
S \approx \frac 1{N\,g} \,\Tr (F_r)^2 \approx \frac 1{4\pi\, g}
\,\int_{S^2}\,\ddd\Omega~ (F_r)^2 \ .
\ee

\subsection{Symplectic geometry of the configuration
  space\label{SymplStruct}}

The standard Kirillov-Kostant construction makes the orbit space
(\ref{orbit-2}) into a symplectic manifold~\cite{BGVbook}. Given two
tangent vector fields
$V_\phi, V_\psi$ as above with
$\phi,\psi\in\mun(\cN\,)$, the symplectic two-form
$\omega\in\Omega^2(\cO)$ is defined locally through its pairing with
the bivector $V_\phi\wedge V_\psi$ as
\be
\langle\omega,V_\phi\wedge V_\psi\rangle = \ii\Tr\big(C\,[\phi,\psi]
\big) \ .
\label{symplectic-form}
\ee
Using trace manipulations it is easy to see that the kernel of this
pairing coincides with the stabilizer algebra $\mr$, and hence it is
nondegenerate on $\cO(\Xi)$. 
We will derive below 
an explicit form of $\omega$ \eq{omegaexpl}, which allows 
to verify directly the well-known fact that $\omega$ is closed,
\be
\ddd\omega =0 \ .
\label{omegaclosed}\ee
Thus $\omega$ indeed defines an invariant symplectic structure on
$\cO(\Xi)$.

The tangent vectors $V_\phi$ are hamiltonian vector fields, and we
claim that their generator is given by
\be
H_\phi = \Tr(\phi\, C)
\label{Hphigen}\ee
for $\phi\in\mun(\cN\,)$. Indeed, then $\ddd H_\phi = \Tr(\phi~\ddd C)$,
and by using the dual evaluation 
\beq
\langle \ddd C,V_\phi\rangle =\ii[C,\phi]
\label{dualeval}\eeq
one has
\bea
\langle \ddd H_\phi,V_\psi\rangle &=& \ii\Tr\big(\phi\,
[C,\psi]\big)\nn\\[4pt]
 &=& -\ii\Tr\big(C\,[\phi,\psi]\big)\nn\\[4pt] &=&
 -\langle\omega,V_\phi\wedge
V_\psi\rangle ~=~ -\langle \iota_{V_\phi} \omega,V_\psi\rangle
\eea
where $\iota_{V_\phi}$ denotes contraction with the vector field
$V_\phi$. Thus
\be
\ddd H_\phi = -\iota_{V_\phi} \omega 
\label{hamiltonian}
\ee
as claimed. This means that the hamiltonian function (\ref{Hphigen})
defines a periodic flow generated by the action of a one-parameter
subgroup $C\mapsto \e^{\ii t\,\phi}\,C~\e^{-\ii t\,\phi}$,
$t\in\R$. The corresponding equivariant moment map
$\mu:\cO(\Xi)\to\mun(\cN\,)^\vee$ is the inclusion map which has the
pairings
\be
\big\langle\mu(C)\,,\,\phi\big\rangle = H_\phi \ ,
\label{momentmap}
\ee
and it defines a representation of the Lie algebra $\mun(\cN\,)$ through
the Poisson algebra corresponding to $\omega$.

For gauge transformations $\phi = \phi_0 \otimes\sigma^0$, the
moment map $\mu$ reduces to
\beq
\big\langle\mu(C)\,,\,\phi\big\rangle=2\Tr\big(\phi_0\,C_0\big)= 
\Tr\big(\phi_0 \,(\one_{n\,N} +2A_0)\big) \ .
\label{momentmapred}\eeq
In the commutative limit and for abelian gauge fields $n=1$, this
becomes
\beq
\big\langle\mu(C)\,,\,\phi\big\rangle\approx\Tr(\phi_0)-
\frac2N\,\Tr(\phi_0\, F_r) \approx -\frac1{2\pi}\,
\int_{S^2}\,\ddd\Omega~ \phi_0 \,F_r \ 
\label{momentmap-3}
\ee
up to an irrelevant shift,
which is just the anticipated moment map for Yang-Mills theory on the
classical sphere~\cite{witten1}. Given the appropriate symplectic
structure and moment map on the gauge field configuration space $\cO$,
the nonabelian localization principle for two-dimensional Yang-Mills
theory can be applied for the action constructed as the square of the
moment map. This is precisely the Yang-Mills action on $S^2_N$ given
in~\eq{YM-action}. The constant term $\frac 12~\one_{n\,N}$ is just the
first Chern number of a background gauge field configuration and is
of no significance for this discussion. This procedure
will be worked out in detail in Section~\ref{NonabLoc}.

\subsubsection*{\it More about the symplectic form}

For later use, we will now derive some properties of the symplectic
form introduced in (\ref{symplectic-form}). Consider the
$\ii\mun(\cN\,)$-valued one-form on $\cO(\Xi)$ given by
\be
\theta := C^{-1} ~\ddd C \ .
\label{thetaCMdef}\ee
Given the constraints (\ref{constraint}) and using $\ddd C^2 =0$, this
can be rewritten as
\beq
\theta = \mbox{$\frac 4{N^2}$}\, C ~\ddd C = \mbox{$\frac 2{N^2}$}\,
 [C,\ddd C] \ .
\label{thetaCrel1}\eeq
It obeys the constraints
\be
\ddd\theta + \theta^2 =0 \qquad \mbox{and} \qquad \Tr(\theta) =0 \ .
\label{thetaconstrs}\ee
Thus $\theta\in\Omega^1(\cO,\ii\mun(\cN\,))$ is essentially the
canonical invariant Maurer-Cartan one-form, with the additional
property
\be
[C,\theta] = -2 \cJ^2 (\ddd C) = 2~\ddd C
\label{MCformprop1}\ee
where we have used the fact that $\ddd C$ is tangent to the orbit
space and applied the projection property (\ref{cJprojprop}). In
particular, along with the fact that $C^2$ is constant, this implies
that
\beq
C\, \theta + \theta\, C =0 \ .
\label{MCformprop2}\eeq
Using again the constraint (\ref{constraint}), the symplectic two-form
\eq{symplectic-form} can be written as
\be
\omega = -\mbox{$\frac \ii{2 N^2}$}\,\Tr \big(C\,[\ddd C, \ddd C] \big) 
= \mbox{$\frac \ii4$}\,\Tr \big(C\,\theta^2\big) \ .
\label{omegaexpl}
\ee
To see this, we substitute  this expression 
using (\ref{cJantiherm}) and (\ref{cJprojprop}) into
\bea
\langle\omega,V_\phi\wedge V_\psi\rangle &=&
- \mbox{$\frac\ii{N^2}$} \,\Tr\big(C\,[\,[C,\phi]\,,\,[C,\psi]\,]
\big)\nn\\[4pt] 
&=& \ii \Tr\big(C\,[\cJ(\phi),\cJ(\psi)]\big) \nn\\[4pt]
&=& \ii \Tr\big([C,\cJ(\phi)]\,\cJ(\psi)\big) \nn\\[4pt]
& =& -N\,\Tr\big(\cJ^3(\phi)\, \psi\big) \nn\\[4pt]
&=& N\, \Tr\big(\cJ(\phi)\, \psi
\big)\nn\\[4pt] & =&\ii \Tr\big([C,\phi]\, \psi\big)
~=~ \ii\Tr\big(C\,[\phi, \psi]\big)
\label{omegaexpl-2}
\eea
for any $\phi,\psi \in \mun(\cN\,)$,
which coincides with the definition (\ref{symplectic-form}). Using
\eq{MCformprop1} and \eq{MCformprop2}, this
identity gives a simple proof of the closure property
(\ref{omegaclosed}) as
\be
\ddd\omega = \mbox{$\frac \ii4$}\,\Tr \big(\ddd C~\theta^2\big)
= -\mbox{$\frac \ii{8}$}\,\Tr\big( [\theta,C]\,\theta^2\big)
= 0 \ .
\ee

\bigskip

\section{The classical configuration space\label{ClassSols}}

In this section we will investigate in detail the space of classical
solutions of $U(n)$ gauge theory on the fuzzy sphere $S_N^2$ defined
by the action (\ref{YM-action}). Understanding this space will be
crucial for the exact solution of the quantum gauge theory, which as
we will see in the next section is given exactly by its semiclassical
expansion. We will first classify the solutions to the classical
equations of motion, over which the partition function will be
summed. Among these solutions we will find a variety of fluxons and,
as in the case of gauge theory on the noncommutative torus, only a
very small subset of all two-dimensional noncommutative instantons on
$S_N^2$ map into the usual instantons of Yang-Mills theory on $S^2$ in
the commutative limit $N\to\infty$. We will then thoroughly describe
the local symplectic geometry of the configuration space $\cO$ near
each critical point of the Yang-Mills action, as symplectic integrals
over these neighbourhoods will produce the required quantum
fluctuation determinants in the semiclassical expansion.

\subsection{Classical solutions\label{CritPoints}}

The critical points of the Yang-Mills action \eq{YM-action} are easy
to find. Since the most general variation of a gauge field $C \in
\cO$ is given by $\d C = [C,\phi]$, by varying \eq{YM-action} one
finds that the critical points satisfy
\be
0 =  \Tr\big(\d C_0~ (C_0-\mbox{$\frac 12$}~\one_{n\,N})\big) =
\Tr\big([C,\phi] \,C_0\big) = \Tr\big(\phi \,[C_0, C]\big)
\ee
for arbitrary $\phi\in\mun(\cN\,)/\mr$. They are therefore given
by solutions of the equation $[C_0,C]=0$, which agrees
with the known saddle-points in the matrix model formulation
of~\cite{matrixsphere}. This equation is equivalent to
\beq
[C_0,C_i]=0
\label{eom}\eeq
which together with \eq{C2} implies that 
\bea
[C_i, C_j] &=& \ii \epsilon_{ij}{}^{k} \,(2 C_0)\; C_k \ , \nn\\[4pt]
C_0^2 &=& \mbox{$\frac{N^2}{4}$}~\one_{n\,N} - C_i\,C^i \ .
\label{critical-YM}
\eea
For solutions with $C_0\neq 0$, we can use (\ref{eom}) to define
\beq
L_i=\frac1{2C_0}\,C_i
\label{Lidef}\eeq
and rewrite (\ref{critical-YM}) as
\bea
[L_i, L_j] &=& \ii \epsilon_{ij}{}^{k} \, L_k \ , \nn\\[4pt]
L_i\,L^i&=& \big(\mbox{$\frac{N^2}{4C_0^2} - \frac
  14$}\big)~\one_{n\,N} \ .
\label{di-su2}
\eea

These equations mean that the critical points of the Yang-Mills action
correspond to (isomorphism classes of) $(n\,N)\times(n\,N)$ unitary
representations of the isometry group $SU(2)$, i.e. homomorphisms
$\pi_{n\,N}:SU(2)\to U(n\,N)$. Up to isomorphism, for each integer $p\geq1$
there is a unique irreducible $SU(2)$ representation $(p)$ of dimension
$p$. Therefore, there is a one-to-one correspondence between classical
solutions and ordered partitions $(n_1,\dots,n_k)$ of the integer
$n\,N=n_1+\dots+n_k$, with $n_i$ the dimension of the $i$-th
irreducible subrepresentation in the representation $\pi_{n\,N}$
characterizing the given critical point. Each such classical solution
breaks the $U(n\,N)$ gauge symmetry locally to the centralizer
$\prod_i\,U(k_i)$ of the homomorphism $\pi_{n\,N}$, 
where $k_i$ denotes the multiplicity of the blocks. They can be
seen~\cite{matrixsphere} to give precisely the usual two-dimensional
instantons for $U(n)$ Yang-Mills theory on $S^2$. These solutions also
agree with those that can be interpreted as configurations of
D0-branes inside D2-branes~\cite{Alekseev:2000fd}, although the ones
which will survive the large $N$ limit are different.

Therefore, each critical point is labelled (up to gauge
equivalence) by the set of dimensions $n_i$ of the irreducible
representations, supplemented with a sign $s_i$ which is defined by $s_i =
{\rm sgn}(C_0(n_i)) = \pm\, 1$ (in that representation) when $C_0(n_i)
\neq 0$ and $s_i =0$ if $C_0(n_i)=0$. We can thereby label the
{\it critical surfaces}, i.e. the connected components of the moduli
space of classical solutions in $\cO$, as
\be
\cC_{(n_1,s_1),\dots,(n_k,s_k)} \qquad \mbox{with} \qquad
n_i \in \N \quad \mbox{and} \quad s_i \in \{\pm \,1,0\}
\ee
with the constraints
\be
1 \leq n_1 \leq n_2 \leq \cdots\leq n_k \ , \quad
\sum_{i=1}^k\, n_i = n\,N \quad \mbox{and} \quad
\sum_{i=1}^k\, s_i =n \ ,
\label{partitionconstrs}\ee
and $s_i =0$ only if $n_i=1$. Any non-trivial irreducible
representation with $n_i>1$ and $C_0\neq 0$ gives a contribution $\pm N$
to the trace $\Tr(C)$, which must be balanced in order to satisfy the
eigenvalue multiplicity constraint (\ref{UnTrconstr}). This is the
role of the condition $\sum_i\, s_i =n$ in
(\ref{partitionconstrs}). Note that one can change the sign of any
individual irreducible representation.

The meaning of the blocks $(n_i,s_i)$ can be described as follows:
\begin{itemize}
\item  \underline{$s_a = \pm\, 1~$:} ~ In this case $C_0 \neq 0$, and
  hence $\|C_0\| > \frac 12$ due to \eq{di-su2}. These solutions come
  with two signs. Note that any irreducible representation with small
  dimension will be highly suppressed in the large $N$ limit.
The most extreme case is a sum of trivial representations, with $n_a
=1$, for which
\be
C_i =0 \qquad \mbox{and} \qquad C_0(n_a=1)= s_a\,\mbox{$ \frac N2$} \ .
\ee
\item \underline{$s_a=0~$:} ~ In this case $C_0=0$ and $n_a=1$, which  
implies that $C_i = c_i$ with $c_i\in\R$ and
$\frac{N^2}{4} = c_i\,c^i$. These solutions are also suppressed at
large $N$ but less so than those with $C_i=0$ above. 
They correspond to {\em fluxons}~\cite{Gross:2000ss} whose positions
on $S^2$ are determined by the vector $c_i$.
\end{itemize}
Note that each such saddle-point (or more generally any gauge field
configuration $C$) defines a projective module over the fuzzy sphere
algebra $S^2_N$, obtained by writing $C$ in $2n\times 2n$ block-matrix
form. The module then corresponds to a projector 
$\Pi_{(n_1,s_1),\dots,(n_k,s_k)} \in{\rm Mat}_{2n}(S^2_N)$. Let us
describe some of these critical points explicitly.

\subsubsection*{\it Ground state}

The vacuum solution has $k=n$ and is given by the critical surface
$\cC_{(N,1),\dots,(N,1)}$, which implies that $C_0=\frac
12~\one_{n\,N}$. It follows that $C_i\,C^i=\frac{N^2-1}{4}~\one_N$,
which is the quadratic Casimir invariant of the $N$-dimensional
irreducible representation of $SU(2)$. Using a suitable $U(n\,N)$
gauge transformation, it can be written as
\be
C_i = \xi_i\tens \one_n
\label{vacsolnonab}\ee
and we recover the original coordinates of the fuzzy sphere
$S_N^2$. This is equivalent to the vanishing curvature condition $F
=0$. In the abelian case $n=1$, an application of Schur's lemma shows
that the only matrix which commutes with $C$ is the constant matrix
and so the gauge group $U(N)$ acts freely on the moduli space of
vacuum solutions, corresponding simply to a change of basis in this
case. For $n>1$ the solution is a direct sum of $n$ identical
representations. This commutes with the action of $\mun(n)$, and so
now the gauge group $U(n\,N)$ contains a non-trivial stabilizer. The
moduli space of flat connections is therefore isomorphic to the smooth
manifold $U(n\,N)/U(n)$ in the nonabelian case. Note that any
configuration near the vacuum, with small but finite action, is given
by a small deformation of an irreducible $SU(2)$ representation
describing $S_N^2$, and in particular the gauge field fluctuations
$A_\mu$ are ``small''. It is in this sense that the quantum gauge
theory will describe a fluctuating theory of noncommutative fuzzy
sphere geometries.

\subsubsection*{\it Fluxons}

At the other extreme, if $C_0$ has several zero eigenvalues,
i.e. several fluxons, the situation is much more complicated. For
example, when $C_0=0$ and $n=1$ we obtain a fuzzy version of the
moduli space of constant curvature connections in genus~$0$ provided
by the critical surface
\be
\mu^{-1}(C_0=0)=\big\{C_i\in \mun(N)~\big|~C_i\,C^i=
\mbox{$\frac{N^2}4$}~\one_N~,~[C_i,C_j]=0\big\}
\label{mu0}\ee
along with the condition (\ref{UnTrconstr}) on the multiplicities of
the eigenvalues of $C_i\otimes\sigma^i$. The action of the $U(N)$
gauge group on (\ref{mu0}) can be used to simultaneously diagonalize
the three matrices $C_i$. The Marsden-Weinstein symplectic reduction
of the orbit space $\cO(\Xi)$ is then essentially a symmetric product
orbifold of the classical sphere $S^2$ given by
\be
\cM_0:=\mu^{-1}(C_0=0){/\!/}\,U(N)\cong{\rm Sym}^N\big(S^2\big) \ ,
\label{calM0}\ee
where ${\rm Sym}^N(S^2):=(S^2)^N/\mS_N$ and the quotient by the Weyl
group $\mS_N\subset U(N)$ is the residual gauge symmetry acting by
permutations of the real eigenvalues of the hermitian matrices $C_i$
representing the positions of the fluxons on $S^2$, which are
indistinguishable. The fluxon moduli space $\cM_0$ contains
orbifold singularities arising from the fixed points of the
$\mS_N$-action on $(S^2)^N$, which occur whenever two or more fluxon
locations coincide. This is analogous to the vacuum solution of
two-dimensional $U(N)$ gauge theory on a noncommutative torus wherein
the moduli space of constant curvature connections is the symmetric
product orbifold ${\rm Sym}^N(T^2)$~\cite{szabo}, and there is a natural
correspondence between two-dimensional noncommutative instantons and
fluxons~\cite{Griguolo:2004jp}. In the present case the $U(N)$ action
on the fluxon configuration space (\ref{mu0}) also has additional
fixed points. Note that the restriction of the symplectic two-form
(\ref{omegaexpl}) to the moduli space $\cM_0$ is given by
\be
\omega\big|_{\cM_0}=-\frac{4\ii}{N^2}\,\sum_{a=1}^N\,\epsilon^{ijk}
\,c_i^a~\ddd c_j^a\wedge\ddd c_k^a
\label{omegaM0}\ee
where $c_i^a\in\R$ are the eigenvalues of $C_i$ with
$\sum_i\,(c_i^a)^2=\frac{N^2}4$ for each $a=1,\dots,N$. With the
usual embedding of the two-sphere $S^2\hookrightarrow\R^3$, this is
just the standard round symplectic two-form on the K\"ahler
manifold~$(S^2)^N$.
Each fluxon contributes a suppression factor $\e^{-\frac{N}{4g}}$ due
to  \eq{YM-action}. 

\subsubsection*{\it Instantons on $S^2$}

The configurations which will dominate the path integral in the large
$N$ classical limit are the low-energy solutions with small
actions. These are solutions with $n$ partitions and critical surfaces
$\cC_{(n_1,1), \dots, (n_n,1)}$ with $n_i \approx N$. They correspond
to the usual instantons of $U(n)$ gauge theory on $S^2$ 
with vanishing $U(1)$ flux, as shown in~\cite{matrixsphere}. 
These solutions may also contain additional
fluxons, which behave like localized flux tubes which ensure that the
total $U(1)$ flux vanishes. Their 
contributions are suppressed by factors of at least
$\e^{-\frac{N}{4g}}$, however they do contribute in the
double scaling, quantum plane limit wherein $S_N^2$ becomes
noncommutative $\R^2$~\cite{Chu:2001xi,Behr:2005wp}.

\subsubsection*{\it Monopoles}

As shown in~\cite{matrixsphere,Karabali:2001te}, an 
irreducible representation with $n_i = N-m_i$ corresponds to the gauge
field of a monopole with magnetic charge $m_i\in\Z$. 
Configurations with non-trivial $U(1)$ monopole number can 
therefore be obtained by relaxing the constraint
(\ref{UnTrconstr}) and replacing it by
\be
\Tr(C) = n\,N - {\sf c}_1
\label{tracemodify}\ee
where ${\sf c_1}=\sum_i\,m_i \in \Z$ is the first Chern
number. 
In order to maintain the constraint $C^2 = \frac{N^2} 4~\one_\cN$,
the matrix dimension (\ref{cN2nN}) must then be replaced 
with $\cN=2(n\,N-{\sf c}_1)$. 

Some of these nontrivial $U(1)$ bundles are realized within the
original configuration space \eq{nonaborbit}, 
in the presence of trivial blocks with $n_a=1, s_a = \pm\, 1$. 
For example, in the abelian case $n=1$ the solutions in
$\cC_{(N-2,1),(1,1),(1,-1)}$ are naturally interpreted as monopoles with 
charge $m=2$. The blocks $(1,\pm\, 1)$ have vanishing field
strength $F_i=0$, and are naturally interpreted as Dirac strings.
They are suppressed by  factors of at least
$\e^{-N^3/g}$.
Replacing the trivial blocks with fluxons leads to vanishing 
global $U(1)$ flux as discussed above.

\subsection{The classical action\label{ClassAction}}

The values of the Yang-Mills action (\ref{YM-action}) on the classical
solutions obtained in Section~\ref{CritPoints} above will determine
the classical contributions to the path integral in the next
section. The action at these critical points can be
evaluated as follows. Note that for each $p$-dimensional irreducible
representation $L_i$ of the isometry group $SU(2)$, one has
$L_i\,L^i=\frac{p^2-1}4~\one_p$ and hence from \eq{di-su2} it follows
that
\be
\mbox{$\frac{N^2}{p^2}$}~\one_p =  4C_0(p)^2
\ee
on that representation, so that $C_0(p) = \pm \,\frac{N}{2p}~\one_p$.
Consider the reduced Yang-Mills action 
\beq
S':=\mbox{$\frac Ng$} \,\Tr\big(C_0^2\big) = S + \mbox{$\frac Ng$}\,
\Tr(C_0) - \mbox{$\frac N{4g}$} \Tr(\one_{n\,N})=S +  \mbox{$
\frac {n\,N^2}{4g}$}
\label{YM-action-prime}
\eeq
which is somewhat easier to manipulate than $S$. For a dominant
solution with critical surface $\cC_{(n_1,1),\dots, (n_n,1)}$ and
$n_i >1$, the action $S'$ is given by
\be
S'\big((n_1,1)\,,\,\dots\,,\, (n_n,1)\big) = \frac N{g}\,
\sum_{i=1}^n\, n_i\, \frac{N^2}{4 n_i^2} 
 = \frac{N^3}{4g}\, \sum_{i=1}^n\, \frac{1} {n_i} \ .
\label{action-eval}
\ee
While possible fluxon blocks with $n_i=1$ 
do not contribute at all to $S'$, they do contribute 
$\frac{N}{4g}$ to the original action $S$ (\ref{YM-action}). 
Their total contributions to $S$ is proportional
to the fluxon charge, i.e. the total number of blocks with $n_i=1$,
and agrees with the usual fluxon action~\cite{Gross:2000ss} in the
quantum plane limit of $S_N^2$~\cite{Chu:2001xi}.

The dominant configurations in the classical limit are therefore those
with
\be
n_i = N - m_i\qquad \mbox{and} \qquad \sum_{i=1}^n\, m_i=0
\label{nidomclass}\ee
with small $m_i\in\Z$, for which
\be
C_0(n_i) = \mbox{$ \frac{N}{2(N-m_i)}~\one_{n_i} \approx \frac 12
  \,\big(1+\frac{m_i}N\big)~\one_{n_i}$} \ .
\ee
Note that then
\beq
\Tr(C_0) = \sum_{i=1}^n\, (N-m_i)\, \frac{N}{2(N-m_i)} = \frac{n\,N}2
\label{TrC0dom}\eeq
as required. It follows that
\beq
S\big((n_1,1)\,,\,\dots\,,\, (n_n,1)\big)
\approx\frac Ng\, \sum_{i=1}^n\, (N-m_i)
\,\left(\frac{m_i}{2N}\right)^2 +O\big(\mbox{$\frac1N$}\big)
\approx\frac 1{4g}\, \sum_{i=1}^n\,  m_i^2 \ ,
\label{action-eval-2}
\eeq
which is the usual expression~\cite{Minahan:1993tp,Gross:1994mr} for
the classical action of $U(n)$ Yang-Mills theory on the sphere $S^2$
with trivial gauge bundle evaluated on the two-dimensional instanton
on $S^2$ corresponding to a configuration of $n$ Dirac monopoles of
magnetic charges $m_i\in\Z$. Non-trivial gauge bundles over $S^2$ of
first Chern class ${\sf c}_1\in\Z$ are obtained by modifying the trace
constraint as in \eq{tracemodify}.

\subsection{Local symplectic geometry of the configuration
  space\label{LocalGeomO}}

We will now develop the local symplectic geometry of the
configuration space of gauge fields near each Yang-Mills critical
point. This is done by analysing in more detail the map
(\ref{calJmapdef}), satisfying (\ref{cJprojprop}). We want to find
a useful description of the tangent space $T_C \cO \cong{\rm
  im}(\cJ)$, i.e. of the local geometry of the orbit space
$\cO$. Since $\cJ$ is an anti-hermitian operator with respect to the
Cartan-Killing form on $\mun(\cN\,)$ (see (\ref{cJantiherm})), it
follows that the space $\mun(\cN\,)$ splits into two orthogonal
subspaces as
\be
\mun(\cN\,) = \ker(\cJ) \oplus\ker\big(\cJ^2+\one_\cN\big)
\label{ucN-decomp}
\ee
where $\ker(\cJ) = \mr=\mun(n\,N_+)\oplus\mun(n\,N_-)$ is the
stabilizer subalgebra, while $\ker(\cJ^2+\one_\cN) \cong T_C \cO$ is
the tangent space to the configuration space at $C\in\cO$. In
particular, $\cJ$ defines a complex structure on $T_C \cO$, 
and \eq{ucN-decomp} is just the Cartan decomposition 
of $\mun(\cN\,)$ corresponding to
the
symmetric space $\cO$. This follows immediately by noticing that the
involutive automorphism
\beq
{\sf j}\,:\, \mun(\cN\,) ~\longrightarrow~ \mun(\cN\,) \ , \qquad
\phi~\longmapsto~  C \,\phi\, C^{-1} 
\eeq
is $\one_\cN$ on $\ker(\cJ)$ and $-\one_\cN$ on $\ker(\cJ^2+\one_\cN)$
upon using the constraints (\ref{constraint}). Moreover, for any
$V_\phi,V_\psi\in T_C\cO$, from (\ref{omegaexpl}) one has
\be
\langle\omega,V_\phi\wedge V_\psi\rangle
=  \mbox{$\frac\ii{N^2}$}\, \Tr\big([C,V_\phi]\, V_\psi\big)
=  \mbox{$\frac1{N}$}\, \Tr\big(\cJ(V_\phi)\, V_\psi\big)
\label{symplectic-eval}
\ee
and
\be
\big\langle\omega\,,\,V_\phi\wedge \cJ(V_\psi)\big
\rangle  =  \mbox{$\frac1N$}\, \Tr(V_\phi\, V_\psi) \ ,
\label{Kahlerderiv}\ee
expressing the fact that the symplectic two-form $\omega$ makes the
configuration space $\cO$ into a K\"ahler manifold with respect to the
complex structure (\ref{calJmapdef}). All of these properties are just
standard features of hermitian symmetric spaces~\cite{helgason1}, as
will be exploited at length in this paper.

Consider the restriction of the map $\cJ$ to the gauge algebra
$\mg = \mun(n\,N)\subset\mun(\cN\,)$ containing elements of the form
$g=\phi\otimes\sigma^0$. Since $\cJ(\phi)$ is the infinitesimal gauge
transformation of the gauge field $C$ generated by $\phi$, it
describes the orbits of the gauge group $G = U(n\,N)$ acting on the
configuration space $\cO$, in $T_C\cO$. Generically this action is
free (apart from the trivial $\mun(1)$), 
but not for certain critical points. For example, for the vacuum
solution (\ref{vacsolnonab}) the subalgebra $\one_N\tens \mun(n)$
commutes with $C$. The higher critical points in the nonabelian case
generically have a smaller $\mun(1)^{n}$ centralizer algebra.

More precisely, consider the kernel of $\cJ$ at $C$ restricted to the
gauge algebra $\mg$,
\be
\ms:=\ker(\cJ) \cap {\mg} \ ,
\label{kerJmgdef}\ee
which is the subgroup of the gauge group that stabilizes
$C$. The elements $\phi \in \ms$ are orthogonal to $T_C\cO$ due to
\eq{ucN-decomp}. Hence $\mg$ decomposes into orthogonal subspaces
\be
\mg =  \ms \oplus \mg'
\ee
where $\mg'=\ms^\perp=:\mg\ominus\ms$ contains the ``proper'' gauge
transformations, acting freely near $C$. If $(n_1,\dots,n_n)$ is a
partition of the integer $n\,N$ which does not contain trivial
representations of $SU(2)$ (no fluxons), then $\mg'$ is the tangent
space to the corresponding critical surface
$\cC_{(n_1,1),\dots,(n_n,1)}\subset \cO$,
\be
\cC_{(n_1,1),\dots,(n_n,1)} \cong U(n\,N) / S \ ,
\label{globalcritsurface}\ee
where $S = \exp(\ms)$.

We claim that {\em the subspaces $\cJ(\mg)$ and $\mg$ are linearly
  independent.} For this, assume to the contrary that $\cJ(\mg)$ and
$\mg$ are linearly dependent, i.e. $\cJ(g) \in \mg$ for some $g \in
\mg$. This implies that $[C_i,g] =0$, and therefore
$[C_0^2,g]=0$ due to \eq{C2}. Restricting attention to
critical points $C$ for which the spectrum of $C_0$ is non-negative
(the others being strongly suppressed at large $N$), this implies that
$g$ commutes with the spectral projectors of $C_0$, and
hence also with $C_0$ itself. Together with $[C_i,g] =0$ it follows
that $\cJ(g)=0$. However, $\cJ(\mg)$ and $\mg$ need not be orthogonal
subspaces.

Generically one then has
\be
\cJ^2(\mg) + \cJ(\mg)  \subset T_C\cO \ .
\ee
The two subspaces are not orthogonal in general, since for $g_1,g_2
\in \mg$ one can compute the inner product
\bea
\Tr\big(\cJ^2(g_1)\, \cJ(g_2)\big) &=& \Tr\big(g_1\, \cJ(g_2)\big)
\nn\\[4pt] &=&-\mbox{$\frac{\ii}N$}\,\Tr\big(C\,
[g_1,g_2]\big)\nn\\[4pt] &=&-\mbox{$\frac1N$}\,\langle
\omega,V_{g_1}\wedge V_{g_2}\rangle~=~\mbox{$\frac{\ii}N$}\,
\Tr\big(g_1\, [C_0,g_2]\big)
\label{orthogornot}
\eea
which is non-vanishing in general. For the vacuum solution with
$C_0=\frac12~\one_{n\,N}$, it follows from this expression that the
subspaces are indeed orthogonal, and hence $\cJ^2(\mg) \oplus \cJ(\mg)
\subset T_C\cO$. In fact, one has
\be 
\cJ^2 (\mg) \oplus \cJ(\mg) = T_C\cO \qquad\mbox{if} \quad
C_0 =\mbox{$\frac12$}~\one_{n\,N}
\label{tangent-decomposition}
\ee 
which provides a useful description of the local geometry near the
global minimum. To see (\ref{tangent-decomposition}), note
first that in the abelian case $n=1$ one has $\ms=\mun(1)$, and
\eq{tangent-decomposition} then follows since $\dim (\cO) =
2(N^2-1) = 2 \dim (\mg')$. In the nonabelian case, for the vacuum state
the gauge stabilizer $\ms \cong\mun(n)$ has dimension $n^2$ and
hence $\dim(\cJ^2 (\mg') \oplus \cJ(\mg')) = 2n^2
\,N^2-2n^2 = \dim(\cO)$.

In general, the subspaces $\cJ(\mg) = \cJ(\mg\ominus\ms)$ and
$\cJ^2(\mg)$ are not linearly independent, and we can define
\be
E_0 := \cJ(\mg) \cap \cJ^2(\mg)
\ee
which is generically a non-trivial subspace. Define also the subspaces
$\mh, \tilde \mh \subset \mg\ominus\ms$ with the properties that 
\be
\cJ(\mh) = E_0 = \cJ^2(\tilde \mh) \ .
\ee
Since $\cJ^2(\mh) = -\cJ(\tilde \mh)$ implies that $\mh \subset
\tilde\mh \subset \mh$, we have
\be
\mh = \tilde \mh \qquad \mbox{and} \qquad \cJ(E_0) = E_0 \ .
\ee
We can accordingly decompose the gauge algebra $\mg$ into orthogonal
subspaces as
\be
\mg = \mg_1 \oplus \mh \oplus \ms \ .
\ee
Since $\cJ: \mh \to E_0$ is a bijection, there is a unique map
\beq
j\,:\, \mh~\longrightarrow~ \mh \qquad \mbox{with} \qquad
\cJ^2(h) =\cJ\big(j(h)\big)
\label{juniquemap}\eeq
for all $h\in\mh$ which satisfies $j^2 =-\one_{n\,N}$. Similarly, in
order to span the entire tangent space at $C\in\cO$ we generally have
to introduce another subspace $E_1$, with $ \cJ(E_1) =E_1$, 
which gives the general decomposition
\be
\cJ(\mg\ominus\mh) \oplus \cJ^2(\mg\ominus\mh)
\oplus E_0 \oplus E_1 ~=~ T_C\cO \ .
\label{TCOdecomp}\ee

\subsection{Explicit decomposition at Yang-Mills critical
  surfaces\label{ExplYMDecomp}}

We will now provide an explicit description of the various subspaces
appearing in the decomposition of the tangent space
(\ref{TCOdecomp}). Consider the Yang-Mills critical surfaces
$\cC_{(n_1,1),\dots,(n_n,1)}$ and suppose first that $n_1 \neq
\cdots\neq n_n$ are all distinct integers, corresponding to a
completely nondegenerate solution. The elements $\phi$ of the
subspace (\ref{kerJmgdef}) satisfy $[C,\phi] =0$. This implies that
$\phi$ respects the block decomposition described by the given partition
$(n_1,\dots, n_n)$, and is therefore proportional to $\one_{n_i}$ on
each block. These are thus $\mun(1)^n$ degrees of freedom. If
some $n_i$ are degenerate, this space is enhanced to 
\be
\ms = \mun(k_1) \times \cdots \times \mun(k_l)
\ee
for a critical surface with $C = \bigoplus_i\, C(n_i)\otimes
\one_{k_i}$ and $n_i$ all distinct. For the vacuum this is $\mun(n)$,
corresponding to the maximally degenerate solution, as in
Section~\ref{LocalGeomO} above.

We wish to work out the map $\cJ$ explicitly. For this, we decompose
\be
\phi = \left(\begin{array}{ccc} \phi_{11} & \phi_{12} & \dots \\
                                \phi_{21} & \phi_{22} & \dots \\
                                     \cdots &\cdots &\ddots
\end{array}\right)
\ee
where $\phi_{ij} \in (n_i) \otimes (n_j)$ and as before $(p)$ denotes 
the $p$-dimensional irreducible representation of $SU(2)$. In the
degenerate case, there is another factor corresponding to $\mun(k_j)$.
The non-orthogonality of $\cJ(\mg)$ and $\cJ^2(\mg)$ in
\eq{orthogornot} is now easily understood as being simply due to the
different $\mun(1)$ charges between the $SU(2)$ sectors of
$\ms$. Since $[C,C_0]=0$ at the Yang-Mills critical
surfaces, one has $\cJ([C_0,\phi]) =
[C_0,\cJ(\phi)]$. Thus the hermitian operator
\be
({\rm ad}_{\ii C_0})_{ij} = \ii C_0(n_i) - \ii C_0(n_j) = 
\ii \frac N2 \,\frac{n_j-n_i}{n_i\,n_j} =: \ii c_{ij}
\label{ad-C0-explicit}
\ee
acting on $\phi_{ij} \in (n_i) \otimes (n_j)$ commutes with
$\cJ$. This implies that we can decompose the subspaces
in \eq{TCOdecomp}   such as
$\cJ(\mh) = \cJ^2(\mh)=E_0$ into irreducible representations
of the operator ${\rm ad}_{\ii C_0}$, i.e. into the various $\mun(1)$
blocks. Restricted to the diagonal blocks $C_0(n_i)$ is proportional
to the unit matrix $\one_{n_i}$, so that $\Tr(\cJ(g_1)\, \cJ^2(g_2))
=0$ there as for the vacuum.

\subsubsection*{\it Global $SU(2)$ symmetry}

To proceed further, we need to exploit an additional symmetry that we
have neglected so far, the global rotation group $SU(2)$. Recall from
Section~\ref{CritPoints} above that each saddle-point defines a
representation of $SU(2)$ acting on the representation space $V \cong
\C^{n\,N}$ as (\ref{Lidef}), and trivially on potential fluxon
components. In the abelian case $n=1$, this induces via the adjoint
action the rotations of functions $f\mapsto J_if=[L_i,f]$ in $S^2_N
\cong V \otimes \obar{V}$, but it is a somewhat different symmetry for
the nonabelian instantons. Let us decompose $V$ into irreducible
representations as
\be
V = \bigoplus_{i=1}^n\, (n_i) \ .
\ee
This representation can be extended to the module $V \otimes \C^2$ for
the action of the operators
\be
J_i = L_i + \mbox{$\frac 12$}\, \sigma^i
\ee
which by construction commute with $C$,
\be
[J_i,C] =0 \ ,
\ee
on the critical surfaces. This follows from the fact that
$C_i\otimes\sigma^i$ is an intertwiner for the action of $J_i$ on 
\be
V \otimes \C^2 = \Big(\,\bigoplus_{i=1}^n\, (n_i+1)\Big)~ 
 \oplus~ \Big(\,\bigoplus_{i=1}^n\,(n_i-1)\Big)  =: V^+  \oplus  V^-
\label{C-decomp-1}
\ee
and $C$ has eigenvalues $\pm \,\frac N2$ on the component subspaces
$V^\pm$. This enables one to decompose $C$ further using the projectors
$\Pi_i^{\pm}$ onto the irreducible representations $(n_i\pm1)$ with
\be
\big[C\,,\,\Pi_i^\pm\big] =0 \ ,
\ee
and the constrained covariant coordinates take the simple form
\be
C = \frac N2\,
\left(\begin{array}{cc}\mbox{$\bigoplus\limits_{i=1}^n$}\, \Pi_i^+ & 0
      \\ 0 & -\,\mbox{$\bigoplus\limits_{i=1}^n$}\,
      \Pi_i^-\end{array}\right) \ .
\label{C-explicit}
\ee
In particular, since $C_0\otimes\sigma^0$ is two-fold degenerate it
follows that
\be
C_0\otimes\sigma^0 =
\left(\begin{array}{cc}\mbox{$\bigoplus\limits_{i=1}^n$}\,
C_0(n_i)\, \Pi_i^+ & 0
      \\ 0 & \,\mbox{$\bigoplus\limits_{i=1}^n$}\,C_0(n_i)\,
      \Pi_i^-\end{array}\right)
\label{C0-explicit}
\ee
separates the explicit blocks according to \eq{ad-C0-explicit}.

The complex structure map $\cJ$ respects this $SU(2)$ symmetry,
\be
[J_i,\cJ] =0 \ ,
\ee
which enables one to decompose the tangent space $T_C\cO$
into irreducible representations of the $SU(2)$ isometry group. With
respect to the block decomposition (\ref{C-decomp-1}), the subspace
$\ker(\cJ)\subset\mun(\cN\,)$ consists of block diagonal operators while
$T_C\cO$ consists of block off-diagonal operators, and the action of
$\cJ$ on tangent vectors is given explicitly by
\be
\cJ \left(\begin{array}{cc}0 & X \\
                           X^\dagger & 0\end{array}\right)
= \left(\begin{array}{cc}0 & \ii X \\
                           -\ii X^\dagger & 0\end{array}\right) \ .
\label{J-tangent-explicit}
\ee
This is the obvious complex structure on $T_C \cO$ compatible with
the action of the isometry group. 

The decomposition of the tangent space $T_C\cO$ into irreducible
representations of $SU(2)$ is now provided by
\be
T_C^-\cO \cong \Big(\,\bigoplus_{i=1}^n\, (n_i+1)
\Big) \otimes \Big(\,\bigoplus_{j=1}^n\, (n_j-1)\Big) 
= \bigoplus_{i,j=1}^n\, (n_i+1)\otimes (n_j-1) \ ,
\label{X-decomp-1}
\ee
where $T_C^\pm\cO:=T_C\cO\big|_{V^\pm}$ 
corresponds to the upper-right respectively lower-left blocks in
\eq{J-tangent-explicit}, and the different sectors
$(i,j)$ are separated by the eigenvalues of the operator ${\rm ad}_{\ii
  C_0}$ in the irreducible case. Note in particular that the lowest  
spin component in the Clebsch-Gordan decomposition of
$(n_i+1)\otimes (n_i-1)$  is a spin one field as appropriate for
gauge fields.
This implies $\cJ (\mg_0) =0$, where $\mg_0$
is the subspace of $SU(2)$ singlet components of $\mg$, 
and in fact $\mg_0 = \ms$ by Schur's lemma.

\subsubsection*{\it Global minimum}

Consider first the vacuum surface $\cC_{(N,1),\dots,(N,1)}$. Compare
the $SU(2)$-invariant decomposition of the gauge algebra $\mg$, given
by
\bea
\mg &\cong&  (N) \otimes (N) \otimes \mun(n)\nn\\[4pt]
&=& \big((1)\oplus (3) \oplus \cdots \oplus (2N-1 )\big)
\otimes \mun(n)~=~ \big((1) ~\oplus~ (N+1) \otimes (N-1)\big)
\otimes \mun(n) \ ,
\label{mg-decomp-abel-vac}
\eea
with \eq{X-decomp-1} in the degenerate case $C_0=\frac
12~\one_{n\,N}$. It follows that the image of $\cJ(\mg)$ indeed covers
all modes of $T_C\cO$, and the complexification is achieved by adding
$\cJ^2(\mg)$. This gives another proof of the decomposition
(\ref{tangent-decomposition}). The singlet subspace of
(\ref{mg-decomp-abel-vac}) is
$\mg_0=(1)\otimes\mun(n)\cong\mun(n)=\ms$.

\subsubsection*{\it Maximally irreducible saddle points}

Now consider a generic, completely non-degenerate critical surface
$\cC_{(n_1,1),\dots,(n_n,1)}$, and the corresponding decomposition of
$T_C\cO =T_C^-\cO \oplus T_C^+\cO$ given by \eq{X-decomp-1}. The
different sectors $(i,j)$ are distinguished by the eigenvalues of the
operator ${\rm ad}_{\ii C_0}$. Hence we can pick some fixed pair $n_i
>n_j$, and decompose
\be
(n_i+1)\otimes (n_j-1) \cong 
\big(|n_i-n_j|+3\big)_{\ii c_{ij}}\oplus\big(|n_i-n_j|+5
\big)_{\ii c_{ij}} \oplus \cdots \oplus \big(n_i+n_j-1
\big)_{\ii c_{ij}}\subset T_C \cO
\label{X-1}
\ee
which has eigenvalue given by \eq{ad-C0-explicit} as indicated by the
subscripts. Similarly, one has
\be
(n_j+1)\otimes (n_i-1) \cong 
\big(|n_i-n_j|-1\big)_{\ii c_{ji}} \oplus\big(|n_i-n_j|+1
\big)_{\ii c_{ji}} \oplus \cdots \oplus \big(n_i+n_j-1
\big)_{\ii c_{ji}} \subset T_C \cO
\label{X-2}
\ee
(where $(0)$ is omitted) with $\ad_{\ii C_0}$ eigenvalue $\ii
c_{ji}= -\ii c_{ij}$. The corresponding conjugate matrix
decompositions $(n_j-1)\otimes(n_i+1)$ and $(n_i-1)\otimes(n_j+1)$ are
determined by hermiticity. They are given respectively by \eq{X-1}
with eigenvalue $\ii c_{ji}=-\ii c_{ij}$ and by \eq{X-2} with
eigenvalue $\ii c_{ij}$.

We denote the tangent space decomposition (\ref{X-decomp-1})
determined by \eq{X-1} and \eq{X-2} as
\be
T_C \cO := \bigoplus_{i,j=1}^n \, \C|n;n_i+1,n_j- 1; \ii c_{ij},l
\rangle_{T_C\cO} \ 
\label{TO-basis}
\ee
where $n$ denotes the dimension of $(n)$, and 
we will drop its magnetic quantum number $l$ from now on.
This defines a natural basis for $T_C \cO$, in which the action of 
$\cJ$ is given by {\em block-wise} multiplication with
\beq
\cJ=\sigma^2=\left(\begin{array}{ccc}0 & \ii\\ -\ii & 0\end{array}\right)
\label{cJblockwise}\eeq
as in \eq{J-tangent-explicit}, 
and the action of $\ad_{\ii C_0}$ by  
\beq
\big(\ad_{\ii C_0}\big)_{ij}= |c_{ij}|\,\left(\begin{array}{ccc} 0 & 
\sigma^2 \\ \sigma^2 & 0\end{array}\right) \ 
\label{adiC0blockwise}\eeq
since its sign depends on $n_i \gtrless n_j$.
In particular, by virtue of \eq{symplectic-eval} the tangent space
$T_C \cO$ is naturally a symplectic vector space with symplectic form
of type~$(1,1)$ with respect to the complex structure $\cJ$. This
construction thereby defines a local symplectic model for the
neighbourhood of the Yang-Mills critical point $C$ in the K\"ahler
manifold $\cO$. In the next section this model space will be used to
evaluate fluctuation integrals over tubular neighbourhoods of the
critical surfaces. In particular, all pertinent one-forms can be
explicitly evaluated on $T_C \cO$ by using the explicit expressions
for $C$ and $C_0$ in \eq{C-explicit} and \eq{C0-explicit}.

Let us now look at the $SU(2)$-invariant decomposition of the gauge
algebra $\mg$ given by
\bea\label{mg-decomp-abel}
\mg &\cong& \bigoplus_{i,j=1}^n\, (n_i) \otimes (n_j) \\[4pt]
&=& \bigoplus _{i,j=1}^n\, \big( (|n_i-n_j|+1) \oplus(|n_i-n_j|+3)
\oplus \cdots\oplus(n_i+n_j-1)\big) 
~=:~ \bigoplus_{i,j=1}^n\, \C|n;n_i,n_j;\ii c_{ij}\rangle_\mg \ . \nn
\eea
This can be compared with the $SU(2)$-invariant decomposition of the
tangent space $T_C\cO$ in \eq{TO-basis} above, whose higher modes
match perfectly with those of $\mg$ except for a doubling due to the
complex structure $\cJ$. There is, however, some mismatch in the low
lying modes. In particular, $T_C\cO$ contains the extra subspace
\be
E_1 := \bigoplus_{i>j}\, \C|n_i-n_j-1;n_j +1,n_i-1;-\ii
c_{ij}\rangle_{T_C\cO}
\label{E_1}
\ee
which is not contained in $\cJ(\mg)$. On the other hand, the modes in
the subspace
\be
E_0 := \bigoplus_{i>j}\, \C|n_i-n_j+1;n_j +1,n_i-1;-\ii
c_{ij}\rangle_{T_C\cO}
\label{E_0}
\ee
occur only {\em once} in $T_C\cO$, which means that they are already
spanned by the image $\cJ(\mg)$ since $\cJ \neq 0$ on the non-trivial
modes. This implies that $E_0 = \cJ(E_0) = \cJ(\mh)$ where 
\be
\mh = \bigoplus_{i\neq j}\,\C \big||n_i-n_j|+1\,;\,n_i,n_j\,;\,\ii
c_{ij}\big\rangle_\mg \ .
\label{mh-explicit}
\ee
The linear independence of the subspaces $\cJ(\mg \ominus \mh)$ and
$\cJ^2(\mg \ominus \mh)$ follows from the explicit embedding
$T_C\cO\hookrightarrow\mun(\cN\,)$ given below. Therefore 
$\cJ(\mg\ominus \mh) \oplus \cJ^2(\mg\ominus \mh) $
spans the entire tangent space $T_C \cO$ except for the subspace
$E_1$, which gives the decomposition \eq{TCOdecomp} with
the various subspaces now explicitly identified. We have $\cJ(E_0) =
E_0$ and $\cJ(E_1) = E_1$, with the action of $\cJ$ given by diagonal
eigenvalues $\pm \ii$ on the two components in \eq{E_0} and
\eq{E_1}. On the remaining space $T_C \cO \ominus E_0 \ominus E_1$ the
action of $\cJ$ is obtained by exchanging the two components in
\eq{TCOdecomp}.

To complete this analysis, we need to explicitly embed $T_C \cO$
into the space $\mun(\cN\,)$, which 
admits the $SU(2)$-invariant decomposition 
\bea
\mun(\cN\,) ~\cong~ \mg \otimes \big((2)\otimes (2)\big) &=&
\bigoplus_{i,j=1}^n\, \big((n_i+1)\otimes (n_j+1)~\oplus~
(n_i-1)\otimes (n_j-1) \nn\\ && \quad\qquad
\oplus~(n_i+1)\otimes (n_j-1) ~\oplus~(n_i-1)\otimes (n_j+1) \big) 
\label{ucN-decomp-2}
\eea 
corresponding to \eq{C-decomp-1}.
Since we know the action of $\cJ$ on the rhs, we can 
determine the map 
\be
\cJ\,:\, \mg ~\longrightarrow~ T_C \cO ~\hookrightarrow~ \mg \otimes
\big((2)\otimes (2)\big)
\ee
using
\be
\cJ\big(|n;n_i,n_j;\ii c_{ij}\rangle_\mg\big) = \sum_{k,l=1}^n\,
\cJ\big(|n;n_k+ 1,n_l- 1;\ii c_{kl}\rangle_{T_C\cO}\big)~
{}_{T_C\cO}\langle n;n_k+ 1,n_l- 1;\ii c_{kl} |n;n_i,n_j;
\ii c_{ij}\rangle_\mg \, + h.c. 
\label{J-phi-action}
\ee
The non-vanishing inner products in this expression can be
written in terms of Wigner $6j$-symbols for the group $SU(2)$, which
are known explicitly. This also enables one to compute the projection
\bea
\Pi_0\,:\,T_C \cO  ~\longrightarrow~ \mg  \ , \qquad
V_0 \otimes\sigma^0+ V_i \otimes\sigma^i ~\longmapsto~ V_0
\label{p0-def}\eea
as
\beq
\Pi_0 |n;n_i+ 1,n_j- 1;\ii c_{ij}\rangle_{T_C\cO} =\sum_{k,l=1}^n\,
|n;n_k,n_l;\ii c_{kl}\rangle_\mg ~
{}_\mg\langle n;n_k,n_l;\ii c_{kl}|n;n_i+ 1,n_j- 1;\ii c_{ij}
\rangle_{T_C\cO} \ .
\eeq
In the basis \eq{TCOdecomp}, one has the useful explicit formula
\be
\Pi_0\,\cJ(\mg)=\ad_{\ii  C_0}(\mg)
\label{pi0-explicit}
\ee
which is of order $\frac1N$ and can also be used for $E_0$, 
while $\Pi_0\,\cJ^2(\mg)$ is of order
$\frac1{N^2}$ and $\Pi_0(E_1)= \{0\}$. 

\subsubsection*{\it General solutions}

The case where some of the irreducible representations $(n_i)$ have
multiplicity $k_i>1$ is a combination of the structures above for the
vacuum state and for the nondegenerate case. Now the
basis \eq{TO-basis} acquires additional labelling reflecting the
$\mun(k_i)$ degrees of freedom, and it takes the symbolic form
\beq
T_C \cO =\bigoplus_{i,j=1}^l \, \C\big|n\,;\,(n_i+ 1,a_i)\,,\,
(n_j- 1,a_j)\,;\,\ii c_{ij}\big\rangle_{T_C\cO} \ .
\label{TO-basis-general}
\ee
In particular, one can now easily compute the symplectic form on 
$T_C\cO$ using \eq{symplectic-eval}. It is essentially given by the
complex structure $\cJ$.

\subsection{Fluctuations around the critical surfaces\label{Fluct}}

We conclude this section with a summary of the salient features of the
decompositions in Sections~\ref{LocalGeomO} and~\ref{ExplYMDecomp}
above, as pertaining to how they will be exploited in the next section
to evaluate fluctuation integrals over the local neighbourhoods of
Yang-Mills critical points. Recall that globally the critical surface
(with no fluxons) through some critical point $C$ is given by the
space of gauge transformations acting on $C$, as in
\eq{globalcritsurface}. Its tangent space is embedded locally as
\be
T_C\cC_{(n_1,s_1),\dots,(n_k,s_k)} = \cJ(\mg\ominus\ms)
\subset T_C \cO \ ,
\label{critical-embedding}
\ee
which can be determined explicitly using \eq{J-phi-action}.
Recall also that the gauge stabilizer $\ms$ of $C$ consists of the
$SU(2)$ singlets in $\mg$. It is given by $\ms \cong\mun(n)$ for
the vacuum, and $\ms\cong\mun(1)^{n}$ for completely irreducible
saddle-points. In particular, $\ms$ is never trivial, 
quite unlike the situation in ordinary
two-dimensional Yang-Mills theory~\cite{witten}. The global symmetry
cannot be disentangled in the noncommutative case, and the nonabelian
localization even at the global minimum is akin to that at higher
critical points of two-dimensional Yang-Mills theory or more precisely
at the flat connections of Chern-Simons gauge theory on a Seifert
fibration~\cite{witten}. The non-trivial part of the localization at
higher critical points will therefore be given by fluctuation
integrals over the spaces $E_0$, $E_1$ and $\ms$. The only effect of
the remaining part $\cJ(\mg \ominus \mh)\oplus \cJ^2(\mg \ominus \mh)
$ will be to induce normalization terms as for the vacuum critical
point. In particular, the subspaces $\cJ(\mg\ominus\ms)$ and
$\cJ^2(\mg\ominus\ms)$ locally model the tangent space $T_C\cO$ near
the vacuum.

To understand the physical meaning of the subspace $E_1$, note that
the gauge field strength remains constant for variations along 
$\phi=X\in E_1$, since
$\delta C_0\big|_{E_1}= i [C,\phi]_0\in \Pi_0(E_1) =\{0\}$. Let us
compute the second order variation of the Yang-Mills action, given by
\bea
\Tr\big(C_0~\delta^2C_0\big) &=& -\Tr\big( C_0\,[\,[C,\phi]\,,\,\phi]
\big)\nn\\[4pt]
&=& \Tr \big([C_0,\phi]\,[C,\phi]\big) ~=~ -N\,
\Tr \big(\ad_{\ii C_0}(\phi)\, \cJ(\phi)\big) \ .
\eea
Restricting to fluctuations $\phi=X\in E_1$ with respect to the
decomposition (\ref{E_1}) one has
\beq
\Bigl.\Tr\big(C_0~\delta^2C_0\big)\Bigr|_{E_1}=
-N\, \sum_{i>j}\,\Tr\big(\ad_{\ii C_0}
(X_{ji}^\dagger)\, \cJ(X_{ji})\big)= -2N\,\sum_{i>j}\,
|c_{ij}| \, \Tr \big(X_{ji}^\dagger\, X_{ji}\big)
\eeq
by using the actions (\ref{cJblockwise}) and
(\ref{adiC0blockwise}), cf. \eq{J-ad-id}. For the maximally nondegenerate
saddle-points, this fluctuation is thus negative, demonstrating that
the two-dimensional instantons on the fuzzy sphere $S_N^2$ are
generically {\it unstable}. On the other hand, since the subspace
$E_0= \cJ(\mh)$ is obtained through gauge transformations, it produces
flat directions for the Yang-Mills action.

\bigskip

\section{Nonabelian localization\label{NonabLoc}}

This section is the crux of the present paper, wherein we shall derive
the semiclassical expansion of the partition function for Yang-Mills
theory on the fuzzy sphere $S_N^2$ and show that it agrees with the
known instanton expansion of quantum gauge theory on $S^2$ in the classical
limit $N\to \infty$. We will begin by describing the nonabelian
localization principle, adapted to our specific gauge theory. 
We will then explicitly evaluate the contributions from
two extreme classes of Yang-Mills critical points, the vacuum and the
maximally irreducible solutions, and show that they give the
expected contributions to the path integral at large $N$. The
intermediate contributions from degenerate solutions, which we do not
treat in detail here, are somewhat more involved but can in principle
be evaluated using our techniques. The contribution from the vacuum to
the partition function could be expressed in terms of the abstract
cohomological formula of~\cite{witten1} given by intersection pairings
on the vacuum moduli space, or by using the more explicit residue
formula of~\cite{JK1}. The contributions from some higher unstable
critical points to the nonabelian localization formula are formally
described in~\cite{Paradan1,Woodward:2004xz,witten}, but the general
cases that we need (including reducible saddle points) are not
explicitly treated in full generality. Here we will directly evaluate,
following~\cite{witten}, the explicit quantum fluctuation integrals
near the critical points using the local symplectic geometry of the
previous section.

\subsection{Equivariant cohomology and the localization
  principle\label{LocPrinc}}

The goal of this section is to compute the partition function of
quantum Yang-Mills theory on the fuzzy sphere defined by the action
(\ref{YM-action}) on the configuration space (\ref{orbit-2}) of gauge
fields. After an irrelevant shift of the covariant coordinates
\eq{covcoordsdef} which is equivalent to working with the reduced
Yang-Mills action \eq{YM-action-prime}, it is defined by 
\bea
Z&:=&\frac1{\vol(G)}\,\left(\frac{g}{4\pi\,N}\right)^{\dim(G)/2}\,
\int_{\cO}\, \ddd C~ \exp\Big(-\mbox{$\frac{N}{g}$}\,
\Tr\big(C_0^2\big)\Big) \nn\\[4pt] &=&
\frac1{\vol(G)}\,\left(\frac{g'}{2\pi}\right)^{\dim(G)/2}\,
\int_{\cO}\, \exp\Big(\omega -\mbox{$\frac{1}{2g'}$}\,
\Tr \big(C_0^2\big)\Big)
\label{Z-1}
\eea
where we have used the fact that the symplectic volume form
$\omega^d/d!$, with $d:=\dim_\C(\cO)$, defines the natural gauge
invariant measure on $\cO$ provided by the Cartan-Killing riemannian
volume form (up to some irrelevant normalization). This
follows from the fact that the natural invariant metric on $\cO$ is a
K\"ahler form. We have divided by the volume of the gauge group
$G=U(n\,N)$ with respect to the invariant Cartan-Killing form and by
another normalization factor for later convenience, and also
introduced the rescaled gauge coupling
\be
g' = \frac g{2N} \ .
\label{gprime}\ee
We will now describe, following~\cite{witten1,witten}, how the
technique of nonabelian localization can be applied to evaluate the
symplectic integral (\ref{Z-1}) exactly.

We begin by using a gaussian integration to rewrite \eq{Z-1} as
\be
Z = \frac1{\vol(G)}\,\int_{\mg\times \cO}\, \Big[\,\frac{\ddd\phi}{2\pi}\,
\Big]~\exp\Big(\omega -\ii \Tr(C_0\, \phi) -\mbox{$
\frac{g'}{2}$}\, \Tr \big(\phi^2\big)\Big) \ ,
\label{Z-2}\ee
where the euclidean measure for integration over the gauge algebra
$\phi\in\mg=\mun(n\,N)$ is determined by the invariant
Cartan-Killing form. Since the moment map for the $G$-action on $\cO$
is given by \eq{momentmapred}, by \eq{hamiltonian} we have
\be
\ddd \Tr(C_0\, \phi) = -\iota_{V_\phi} \omega \ .
\label{moment-property}
\ee
Introduce the BRST operator
\be
Q= \ddd - \ii \iota_{V_\phi} \ ,
\ee
where $\ddd$ is the exterior derivative on $\Omega(\cO)$ and the
contraction $\iota_{V_\phi}$ acts trivially on $\phi$. It preserves
the gradation if one assigns charge~$+2$ to the elements $\phi$ of
$\mg$, and it satisfies
\be
Q^2 = -\ii\{\ddd,\iota_{V_\phi}\} = -\ii {\cal L}_{V_\phi}
\ee
where ${\cal L}_{V_\phi}$ is the Lie derivative along the vector field
$V_\phi$. Thus $Q^2=0$ exactly on the space
\beq
\Omega_G(\cO):=\big(\C[[\mg]]\otimes\Omega(\cO)\big)^G
\eeq
consisting of gauge invariant differential forms on $\cO$ which take
values in the ring of symmetric functions on the Lie algebra $\mg$.

By construction one has
\be
Q\big(\omega -\ii \Tr(C_0 \,\phi)\big) =0
\ee
using \eq{omegaclosed} and \eq{moment-property}, and
\be
Q\Tr \big(\phi^2\big) =0 \ .
\ee
Therefore, the integrand of the partition function (\ref{Z-2}) defines
a $G$-equivariant cohomology class in $H_G(\cO)$, and the value of $Z$
depends only on this class. The integral of any $Q$-exact equivariant
differential form in $\Omega_G(\cO)$ over $\mg\times\cO$ is
clearly~$0$, as is the integral of any $\iota_{V_\phi}$-exact form
even if its argument is not gauge invariant. Thus $Z$ is unchanged by
adding any $Q$-exact form to the action, which will fix a gauge for
the localization. Hence we can replace it by
\be
Z = \frac1{\vol(G)}\,\int_{\mg\times\cO}\,
\Big[\,\frac{\ddd\phi}{2\pi}\,\Big]~ 
\exp\Big(\omega -\ii \Tr(C_0\, \phi) -\mbox{$\frac{g'}{2}$}\, \Tr 
\big(\phi^2\big)+ t~ Q\alpha\Big) \ ,
\label{Z-3}
\ee
which is independent of $t\in\R$ for any $G$-invariant one-form
$\alpha$ on $\cO$, where
\be
Q\alpha = \ddd\alpha - \ii \langle\alpha,V_\phi\rangle \ .
\label{Qalpha}\ee
The independence of \eq{Z-3} on the particular representative
$\alpha\in\Omega(\cO)^G$ of its equivariant cohomology class will play
a crucial role in our evaluation of the partition function.

Expanding the integrand of (\ref{Z-3}) by writing $\exp(t~\ddd\alpha)$
as a polynomial in $t$ and using the fact that the configuration space
$\cO$ is compact, it follows 
that for $t \to\infty$ the integral localizes at the stationary points
of $\langle\alpha,V_\phi\rangle$ in $\mg\times\cO$. By writing $V_\phi
= V_a ~\phi^a$, where $\phi^a$ is an orthonormal basis of $\mg^\vee$,
we have $\langle\alpha,V_\phi\rangle = \langle\a,V_a\rangle~\phi^a$
and the critical points are thus determined by the equations
\bea
 \langle\alpha,V_a\rangle &=& 0 \ , \label{crit-1}\\[4pt]
\phi^a ~\ddd\langle\alpha,V_a\rangle &=& 0 \ . \label{crit-2}
\eea
Since \eq{crit-2} is invariant under rescaling of $\phi$ and the Lie
algebra $\mg$ is contractible, the homotopy
type of the space of solutions in $\mg\times\cO$ is unchanged by
restricting to $\phi=0$ and the saddle-points reduce to the zeroes
of $\langle\alpha,V_a\rangle$ in $\cO$.

Given the reduced Yang-Mills function \eq{YM-action-prime}, let
us consider explicitly the invariant one-form $\alpha$ given
by~\cite{witten,szabo}
\beq
\a=-\ii\Tr\big(C_0 \,[C,\ddd C]_0\big)= g'\,\cJ\big(\ddd S'\,\big) \ .
\label{loc1form}\eeq
We claim that the vanishing locus of $\langle\alpha,V_a\rangle$ in
this case coincides with the critical surfaces of the original
Yang-Mills action (\ref{YM-action}) as found in
Section~\ref{CritPoints}. To see this, we note that the condition
\beq
0 = \langle\alpha,V_a\rangle =\Tr\big(C_0\,[C\,,\,[C,\phi^a]\,]_0
\big)=-\Tr\big([C,C_0]\,[C,\phi^a]\big)
\eeq
certainly holds whenever $[C,C_0]=0$. On the other hand, by
setting $\phi = C_0$ it implies
\be
0=\langle\alpha,V_\phi\rangle = -\Tr\big([C,C_0]^2\big)
\ee
which by nondegeneracy of the inner product implies that $[C,C_0] =
0$. Therefore the action in \eq{Z-3} has indeed the same critical
points as the Yang-Mills action \eq{YM-action}.

Let us now explicitly establish, following~\cite{szabo}, the
localization of the partition function onto the classical solutions of
the gauge theory. Plugging (\ref{loc1form}) and (\ref{Qalpha}) into
\eq{Z-3} and carrying out the integration over $\phi\in\mg$ gives
\bea
Z &=& \frac1{\vol(G)}\,\int_{\mg\times \cO} \,\Big[\,\frac{\ddd\phi}
{2\pi}\,\Big] ~\exp\Big(t~\ddd\a+\omega\Big)\nn\\ &&
\qquad\qquad\qquad \times~
\exp\Big(-\ii \Tr(C_0\, \phi) -\mbox{$\frac{g'}{2}$} \Tr
\big(\phi^2\big)- \ii t  \,\Tr\big([C\,,\,[C,C_0]\,]\,\phi\big)
\Big)\nn\\[4pt] \label{Z-4}
&=& \frac1{\vol(G)}\,\left(\frac{g'}{2\pi}\right)^{\dim(G)/2}\,\int_{\cO}\,
\exp\Big(t~\ddd\a + \omega \Big)\\ && \qquad\qquad\qquad \times~
\exp\Big( -\mbox{$\frac{1}{2g'}$}\, \Tr\big(C_0^2\big) 
+ \mbox{$\frac t{g'}$}\, \Tr\big(C_0\,[C\,,\,[C,C_0]\,]\big)
-\mbox{$\frac{t^2}{2g'}$}\, \Tr\big([C\,,\,[C,C_0]\,]\big)^2\Big)\nn
\eea
where we have used $\Tr(C\,[C,-]) =0$. The only configurations which
contribute to (\ref{Z-4}) in the large $t$ limit are therefore
solutions of the equation
\be
[C\,,\,[C,C_0]\,]=0
\ee
which implies as in~\cite{szabo} that
\be
0 = \Tr\big(C_0 \,[C\,,\,[C,C_0]\,]\big) = - \Tr\big([C,C_0]^2
\big) \ ,
\ee
giving $[C,C_0]=0$ as desired. Therefore the integral \eq{Z-4}
receives contributions only from the solutions of the Yang-Mills
equations \eq{eom}, which establishes the claimed localization.

The local geometry in $\mg\times\cO$ about each critical point, as
analysed in detail in the last section, determines the partition
function as a sum of local contributions involving the values of the
Yang-Mills action evaluated on the classical solutions as in
Section~\ref{ClassAction}. Consider an equivariant tubular
neighbourhood $\cN_{(n_1,s_1),\dots,(n_k,s_k)}$ of a critical surface
$\cC_{(n_1,s_1),\dots,(n_k,s_k)}$ in $\mg\times\cO$. Since the
partition function \eq{Z-3} is independent of $t$, we can consider its
large $t$ limit as above, and this limit will always be implicitly
assumed from now on. Let $\cW$ be a compact subset of $\cO$ with
$\cW\cap\cC=\emptyset$, where
$\cC:=\bigcup_{(n_i,s_i)}\,\cC_{(n_1,s_1),\dots,(n_k,s_k)}$. Then the
integral over $\cW$ in \eq{Z-4} has a gaussian decay in
$t\to\infty$. This means that in expanding $\exp(t~\ddd\a  + \omega)$
into a finite sum of terms of the form $\omega^p\wedge (t~\ddd \a)^{ m}$,
we can disregard all terms which contain $\omega$ since they will
be suppressed by factors of $\frac 1t$ and vanish in the large $t$
limit. The only terms which survive the $t\to\infty$ limit are those
with $p=0,m=d$, and the integral therefore vanishes unless $\omega$ is
replaced by $\ddd\alpha$, except at the saddle point where
$\ddd\alpha=0$. Then one has
\be
Z = \frac1{\vol(G)}\,
\int_{\mg\times\cO}\, \Big[\,\frac{\ddd\phi}{2\pi}\,\Big]~
\exp\Big(t\, \big(\ddd\a - \ii \langle\a,V_\phi\rangle\big)\Big)~ 
\exp\Big(-\ii \Tr(C_0\, \phi) -\mbox{$\frac{g'}{2}$}\, \Tr \big(\phi^2
\big)\Big)
\label{Z-5}
\ee
in the vicinity of any critical point in which $\ddd\a$ is
nondegenerate.

The integral $Z_{(n_1,s_1),\dots,(n_k,s_k)}$ in (\ref{Z-5}) over the
neighbourhood $\cN_{(n_1,s_1),\dots,(n_k,s_k)}$ is determined by the
local behaviour of $\alpha$ and the $G$-action near
$\cC_{(n_1,s_1),\dots,(n_k,s_k)}$. Then
\beq
Z=\sum_{\stackrel{\scriptstyle(n_1,s_1),\dots,(n_k,s_k)}
{\scriptstyle\sum_i\,n_i=n\,N~,~\sum_i\,s_i=n}}\,
Z_{(n_1,s_1),\dots,(n_k,s_k)} \ .
\label{Zlocalsum}\eeq
As expected~\cite{Paradan1}, the sum over critical surfaces in
\eq{Zlocalsum} contains the sum over weights $1\leq n_1\leq
n_2\leq\cdots\leq n_k$ of the gauge group $G=U(n\,N)$. Our explicit
computations will confirm the local behaviour of the partition
function given by~\cite{Paradan1}
\beq
Z_{(n_1,s_1),\dots,(n_k,s_k)}=\big(g'\,\big)^{-\dim(G)}~
\e^{-\frac1{2g'}\,\sum_i\,n_i^2}~H_{(n_1,s_1),\dots,(n_k,s_k)}
\big(\,\sqrt{g'}~\big) \ .
\label{ZParadan}\eeq
The smooth function $H_{(n_1,s_1),\dots,(n_k,s_k)}:\R\to\C$, which is
bounded by a polynomial at infinity, is determined by the equivariant
Euler class of the fixed point locus corresponding to the weight
$(n_1,\dots,n_k)$ after reducing the integral over $\mg$ to its Cartan
subalgebra, as we do explicitly in the next section. 

\subsection{Explicit evaluation of the localization
  forms\label{LocFormProps}}

The explicit computation of the local contributions
$Z_{(n_1,s_1),\dots,(n_k,s_k)}$ to the Yang-Mills partition function
on $S_N^2$ will rely on the local behaviour of the invariant one-form
$\alpha$ introduced in \eq{loc1form} near the Yang-Mills critical
points. We will now pause to derive explicit expressions for the BRST
transformations (\ref{Qalpha}) on the subspaces appearing in the
tangent space decomposition \eq{TCOdecomp}. Given the invariant
Maurer-Cartan one-form \eq{thetaCMdef} and the projector \eq{p0-def},
consider the $\mun(n\,N)$-valued one-form
\be
\theta_0 := \Pi_0(\theta) = \mbox{$\frac 12$}\, \tr_\sigma (\theta)
\ee
where $\tr_\sigma$ denotes the partial trace over the spin matrices
$\sigma^\mu$. It is given explicitly by
\be
\theta_0=\mbox{$\frac4{N^2}$}\,\big(C~\ddd C\big)_0 
= \mbox{$\frac4{N^2}$}\,\big(C_i~\ddd C^i + C_0~ \ddd C_0\big)
\ee
and satisfies
\be
\ddd\theta_0 =  - \mbox{$\frac 12$}\,\tr_\sigma\big(\theta^2\big)
= - \Pi_0\big(\theta^2\big) \ .
\ee
One has
\beq
\langle \theta,V_\phi\rangle=\mbox{$\frac{2}{N^2}$}\,  [C,V_\phi]
  = -\mbox{$\frac{2\ii}{N}$}\, \cJ (V_\phi) \qquad \mbox{and} \qquad
\langle \theta_0,V_\phi\rangle =  -\mbox{$\frac {2\ii}N$}\,
\Pi_0\,\cJ (V_\phi)
\label{etatilde-eval}
\eeq
for any tangent vector $V_\phi = \ii[C,\phi]$.

Using the identity $C~\ddd C=-\ddd C~C$, the localization one-form
(\ref{loc1form}) can now be written as
\be
\a = -\mbox{$\frac {\ii N^2}2$}\; \Tr(C_0\, \theta) = 
-\mbox{$\frac {\ii N^2}2$}\; \Tr(C\, \theta_0) \ .
\ee
Hence the pairing in (\ref{Qalpha}) is given by
\bea
\langle \a,V_\phi\rangle &=& - N\, \Tr\big(C_0\, \cJ (V_\phi)\big)
\nn\\[4pt] &=&  N\, \Tr\big(\cJ (C_0)\, V_\phi\big)  
~=~ - N^2\, \Tr\big(\cJ^2(C_0)\, \phi\big) \ .
\eea
This vanishes on the critical surfaces, where
$\cJ(C_0)=0$. Furthermore, for any $g\in\mg$ one has 
\bea
\big\langle \a\,,\,\cJ^2(g)\big\rangle &=& - N \,
\Tr\big(C_0\, \cJ^3(g)\big)\nn\\[4pt]
&=& N \,\Tr\big(C_0\, \cJ(g)\big) 
= \ii\Tr\big(C_0\, [C_0,g]\big) ~=~0 \label{alpha-J2}
\eea
while for $e_0\in E_0$ one has
\beq
\big\langle \a\,,\,e_0\big\rangle = \big\langle \a\,,\,\cJ(e_0)
\big\rangle = \big\langle \a\,,\,\cJ^2(h)\big\rangle =0
\label{alpha-E0}\eeq
for some $h\in\mh$. Both identities \eq{alpha-J2} and \eq{alpha-E0}
hold even off-shell. We also note the on-shell relations
\beq
\big\langle \a\,,\,\cJ(g)\big\rangle = - N \,\Tr\big(C_0\,
\cJ^2(g)\big) =0 \qquad \mbox{and} \qquad
\big\langle \a\,,\,e_1\big\rangle =-N \,\Tr\big(C_0\, \cJ
(e_1)\big) =0 \label{alpha-E1}
\eeq
for $e_1\in E_1$.

To evaluate the integral \eq{Z-5} using the stationary phase method, 
we must understand how it behaves near the Yang-Mills critical
points. For this, we will study the local behaviour of the BRST
variation (\ref{Qalpha}), beginning with the pairing
$\langle\alpha,V_\phi\rangle$. Let us write a generic gauge field of
$\cO$ as $C = \obar{C} + \varepsilon \ii[\,\obar C,\Psi] + \frac 12\,
\varepsilon^2\ii[\,\obar C\,,\,\ii[\,\obar
C,\Psi]\,]+O(\varepsilon^3)$, where $\obar{C}$ is the given critical
point, $\Psi \in\ms\mun(\cN\,)$ are the fluctuations around $\obar C$
and $\varepsilon$ is a small real parameter. Then 
\bea
\cJ^2(C_0) &=& 0 + \varepsilon\, \Big(
\cJ^2\big(\ii[\,\obar C,\Psi]_0\big)    + \mbox{$\frac\ii N$}\,\big[\ii
[\,\obar C,\Psi]\,,\,\cJ(\,\obar C_0)\big] 
+ \mbox{$\frac\ii N$}\,\cJ\big([\ii[\,\obar C,\Psi]\,,\,\obar
C_0]\big)\Big) + O\left(\varepsilon^2\right) \nn\\[4pt]
&=& \varepsilon\, \Big(\cJ^2\big((V_\Psi)_0\big) 
+ \mbox{$ \frac\ii N$}\,\cJ\big([V_\Psi,\obar C_0]\big) \Big) + O
\left(\varepsilon^2\right)  \ , 
\label{J2C0explO1}\eea
which for $\phi \in \mg$ gives
\bea
\langle \a,V_\phi\rangle 
&=&  -\varepsilon\, N^2 \,\Tr\big(\cJ^2((V_\Psi)_0)\,\phi
+\mbox{$\frac\ii N$}\,\big[\cJ(V_\Psi),\,\obar C_0\,\big] ,\phi
\big) +O\left(\varepsilon^2\right)\nn\\[4pt] 
&=&  -\varepsilon\, N^2 \,\Tr\big((V_\Psi)_0\,\cJ^2(\phi)
+ \cJ(V_\Psi) \big[\,\mbox{$\frac\ii N$}\,\obar C_0\,,\phi\,\big] 
\big) +O\left(\varepsilon^2\right)\nn\\[4pt] 
&=&  -\varepsilon\, N^2 \,\Tr\Big(V_\Psi\,\big(\cJ^2(\phi)_0
- \cJ(\cJ(\phi)_0) \big)\Big) +O\left(\varepsilon^2\right) \ . 
\label{alphaV-expand1}
\eea  
using \eq{pi0-explicit}. This is non-degenerate for $\phi\in\mg
\ominus \ms \ominus \mh$, i.e. non-vanishing 
for some $V_\Psi \in T_C \cO$. To see this, it is
sufficient to show that $\cJ(\cJ^2(\phi)_0 - \cJ(\cJ(\phi)_0)) \neq 0$.
Indeed, assuming the contrary
$\cJ(\cJ^2(\phi)_0) = \cJ^2(\cJ(\phi)_0)$ would imply that either
$\phi \in \ms$, or $\cJ(\phi)_0 \in \mh$ which is amounts to $\phi \in
\mh\oplus \ms$. On the other hand, this pairing is indeed degenerate
for any $V_\Psi \in E_1$.

For $\phi \in \ms$, the second-order contribution to the
form (\ref{alphaV-expand1}) can be obtained from
\be
V_\phi = \ii[C,\phi] = \ii\varepsilon[V_\Psi,\phi] + O
\left(\varepsilon^2\right)
\ee
and
\be
\cJ(C_0) =\mbox{$\frac\ii N$}\,[C,C_0] = \mbox{$\frac\ii N$} \,
\varepsilon \, \big([V_\Psi,C_0] +[C,(V_\Psi)_0]\big)+O
\left(\varepsilon^2\right) \ .
\ee
It follows that 
\be
\langle \a,V_\phi\rangle =  
-\varepsilon^2 \,\Tr\Big(\ad_\phi(V_\Psi)\, \big(\ad_{C_0}(V_\Psi)+
\ii N\, \cJ((V_\Psi)_0)\big)\Big)+O\left(\varepsilon^3\right) \ .
\label{alpha-eval-second}
\ee
In particular, for $V_\Psi \in E_1$ this pairing simplifies to
\be
\langle \a,V_\phi\rangle =  
-\varepsilon^2 \,\Tr\big(\ad_\phi(V_\Psi)~\ad_{C_0}(V_\Psi)\big)+O
\left(\varepsilon^3\right) \ .
\label{alpha-expand}
\ee
We now turn to the exact part $\ddd\alpha$ of (\ref{Qalpha}). Using
\eq{thetaconstrs}--\eq{MCformprop2}, one finds
\beq
\ddd\a=-\ii\mbox{$\frac{N^2}2$} \,\Tr\big(\ddd C~ \theta_0 -
C_0 \,\theta^2\big)
=-\ii\mbox{$\frac{N^2}2$} \,\Tr\big(C\, \theta\,
\theta_0+C_0~ \ddd\theta\big)  \ . \label{da-2}
\eeq
For flat connections with $F=0$, the second term in the first equality
of \eq{da-2} vanishes and one has
\be
\ddd\a= -\ii\mbox{$\frac{N^2}2$} \,\Tr(\ddd C ~\theta_0) 
=-\ii\mbox{$\frac{N^2}2$} \,\Tr(C\, \theta\, \theta_0) \qquad
\mbox{if}\quad C_0
=\mbox{$\frac12$}~\one_{n\,N} \ .
\ee
{}From \eq{thetaCrel1} and \eq{thetaconstrs} one generally has
$\theta^2= -\frac{4}{N^2}~(\ddd C)^2$, and hence
\bea
\big\langle\Tr(C_0\, \theta^2)\,,\,V_\phi\wedge V_\psi\big\rangle &=&
 \mbox{$\frac{4}{N^2}$} \,\Tr\big(C_0 \,[\,[C,\phi]\,,\,[C,\psi]\,]
\big) \nn\\[4pt]
&=& \mbox{$\frac{4}{N^2}$} \,\Tr\big([C_0\,,\, [C,\phi]\,]\,[C,\psi]
\big) ~=~-\mbox{$\frac{4}{N^2}$}\,\Tr\big(\ad_{C_0}(V_\phi) \,V_\psi
\big)
\label{da1-eval}
\eea
for any pair of tangent vectors $V_\phi = \ii[C,\phi]$ and $V_\psi =
\ii[C,\psi]$. Similarly, one has
\be
\big\langle  \Tr(C\, \theta\,\theta_0) \,,\,V_\phi\wedge V_\psi\big
\rangle = \big\langle  \Tr(\ddd C~ \theta_0) \,,\,V_\phi\wedge V_\psi
\big\rangle= - \mbox{$\frac{2\ii}{N}$}  \,\Tr\big(V_\phi \,
\cJ (V_\psi)_0 - V_\psi\, \cJ (V_\phi)_0\big)
\label{TrCtheta0pair}\ee
which vanishes if any of the arguments belongs to the subspace $E_1$.

If $V_\psi = \cJ(h) \in E_0$ for some $h\in\mh$, then by using the map
(\ref{juniquemap}) along with (\ref{TrCtheta0pair}) one computes
the on-shell pairing
\bea
\big\langle\Tr(C\, \theta\,\theta_0) \,,\,V_\phi\wedge V_\psi
\big\rangle &=&  - \mbox{$\frac{2\ii}{N}$}\,\Tr\big(V_\phi\,\cJ(j(h))_0
-\cJ(h)_0 \,\cJ( V_\phi)\big)  \nn\\[4pt]
&=& - \mbox{$\frac{2}{N^2}$}\,\Tr\big(\ad_{C_0}(V_\phi)\,j(h) +
\ad_{C_0}(h)\, \cJ(V_\phi)\big)  \nn\\[4pt]
&=& - \mbox{$\frac{2}{N^2}$}\,\Tr\big( N~\ad_{C_0}(\cJ(\phi))\,
j(h)+ \ad_{C_0}(h)\, \cJ(V_\phi)\big)  \nn\\[4pt]
&=& - \mbox{$\frac{2}{N^2}$}\,\Tr\big(-N~\ad_{C_0}(\phi)\, \cJ^2(h)
+ \ad_{C_0}(h)\, \cJ(V_\phi)\big)  \nn\\[4pt]
&=& - \mbox{$\frac{2}{N^2}$}\, \Tr\big(N~\ad_{C_0}(\cJ(\phi))\,
\cJ(h) - \ad_{C_0}(\cJ(h))\, V_\phi\big)\nn\\[4pt]
&=&-\mbox{$\frac{2}{N^2}$}\,
\Tr\big(\ad_{C_0}(V_\phi)\, V_\psi - \ad_{C_0}(V_\psi)\, V_\phi\big) \
.
\eea
This coincides with \eq{da1-eval}, and in particular it vanishes
unless the vector field $V_\phi$ also belongs to the subspace
$E_0$. In summary, we have the on-shell evaluations
\beq
\langle \ddd\a,V_\phi\wedge V_\psi\rangle =
2\ii\Tr\big(V_\phi~\ad_{C_0}(V_\psi)\big) \qquad \mbox{if} \quad
V_\psi \in E_1 \label{da-E1}
\eeq
and
\beq
\langle \ddd\a,V_\phi\wedge V_\psi\rangle =  0 \qquad \mbox{if} \quad
V_\psi \in E_0 \ .
\label{da-E0}
\eeq

\subsection{Localization at the vacuum moduli
  space\label{LocVacSurface}}

We will now compute the localized partition function
$Z_0:=Z_{(N,1),\dots,(N,1)}$ at the vacuum critical surface. We denote
this gauge orbit as
\be
\cO_0 := \cC_{(N,1),\dots,(N,1)} 
= \big\{g\, C \,g^{-1}~\big|~ g \in U(n\,N)\big\}
\cong U(n\,N)/U(n) \ .
\ee
In this case the subspaces $E_0$ and $E_1$ in (\ref{TCOdecomp}) are
trivial. Localization implies that we can restrict ourselves to a
$G$-equivariant tubular neighbourhood $\cN_0=\cN_{(N,1),\dots,(N,1)}$
of the critical surface, under the action of the gauge group $G =
U(n\,N)$. The neighbourhood $\cN_0$ has an equivariant
retraction~\cite[Chap.~27]{GSbook} by a local equivariant
symplectomorphism onto the {\it local symplectic model} ${\cal F}_0$,
defined to be an equivariant symplectic vector bundle over $\cO_0$ with fibre
$\cJ^2(\mg\ominus\ms)$ which is a sub-bundle of the tangent bundle
$T\cO$ restricted to $\cO_0$. This means that the tangent space to
${\cal F}_0$ at the vacuum critical point $C$ in \eq{vacsolnonab} is
given by $T_C \cO_0 \oplus\cJ^2(\mg\ominus\ms)
\cong\cJ(\mg\ominus\ms)\oplus\cJ^2(\mg\ominus\ms) = T_C \cO$, the
symplectic two-form on ${\cal F}_0$ is simply $\omega$, and the
hamiltonian $G$-action on ${\cal F}_0$ descends from the moment map
$\mu$. In physical terms, the gauge fields are decomposed along the
vacuum moduli space $\cO_0$ plus infinitesimal non-gauge variations in
the subspace $\cJ^2(\mg\ominus\ms)$.

Due to the presence of the localization form $\a$ in the path
integral, we can restrict ourselves to this model ${\cal F}_0$ and use
it to replace the open neighbourhood $\cN_0$~\cite{witten}. Indeed,
because ${\cal F}_0$ is an equivariant retraction from $\cN_0$, the
$G$-equivariant cohomology of $\cN_0$ is the same as that of ${\cal
  F}_0$. Furthermore, since the fibres of the bundle ${\cal F}_0$ are
contractible, its $G$-equivariant cohomology is identified under
pullback with the $S$-equivariant cohomology of its base space
$\cO_0$, so that $H_G(\cN_0)\cong H_S(\cO_0)$. Since $S$ acts
trivially on $\cO_0$, one has $H_S(\cO_0)\cong\C[[\ms]]^S\otimes
H(\cO_0)$ and the $S$-equivariant cohomology classes of $\cO_0$
coincide with ordinary cohomology classes of $\cO_0$ valued in the
ring of invariant functions on the stabilizer $\ms$. Putting
everything together gives an isomorphism
$H_G(\cN_0)\cong\C[[\ms]]^S\otimes H(\cO_0)$ which reduces the
equivariant integral over $\mg\times\cN_0$ in \eq{Z-5} to an ordinary
integral over $\ms\times\cO_0$. This is precisely the nonabelian
localization that is formally carried out in~\cite{Woodward:2004xz},
and will turn out to be very much like the localization at the trivial
connection of Chern-Simons theory on a Seifert homology
sphere~\cite{witten}. In the present case, the integral over
$\phi\in\ms$ will then give the interesting non-trivial quantum
fluctuation determinants about the classical solution. We will now
carry out this reduction explicitly.

Let $g'_i$ be an orthonormal basis of  $\mg' = \mg\ominus\ms$, and consider the
corresponding basis
\be
J_i = \cJ(g'_i) \qquad \mbox{and} \qquad \tilde J_j = \cJ^2(g'_j)
\label{J-i}
\ee
of $T_C\cO=\cJ(\mg\ominus\ms) \oplus \cJ^2(\mg\ominus\ms)$, with the dual
basis $\lambda^i, \tilde \lambda^j$ defined by
\be
\big\langle\lambda^i\,,\, J_j\big\rangle = \delta^i{}_j \ ,
\quad \big\langle\,\tilde\lambda^i\,,\, \tilde J_j\big\rangle =
\delta^i{}_j \qquad \mbox{and} \qquad \big\langle\lambda^i\,,\,\tilde
J_j\big\rangle =\big\langle\,\tilde\lambda^i\,,\, J_j\big\rangle = 0 \ .
\ee
Introduce the functions
\be
f_i = \langle\a,J_i\rangle
\ee
which vanish on-shell but have non-degenerate
derivatives $\ddd f_i$ due to \eq{alphaV-expand1}. 
Then by expanding $\phi = \phi^i~ g_i + \phi^a~ s_a$ into components
$\phi^i$ along $\mg\ominus\ms$ and $\phi^a$ along $\ms$, we have
\be
\langle\a, V_\phi\rangle =
N\,\big\langle\a\,,\, \cJ(\phi)\big\rangle = N \,f_i \,\phi^i \ .
\label{alpha-Vphi}
\ee
It follows that the localization one-form can be expanded as
\be
\a =  f_i~ \lambda^i
\ee
with
\be
\ddd\a = \ddd f_i\wedge\lambda^i + f_i ~\ddd\lambda^i \ .
\ee
In particular, one has
\be
\frac{(\ddd\a)^d}{d!} =
\bigwedge_{i=1}^d\, \left(\ddd f_i \wedge \lambda^i\right)  + f_j~
\Upsilon^j
\ee
where $d=\dim_\C(\cO)=n^2\,(N^2-1)$ is the (real) dimension of the
vacuum orbit $\cO_0$. The forms $f_j~\Upsilon^j$ vanish on-shell, and are
killed by localization in the large $t$ limit. For example, inner
products of the form $\langle\a, \cJ(s)\rangle$, $s\in\ms$ are
non-vanishing off-shell at second order due to \eq{alpha-eval-second},
but these higher-order terms do not contribute because of the
localization in the large $t$ limit. This can be seen explicitly by
rescaling $f_i = \sqrt{t}\,f_i'$.

The corresponding local contribution to the partition function
(\ref{Z-5}) for $t\to\infty$ is then given by
\bea
Z_0 &=&\frac1{\vol(G)}\,
\int_{\mg\times {\cal F}_0}\,\Big[\,\frac{\ddd\phi}{2\pi}\,\Big]~
\frac{t^d}{d!}\,(\ddd\a)^{d}~
\e^{-\ii t\,\langle\a,V_\phi\rangle-\ii \Tr(C_0\, \phi) -\frac{g'}{2}\,
\Tr (\phi^2)}\nn\\[4pt]
&=& \frac1{\vol(G)}\,
\int_{\mg\times {\cal F}_0}\,\Big[\,\frac{\ddd\phi}{2\pi}\,\Big]~
t^d~\bigwedge_{i=1}^d\, \left(\ddd f_i \wedge \lambda^i\right)~
\e^{-\ii N\,t\, f_i \,\phi^i-\ii \Tr(C_0\, \phi) -\frac{g'}{2} \,
\Tr (\phi^2)} \nn\\[4pt]
&=& \frac1{\vol(G)}\,
\int_{\ms}\,\Big[\,\frac{\ddd\phi}{2\pi}\,\Big]~
\e^{-\ii \Tr(C_0\, \phi) -\frac{g'}{2}\,\Tr (\phi^2)}~
\frac 1{N^{d}}\,\int_{\cO_0}\, \bigwedge_{i=1}^{d}\,\lambda^i \ .
\label{Z0-1}\eea
Here the $f_i$ integrals over the fibre $\cJ^2(\mg\ominus\ms)$
have produced delta-functions setting $\phi^i=0$ in
$\mg\ominus\ms$. We can carry out the integral over the moduli space
$\cO_0$ in \eq{Z0-1} by observing that
\be
\frac 1{N^{d}}\,\int_{\cO_0}\,\bigwedge_{i=1}^{d}\,\lambda^i=\int_{G/S}\,
\bigwedge_{i=1}^{d}\,\eta^i= \frac{{\rm vol}(G)}{{\rm vol}(S)} \ ,
\label{intcO0}\ee
where the pullbacks $\cJ^*(\lambda^i) = \eta^i$ define left-invariant
one-forms on the gauge group $G$ dual to $g'_i$, 
with the map $N\,\cJ$ regarded as the
derivative of the diffeomorphism
\be
G/S ~\longrightarrow~ \cO_0 \ , \qquad g ~\longmapsto~ g\, C \,g^{-1}
\ .
\ee

To evaluate the remaining integral over the gauge stabilizer algebra
$\ms\cong\mun(n)$ in \eq{Z0-1}, we note that, for the vacuum critical
point with $C_0=\frac12\,\one_{n\,N}$, the integrand defines a gauge
invariant function $f:\mun(n)\to\R$. We may thus apply to it the Weyl
integration formula which reduces its integral over $\mun(n)$ to an
integral over the Lie algebra $\mun(1)^n$ of the maximal torus
$U(1)^n$ of $U(n)$. It is given by
\be
\int_{\mun(n)}\,[\ddd\phi]~f(\phi)=\frac{{\rm vol}\big(U(n)\big)}
{n!\,(2\pi)^n}\,\int_{\R^n}\,[\ddd s]~\Delta(s)^2~f(s) \ ,
\label{Weylint}\ee
where we have identified $\mun(1)^n\cong\R^n$ in a basis where the
Cartan subalgebra of $U(n)$ is represented by diagonal $n\times n$
matrices $s={\rm diag}(s_1,\dots,s_n)$ by mapping them onto $n$-vectors
$s=(s_1,\dots,s_n)\in\R^n$. Here
\be
\Delta(s)=\prod_{i<j}\,(s_i-s_j)=\det_{1\leq i,j\leq n}\,
\big(s_i^{j-1}\big)
\ee
is the Vandermonde determinant, which is the Weyl determinant for
$U(n)$ arising as the jacobian for the diagonalization of hermitian
matrices on the left-hand side of (\ref{Weylint}). The factor $n!$ is
the order of the Weyl group $\mS_n$ of $U(n)$ acting by permutations
of the components $s_i$ of $s\in\R^n$, while $(2\pi)^n$ is the
volume of the maximal torus $U(1)^n$ with respect to the chosen
invariant Haar measure.

\subsubsection*{\it An integral identity}

We will make use here and in Section~\ref{LocMaxNon-Deg} below of the
integral identity
\bea
&& \int_{\R^n}\, [\ddd s]~\Delta(s)^2~
\e^{-\ii \frac N2\, \sum_i\, s_i + \frac\ii4\, \sum_i\, m_i\, s_i
 -\frac{g}{4}\, \sum_i\, s_i^2} \nn\\
&& \qquad\qquad\qquad~=~ \e^{-\frac{n\,N^2-m\,N}{4g}}\,
\int_{\R^n}\,[ \ddd s]~\Delta(s)^2~
\e^{\frac\ii4\, \sum_i\, m_i\,s_i -\frac{g}{4}\, \sum_i\, s_i^2} 
\label{Lemma}\eea
where  $m = \sum_i\, m_i$. To derive \eq{Lemma}, we set $s = \sum_i\,
s_i$ and $t_i = s_i - \frac 1n\, s$ so that $\sum_i\, t_i =0$. Then
\bea
&& \int_{\R^n}\,[ \ddd s]~\Delta(s)^2~
\e^{-\ii \frac N2\, \sum_i\, s_i + \frac\ii4\, \sum_i\, m_i\, s_i
 -\frac{g}{4}\, \sum_i\, s_i^2} \nn\\
&&\qquad\qquad\qquad~=~ \int_{\R}\,\ddd s~
\e^{-\ii \frac N2\, s + \ii\frac{m}{4n}\, s}~
\int_{\R^n}\, [\ddd t]~\Delta(t)^2~
\e^{\frac\ii4\, \sum_i\, m_i\, t_i-\frac{g}{4}\, \sum_i\,
(t_i + \frac 1n\, s)^2} \nn\\[4pt]
&&\qquad\qquad\qquad~=~ \int_{\R}\,\ddd s~
\e^{-\ii \frac N2\, s + \ii\frac{m}{4n}\, s- \frac{g}{4n}\, s^2}~
\int_{\R^n}\, [\ddd t]~\Delta(t)^2~
\e^{\frac\ii4\, \sum_i\, m_i\, t_i-\frac{g}{4}\, \sum_i\, t_i^2}
\nn\\[4pt]
&&\qquad\qquad\qquad~=~ 2\,\sqrt{\mbox{$\frac{ \pi\,n}{g}$}}~
\e^{-\frac{(2n\,N -m)^2}{16n\, g}}\,
\int_{\R^n}\, [\ddd t]~\Delta(t)^2~
\e^{\frac\ii4\, \sum_i\, m_i\, t_i-\frac{g}{4}\, \sum_i\, t_i^2} \ .
\eea
On the other hand
\bea
\int_{\R^n}\,[ \ddd s]~\Delta(s)^2~
\e^{\frac\ii4\, \sum_i\, m_i\, s_i -\frac{g}{4}\, \sum_i\, s_i^2}&=&
\int_\R\,\ddd s~\e^{\ii\frac{m}{4n}\, s}~
\int_{\R^n}\, [\ddd t]~ \Delta(t)^2~
\e^{\frac\ii4\, \sum_i\, m_i\, t_i -\frac{g}{4}\, \sum_i\,
(t_i + \frac 1n\, s)^2} \nn\\[4pt]
&=&\int_{\R}\, \ddd s~\e^{\ii\frac{m}{4n}\, s -
\frac{g}{4n}\, s^2}~\int_{\R^n}\,[ \ddd t]~\Delta(t)^2~
\e^{ \frac\ii4\, \sum_i\, m_i\, t_i-\frac{g}{4}\, \sum_i\, t_i^2}
\nn\\[4pt]
&=&2\,\sqrt{\mbox{$\frac{ \pi\,n}{g}$}}~
\e^{-\frac{m^2}{16n\, g}}~
\int_{\R^n}\,[ \ddd t]~\Delta(t)^2~
\e^{\frac\ii4\, \sum_i\, m_i\, t_i-\frac{g}{4}\, \sum_i\, t_i^2} \ .
\nn\\
\eea

\subsubsection*{\it Final reduction}

{}From \eq{Z0-1}, \eq{intcO0} and \eq{Weylint} we obtain 
\bea
Z_0 &=& \frac1{\vol(S)}\, \int_{\ms}\,
\Big[\,\frac{\ddd\phi}{2\pi}\,\Big]~
\e^{-\ii \Tr(C_0\, \phi) -\frac{g'}{2}\,\Tr (\phi^2)} \nn\\[4pt]
&=&\frac{1}{n!}\, \frac 1{(2\pi)^{n^2}}\,
\int_{\R^n}\, \Big[\,\frac{\ddd s}{2\pi}\,\Big]~
\Delta(s)^2~\e^{-\ii \frac N2\, \sum_i\, s_i  -\frac{g}{4}\,
\sum_i\, s_i^2}
\label{Z0-2}\eea
where we have substituted \eq{gprime} and used $\vol(S) =
{N}^{N^2/2}~\vol(U(n))$ with respect to the Cartan-Killing metric on
$\ms$, since $S = U(n) \otimes \one_N$. Applying the integral identity
\eq{Lemma} therefore allows us to finally write the partition function
as
\beq
Z_0 = \frac{1}{n!}\,\frac 1{(2\pi)^{n^2+n}}~
\e^{-\frac{ n\,N^2}{4 g}}\,
\int_{\R^n}\, [\ddd s]~\Delta(s)^2~
\e^{ -\frac{g}{4}\, \sum_i\, s_i^2} \ .
\label{Z0final}\eeq
The exponential prefactor in the above expression
 is the Boltzmann weight of
the action \eq{YM-action-prime} evaluated on the vacuum solution. The
remaining quantum fluctuation integral is the standard
expression~\cite{Minahan:1993tp} for the contribution from the global
minimum of the Yang-Mills action on $S^2$ to the $U(n)$ sphere
partition function. It arises from the trivial instanton configuration
with vanishing monopole charges $m_i=0$ in \eq{nidomclass}.

\subsection{Localization at maximally irreducible saddle
  points\label{LocMaxNon-Deg}}

We now turn to the opposite extreme and look at the local contribution
to the partition function \eq{Z-5} from a generic maximally
non-degenerate critical surface. We denote this gauge orbit by
\be
\cO_{\rm max}:= \cC_{(n_1,1),\dots,(n_n,1)} 
= \big\{g\, C \,g^{-1}~\big|~ g \in U(n\,N-{\sf c}_1)\big\}\cong
U(n\,N-{\sf c}_1)/U(1)^n
\ee
and assume that the integers $n_1 > n_2 > \cdots > n_n$ are explicitly
specified. Here we allow also ${\sf c}_1 \neq 0$ which describes sectors 
with non-vanishing $U(1)$ monopole number \eq{tracemodify}.
We want to compute the integral $Z_{\rm max}$ in \eq{Z-5}
over a local neighbourhood $\cN_{\rm max}$ of $\cO_{\rm max}$, which is
independent of $t$ in the large $t$ limit. 

We first need to find a suitable basis for the tangent space $ T_{C}
\cO$ at the irreducible critical point $C$. The definition of the
basis $J_i,\tilde J_i$ introduced in \eq{J-i} naturally extends
to include the non-trivial subspaces $E_0, E_1$ in this case with
\be
J_i = \cJ(g'_i) \ , \qquad \tilde J_j = \cJ^2(g'_j) \ , \qquad
H_i = \cJ(h'_i) \,\in\, \cJ(\mh) = E_0 \qquad \mbox{and} \qquad K_i \in E_1 \ .
\ee
for $g'_i$ and $h'_i$ an orthonormal basis of
$\mg\ominus\mh\ominus\ms$ and of $\mh \ominus \ms$, respectively.
The elements $K_i$ are assumed to form an orthonormal basis of $E_1$,
orthogonal to $\cJ(\mg) \oplus \cJ^2(\mg)$.
Recall from Section~\ref{ExplYMDecomp} that $E_0$ and $E_1$ are
naturally complex vector spaces, whose generators are embedded into
the tangent space decomposition \eq{TCOdecomp} as
\be
K_{i} = \left(\begin{array}{ccccc}0&0 &\vline & 0 & 0 \\
                                      0&0 &\vline& X_{i} & 0 \\
\hline
                           0 & X_{i}^\dagger&\vline & 0&0  \\
                           0 & 0&\vline & 0&0 
\end{array}\right)
\label{E_1-explicit-2}
\ee
and similarly for $H_i$. The complex structure is given by the map $\cJ$, 
which amounts to multiplying $X_i$ by $\ii$. We accordingly take the
real basis $K_i$ to be ordered as $\{K_i\} = \{(\tilde K_i, \cJ(\tilde
K_i))\}$, and similarly for $H_i$. As matrices, all of the generators
$H_i, K_j$ are hermitian. 
The corresponding dual one-forms
$\beta^i,\gamma^i$ are defined as usual by
\be
\big\langle \b^i\,,\, H_j\big\rangle = \delta^i{}_j \qquad \mbox{and}
\qquad \big\langle \g^i\,,\, K_j\big\rangle = \delta^i{}_j
\ee
with all other pairings equal to~$0$.

We need to evaluate the pairing $\langle \a,V_\phi\rangle$. It
vanishes on-shell, and identically on $\cJ^2(\mg)$. Its evaluation on
$\cJ(\mg\ominus\mh\ominus\ms)$ has the form $\langle\a,
\cJ(g'_i\,)\rangle =f_i$, and as before this implies
\eq{alpha-Vphi}. Together with \eq{alpha-E0} and \eq{alpha-E1}, it
follows that the localization one-form $\alpha$ admits an expansion
\be
\a = f_i\, \lambda^i + g_i \,\b^i + k_i\, \g^i
\ee 
where $f_i, g_i, k_i$ vanish on-shell. We can evaluate
\be
\ddd\a = \ddd f_i\wedge \lambda^i + f_i ~\ddd\lambda^i  + \ddd
g_i\wedge \b^i + g_i~ \ddd\b^i + \ddd k_i\wedge \g^i+ k_i ~\ddd\g^i
\ee
using \eq{da-E0} and \eq{da-E1} to get
\be
\langle \ddd\a,H_i\wedge H_j\rangle =0 \qquad \mbox{and} \qquad
\langle \ddd\a,K_i \wedge K_j\rangle = A_{ij} \ ,
\ee
where
\be
A_{ij} =  2\ii \Tr\big(K_i~\ad_{C_0}(K_j)\big)
\label{Aijdef}\ee
is an antisymmetric matrix. Furthermore, $\ddd\a$ vanishes when
evaluated on mixed terms of the form $K_i\wedge \cJ(g), K_i\wedge
\cJ^2(g)$, $H_i\wedge \cJ(g'\,)$ and $H_i\wedge \cJ^2(g'\,)$ with
$g\in\mg,\, g'\in\mg\ominus\mh\ominus\ms$. Therefore
\be
\ddd\a = \ddd f_i\wedge \lambda^i 
+ \mbox{$\frac 12$} \, A_{ij}~ \g^i \wedge \g^j + O_f
\ee
where $O_f$ denotes contributions which vanish on-shell such as $f_i
~\ddd\lambda^i$. One then has
\be
\frac{(\ddd\a)^{d-d_0}}{(d-d_0)!} = 
\pfaff( A)~\Big(\,\bigwedge_{i=1}^{2d_1}\,\g^i\Big)~ \wedge~
\Big(\,\bigwedge_{j=1}^{d-d_0-d_1}\,\ddd f_j \wedge \lambda^j\Big)
+ O_f
\ee
where $d_0$ (resp. $d_1$) is the complex dimension of the vector space
$E_0$ (resp. $E_1$), and
\be
\pfaff(A)= \epsilon^{i_1\cdots i_{2d_1}}\, 
A_{i_1 i_2}\cdots A_{i_{2d_1-1} i_{2d_1}}
\ee
is the pfaffian of the antisymmetric matrix $A=(A_{ij})$.

Let us now recall the local geometry and define its symplectic model.
The $G$-equivariant tubular neighbourhood $\cN_{\rm max}$ of $\cO_{\rm max}$ 
has an equivariant
retraction~\cite{GSbook} by a local equivariant symplectomorphism onto
the local symplectic model ${\cal F}_{\rm max}$, defined to be an
equivariant symplectic vector bundle over $\cO_{\rm max}$ with fibre
$\cJ^2(\mg\ominus\mh\ominus\ms)\oplus E_1$ which is a sub-bundle of
the tangent bundle $T \cO$ restricted to $\cO_{\rm max}$. This means
that the tangent space to ${\cal F}_{\rm max}$ is given by
\beq
T_C \cO_{\rm max}
\oplus \cJ^2(\mg\ominus\mh\ominus\ms)\oplus E_1 ~\cong~ E_0 \oplus
\cJ(\mg\ominus\mh\ominus\ms) \oplus\cJ^2(\mg\ominus\mh\ominus\ms)\oplus
E_1~ = ~T_C \cO \ ,
\label{TF-max}
\eeq
the symplectic form on ${\cal F}_{\rm max}$ is simply $\omega$, and the
hamiltonian $G$-action on ${\cal F}_{\rm max}$ descends from the
moment map $\mu$. In physical terms, the gauge fields are split along
the moduli space $\cO_{\rm max}$, plus infinitesimal non-gauge variations
belonging to $\cJ^2(\mg\ominus\mh\ominus\ms)$ and unstable modes in
the subspace $E_1$. Due to the presence of the localization form $\a$
in the action, we can restrict ourselves to this model ${\cal F}_{\rm max}$
replacing $\cN_{\rm max}$. Identically to the case of
Section~\ref{LocVacSurface} above, the canonical symplectic integral
over $\mg\times\cN_{\rm max}$ will in this way reduce to an integral
over $\ms\times\cO_{\rm max}$ and the localization now resembles that
at an irreducible flat connection of Chern-Simons
theory~\cite{witten}.

We may now proceed to calculate
\bea
Z_{\rm max} &=& \frac1{\vol(G)}\,\int_{\mg\times \cN_{\rm max}}\,
\Big[\,\frac{\ddd\phi}{2\pi}\,\Big]~
\exp\Big(\omega + t \,\big(\ddd\a-\ii\langle\a,V_\phi\rangle
\big)-\ii \Tr(C_0\, \phi) -\mbox{$\frac{g'}{2}$}\, \Tr \big(\phi^2
\big)\Big) \nn\\[4pt]
&=&\frac1{\vol(G)}\,
\int_{\mg\times \cO_{\rm max} \times \cJ^2(\mg\ominus\mh\ominus\ms)
\times E_1} \,
\Big[\,\frac{\ddd\phi}{2\pi}\,\Big]~ \frac{(t~\ddd\a)^{d-d_0}}
{(d-d_0)!}\wedge\frac{\omega^{d_0}}{d_0!}~
\e^{-\ii t\,\langle\a,V_\phi\rangle-\ii \Tr(C_0 \,\phi) -
\mbox{$\frac{g'}{2}$}\, \Tr(\phi^2)} \nn\\[4pt]
 &=& \frac1{\vol(G)}\,
\int_{(\mg\ominus\mh\ominus\ms) \oplus \mh \oplus \ms}\,
\Big[\,\frac{\ddd\phi}{2\pi}\,\Big]~\pfaff (A)~
\nn\\ && \qquad\qquad\qquad
\times~\int_{\cO_{\rm max} \times \cJ^2(\mg\ominus\mh\ominus\ms)
\times E_1}\,t^{d-d_0}~\Big(\,\bigwedge_{i=1}^{2d_1}\,\g^i
\Big)~ \wedge~\Big(\,\bigwedge_{j=1}^{d-d_0-d_1}\,\ddd f_j
\wedge \lambda^j\Big)~\wedge~\frac{\omega^{d_0}}{d_0!}
\nn\\ && \qquad\qquad\qquad
\times~\e^{-\ii t\, (N \,f_i\,\phi^i+ \langle \a,V_{\phi'}\rangle)
-\ii \Tr(C_0\, \phi) -\frac{g'}{2}\, \Tr (\phi^2)}
\label{Zmaxproceed}\eea
with $\phi'\in\mh\oplus\ms$. In the second line we have used the fact
that $\ddd\a$ vanishes when evaluated on the subspace $E_0$, and
therefore we need $d_0$ powers of $\omega$ to yield a non-trivial
volume form. Then $(t~\ddd\a)^{d-d_0}\wedge\omega^{d_0}$ is the only
term which survives in the large $t$ limit. We will modify this below
by adding a second localization form $\a'$ in order to write the
localization integral in the generic form (\ref{Z-5}) without the
symplectic two-form $\omega$.

We can now evaluate the integrals in \eq{Zmaxproceed} over $f_i$ in
the fibre $\cJ^2(\mg\ominus\mh\ominus\ms)$ and
$\phi^i\in\mg\ominus\mh\ominus\ms$ as in Section~\ref{LocVacSurface}
above, which localizes for $t\to\infty$ to an integral over the
subspace $E_1$ and the gauge orbit $\cO_{\rm max}$ given by
\bea
Z_{\rm max} &=&\frac1{\vol(G)}\,\int_{\mh \oplus \ms}\,
\Big[\,\frac{\ddd\phi}{2\pi}\,\Big]~
\frac{\pfaff (A)}{N^{d-d_0-d_1}}~\int_{\cO_{\rm max} \times E_1}\,
t^{d_1}~\Big(\,\bigwedge_{i=1}^{2d_1}\,\g^i\Big)~ \wedge~
\Big(\,\bigwedge_{j=1}^{d-d_0-d_1}\,\lambda^j\Big)~\wedge~
\frac{\omega^{d_0}}{d_0!}\nn\\ && 
\qquad\qquad\qquad\qquad\qquad\qquad\qquad
\times~\e^{-\ii t\,\langle \a,V_{\phi}\rangle
-\ii \Tr(C_0\, \phi) -\frac{g'}{2}\, \Tr (\phi^2)} \ .
\label{Zdeg-4}
\eea
The gauge invariant volume form for the integration domain whose
tangent space is $E_0$ is given by the symplectic volume form
$\omega^{d_0}/d_0!$, since $\ddd\a$ vanishes on $E_0$, but this will
be modified below. It remains to compute the integral over $E_1$. Upon
evaluating $\langle \a,V_\phi\rangle$ at second order on $E_1$,
i.e. away from the critical surface, we will find below that this
pairing becomes a quadratic form which leads to a localization through
a gaussian integral. However, to evaluate it explicitly it is easier
to first localize the integral over $E_0$, which presently is a
complicated non-gaussian integral which does not admit a gaussian
approximation at $t\to\infty$ and is difficult to evaluate in a closed
analytic form. But this can be done by adapting a trick  taken
from~\cite{witten}, which amounts to adding a further suitable
localization one-form $\a'$, or equivalently a cohomologically trivial
form $Q\a'$, to the action in \eq{Z-5}. Indeed, we may compute $Z_{\rm
  max}$ using any other invariant form $\alpha'$ which is homotopic to
$\alpha$ on the open neighbourhood $\cN_{\rm max}$. The one-form
$\alpha'$ need only be non-vanishing on $E_0\subset\cN_{\rm max}$, as the
other integrals can be directly carried out.

\subsubsection*{\it The localization form $\alpha'$}

In order to evaluate the integrals over $E_0$ and $\mh$,
following~\cite{witten} we introduce an additional localization term
$\exp(t~Q\a'\,)$ in the partition function with
\be
\a' := -\ii\Tr(\theta\, \phi) \Big|_{E_0} = -\mbox{$\frac{2}N$}\,
\cJ~\ddd\Tr(C\, \phi)\Big|_{E_0} \ .
\label{alphaprime}
\ee
The projection onto $E_0$ is equivalent to projecting $\phi\in\mg$
onto $\mh$. This one-form is equivariant on-shell, and it can be
extended to the $G$-equivariant tubular neighbourhood $\cN_{\rm max}$ of
the critical surface $\cO_{\rm max}$ as follows.
On the tangent space $\cJ(\mg\ominus\mh\ominus\ms) \oplus E_0$ of
$T\cO_{\rm max}$ \eq{TF-max} there is an equivariant projection onto the subspace
$E_0$. In this way  $\a'$ is properly defined on the
local model, and can hence be extended to $\cN_{\rm max}$. One could
also define $\a'=-\ii\chi\,\Tr(\theta\, \phi)\big|_{E_0}$ using a smooth
$G$-invariant cutoff function $\chi$ with support near the given
saddle-point and $\chi=1$ in the tubular neighbourhood, which is
globally well-defined over $\cN_{\rm max}$ as an equivariant
differential form. Note that $t_1\,\a+t_2\,\a'$ vanishes only on the
original critical points for any $t_1,t_2\in\R$ with $t_1\neq0$, and
no new ones are introduced. Then our previous computation \eq{Z-4}
would essentially go through, since $\a'$ vanishes on
$\cJ(\mg\ominus\mh\ominus\ms)$ and there are no critical points where
$\ddd\chi\neq0$. It is therefore just as good a localization form to
use as $\a$ is. It follows that the modification of the canonical
symplectic integral over $\cN_{\rm max}$ given by
\be
Z_{\rm max} = \frac1{\vol(G)}\,\int_{\mg\times \cN_{\rm max}}\,\Big[\,
\frac{\ddd\phi}{2\pi}\Big]~\exp\Big(\omega + t_1~Q\a + t_2~Q\a'-
\ii \Tr(C_0\, \phi) -\mbox{$\frac{g'}{2}$}\,\Tr\big(\phi^2\big)\Big)
\label{Zmaxt1t2}\ee
is independent of both $t_1,t_2\in\R$. 
Then $\a'$ will localize the integral over
$\mh\subset\mg$ as well as the integral over the unstable modes in
$E_1$, without the need to expand $\langle\a,V_\phi\rangle$ to higher
order.

\subsubsection*{\it Integration over $\mh$}

The new localization form $\a'$ satisfies 
\be
\ddd\a' = \ii\Tr\big(\theta^2\,\phi\big)\Big|_{E_0} =
-\mbox{$\frac\ii2$}\, \Tr\big(\theta\, [\phi,\theta]\big)\Big|_{E_0}
\ee
and
\be
\langle \a',V_{h_i}\rangle = -\mbox{$\frac{2}N$}\,\Tr\big(
\cJ(V_{h_i})\, \phi\big) = \mbox{$\frac{2}N$}\,\Tr\big(V_{h_i}\,
\cJ(\phi)\big) = 2\Tr\big(H_i\, \cJ(\phi)\big) \ ,
\ee
where $H_{i} = \cJ(h_{i})$ with $h_{i}$ a basis of $\mh$.
This produces a gaussian integral localizing $\mh$ to the gauge
stabilizer algebra $\ms\cong\mun(1)^n$. To evaluate it, we will need
the matrix
\be
M_{ij} :=  \Tr(H_{i}\, H_{j}) 
\label{inner-H}
\ee
which is hermitian since we take $H_{i}$ and $h_{i}$ to be hermitian.
Similarly, one has
\bea
\langle \ddd\a',H_i\wedge H_j\rangle &=& \mbox{$ \frac{4\ii}{N^2}$}\,
\Tr\big(\cJ(H_i)\,[s,\cJ(H_j)]\big) \nn\\[4pt]
&=&-\mbox{$\frac{4\ii}{N^2}$}\,\Tr\big(H_i\,[s,\cJ^2(H_j)]\big)
\nn\\[4pt] &=&\mbox{$\frac{4\ii}{N^2}$}\, \Tr\big(H_i\,[s,H_j]\big)~=:~
\mbox{$\frac{4\ii}{N^2}$}\,\tilde A_{ij}
\label{tildeAdef}\eea
where we have restricted to $\phi=s \in \ms$ using the
localization. This implies that
\be
\ddd\a' =\mbox{$\frac{2\ii}{N^2}$}\,\tilde A_{ij}~\b^i\wedge \b^j 
\qquad \mbox{and} \qquad \frac{(\ddd\a'\,)^{d_0}}{d_0!} 
= \left(\mbox{$\frac{4\ii}{N^2}$}\right)^{d_0}~\pfaff\big(\tilde A\,\big)~
\bigwedge_{i=1}^{2d_0}\, \beta^i \ .
\label{dalphaprime}\ee
To evaluate the matrices $M=(M_{ij})$ and $\tilde A=(\tilde A_{ij})$
above explicitly, we recall that the basis $H_i:=H_{kl;i}$ (where
$k,l$ are block indices) of $E_0$ takes the block form 
\be
H_{kl;i} = \left(\begin{array}{ccccc}0&0 &\vline & 0 & 0 \\
                                      0&0 &\vline& Y_{lk;i} & 0 \\
\hline
                           0 & Y_{lk;i}^\dagger&\vline & 0&0  \\
                           0 & 0&\vline & 0&0 
\end{array}\right) = \cJ(h_{kl;i})
\label{E_0-explicit-2}
\ee
where $h_{kl;i}\in \mh$ is a hermitian block matrix with a similar
block decomposition. They are orthogonal for different $k,l$, and we will
often omit the indices $k,l$. Note that the complex structure on $E_0$
defined by the map $\cJ$ is compatible the natural complex structure
on $\mh$. This basis is particularly useful for evaluating the
pfaffian which appears in \eq{dalphaprime}, because $\ad_s(H_{kl;i})$
for $s\in\ms$ acts as multiplication by $(s_k - s_l)$ in the
upper-right blocks of \eq{E_0-explicit-2}. It follows that
\be
\ii\,\ad_{C_0}(H_{kl;i}) = c_{lk}\, \cJ(H_{kl;i}) \qquad \mbox{and}
\qquad \ii\,\ad_{s}( H_{kl;i}) = (s_l-s_k)\, \cJ(H_{kl;i})
\label{J-ad-id}
\ee
where the eigenvalues $c_{lk} >0$ are defined in
\eq{ad-C0-explicit}. These formulas hold only for $k>l$, and analogous
statements are true for the subspace $E_1$.

We can choose an orthogonal basis $Y_i$ such that $G_{ij} = 2\,
\Tr\big(Y_i\,Y_j^\dagger\big)$ is diagonal, as $G_{ij}$ is a hermitian
matrix. Then
\bea
\Tr\big(H_i\, H_j\big) &=& \Tr\big(Y_i\,Y_j^\dagger + Y_i^\dagger\,
Y_j\big) ~=~ G_{ij} \ ,  \nn\\[4pt]
\Tr\big(H_i\, \cJ(H_j)\big) &=& \Tr\big(\ii Y_i\,Y_j^\dagger -\ii
Y_i^\dagger\, Y_j\big) ~=~0 \ .
\eea
This means that the symmetric matrix $M=(M_{ij})$ in (\ref{inner-H})
has the block decomposition
\be
M = \left(\begin{array}{ccc} G & 0  \\ 0 & G 
\end{array}\right)
\label{Mblock}\ee
in the basis $(\tilde H_i, \cJ(\tilde H_i))$, and similarly the matrix
$\tilde A$ in (\ref{tildeAdef}) is given by
\bea
\tilde A_{ij} &=& \Tr\big(H_i~\ad_{s}(H_j)\big)\nn\\[4pt]
&=& -\ii (s_l-s_k)\,\Tr\big(H_i\,\cJ(H_{j})\big) ~=~
-\ii (s_k-s_l)\, \left(\begin{array}{ccc} 0 & G  \\ -G & 0
\end{array}\right)_{ij} \ .
\label{tildeAblock}\eea
We can read off the pfaffian from this expression and use
(\ref{Mblock}) to write it as
\be
\pfaff\big(\tilde A\,\big) = 
(-\ii)^{d_0}\sqrt{\det(M)}\; \prod_{k > l} \,(s_k-s_l)^{|n_k-n_l|+1} \ .
\label{PfaffA-eval}
\ee

We can now evaluate the localization integral
\beq
\int_{\mh}\,\Big[\,\frac{\ddd\phi}{2\pi}\,\Big]~t_2^{d_0}\,
\frac{(\ddd \a'\,)^{d_0}}{d_0!}~\e^{-\ii t_2 \,\langle \a',V_\phi\rangle } 
=\left(\mbox{$\frac{4\ii}{N^2}$}\right)^{d_0}\,
\int_{\mh}\,\Big[\,\frac{\ddd\phi}{2\pi}\,\Big]~t_2^{d_0}~
\pfaff\big(\tilde A\,
\big)~\e^{-2\ii t_2\, \phi^{i}\, M_{ij}\, \phi^{j}} ~
\bigwedge_{i=1}^{2d_0}\, \beta^i
\eeq
where $\phi = \phi^{i}\, h_{i}= \phi^{{kl;i}}\, h_{{kl;i}}$. The
oscillatory gaussian integral is defined by analytic continuation $t_2
\to t_2-\ii \varepsilon$ for a small positive parameter $\varepsilon$,
which we are free to do as the partition function is formally independent of
$t_2$. With this continuation understood and a suitable orientation of
the vector space $\mh$, we readily compute
\bea
\int_{\mh}\,\Big[\,\frac{\ddd\phi}{2\pi}\,\Big]~t_2^{d_0}\,
\frac{(\ddd \a'\,)^{d_0}}{d_0!}~\e^{-\ii t_2 \,\langle \a',V_\phi\rangle } 
&=& \left(\mbox{$\frac{4\ii}{N^2}$}\right)^{d_0}\,
\left(\mbox{$\frac 1{2\pi}$}\right)^{2d_0}\,
\left(-\mbox{$\frac{\pi}{2\ii}$}\right)^{d_0}~
 \frac{\pfaff\big(\tilde A\,\big)}{\sqrt{\det(M)}}~
\bigwedge_{i=1}^{2d_0}\, \beta^i \nn\\[4pt]
&=& \frac{\ii^{d_0}}{( 2\pi\, N^2)^{d_0}}~
\prod_{k > l}\, (s_k-s_l)^{|n_k-n_l|+1}~
\bigwedge_{i=1}^{2d_0}\, \beta^i \ .
\label{locmhms}\eea
This integral thus produces a measure on $\ms$ which we will use below
to perform the remaining integral over the stabilizer.

\subsubsection*{\it Integration over $E_1$}

Now that the $\phi$-integration in \eq{Zdeg-4} is localized onto
$\ms$, we can proceed to evaluate the integral over $E_1$. This space
has a basis $K_i$ with block decomposition $K_{kl;i}$ similar to
\eq{E_0-explicit-2} for $n\geq k>l\geq1$ (for $k<l$ the $K_{kl;i}$ do
not exist), which are non-vanishing if $n_k>n_l+1$. We need to
evaluate $\langle \a,V_s\rangle$ for $s\in\ms$ up to second order in
the fluctuations about the critical point in $E_1$, which is
non-tangential to the gauge orbit $\cO_{\rm max}$. For this, we
introduce real linear coordinates $x^i, y^i$, $i=1,\dots,d_1$ on $E_1$
such that a generic vector $V_\Psi \in E_1$ is parametrized as
\be
V_\Psi = \big(x^i\, K_i\,,\, y^i\, \cJ(K_i)\big) 
=\left(\begin{array}{ccccc}0&0 &\vline & 0 & 0 \\
                                      0&0 &\vline& z^i\, X_{i} & 0 \\
\hline
                           0 & \obar z^{\,i}\, X_{i}^\dagger&\vline &
                           0&0  \\ 0 & 0&\vline & 0&0 
\end{array}\right)
\ee 
where we have introduced complex coordinates $z^i = x^i + \ii
y^i$. Then $\g^i = \ddd x^i$ and $\g^{i+d_1}=\ddd y^i$ for
$i=1,\dots,d_1$.

As above, we can choose coordinates such that $G_{ij} =
2\,\Tr\big(X_i\, X_j^\dagger\big)$ is diagonal. Then \eq{alpha-expand}
gives
\be
\langle \a,V_s\rangle = -\Tr\big(\ad_s(V_\Psi)~\ad_{C_0}(V_\Psi)
\big)= \big(x^i\,,\,y^i\big) ~\tilde M_{ij}(s) ~
\left(\begin{array}{c} x^j \\ y^j\end{array}\right)
\ee
to second order, where
\bea
\tilde M_{ij}(s) &=& \Tr\big(K_i~\ad_s~\ad_{C_0}(K_j)\big)\nn\\[4pt]
&=& (s_k-s_l)\, c_{kl} \; \Tr(K_i\, K_j)~=~
(s_k-s_l)\, c_{kl}\, \left(\begin{array}{ccc} G & 0  \\ 0 & G 
\end{array}\right)_{ij}
\label{tildeMinblock}\eea
is a symmetric matrix and we have used the obvious analog of
\eq{J-ad-id} for the basis $K_i$. Similarly, the antisymmetric matrix
$A$ in \eq{Aijdef} can be expressed as
\be
A_{ij} = 2\ii \Tr\big(K_{kl;i}~\ad_{C_0}(K_{kl;j})\big)
= 2c_{lk}\,\Tr\big(K_{kl;i}\, \cJ(K_{kl;j})\big)
= 2c_{kl}\, \left(\begin{array}{ccc} 0 & G  \\ -G & 0
\end{array}\right)_{ij} \ ,
\ee
and using \eq{tildeMinblock} its pfaffian is therefore given by
\be
\pfaff (A)
= 2^{d_1}~\sqrt{\det\big(\tilde M(s)\big)}~
\prod_{k > l}\,(s_k-s_l)^{1-|n_k-n_l|} \ .
\label{pfaffAtildeM}\ee
The pfaffians $\pfaff (\tilde A\,)$ and $\pfaff (A)$ represent the
$S$-equivariant Euler classes in $H_S(\cO_{\rm max})$ of equivariant
bundles over $\cO_{\rm max}$ with fibres $E_0$ and $E_1$,
respectively, in terms of the weights $s_k$ for the (trivial)
$S$-action on $\cO_{\rm max}$. They are the typical representatives of
fluctuations in equivariant localization~\cite{szaboloc,BGVbook}, and
they also appear in the nonabelian localization formulas of~\cite{witten} and
of~\cite{Woodward:2004xz}. Using the analytic continuation $t_1 \to
t_1-\ii\varepsilon$ and a suitable orientation of $E_1$ as before, we
can now evaluate the oscillatory gaussian integral
\be
\int_{E_1}\,\prod_{i=1}^{d_1}\,\ddd x^i~\ddd y^i ~ t_1^{d_1} ~
\e^{-\ii t_1\,\langle \a,V_s\rangle}
= \left(\frac{\pi}{\ii}\right)^{d_1}\,\frac 1{\sqrt{\det
\big( \tilde M(s)\big)}} \ .
\label{gaussintE1}\ee

\subsubsection*{\it Symplectic integral over ${\cal F}_{\rm max}$}

Putting the results \eq{Zdeg-4}, \eq{locmhms}, \eq{pfaffAtildeM} and
\eq{gaussintE1} together, we may evaluate the large $t_1,t_2$ limit of
the symplectic integral \eq{Zmaxt1t2} to obtain
\bea
Z_{\rm max} &=& \frac1{\vol(G)}\,\int_{\mg\times {\cal F}_{\rm max}}\,
\Big[\,\frac{\ddd\phi}{2\pi}\,\Big]~
\exp\Big(\ddd(t_1\, \a+t_2\,\a'\,) - \ii\langle t_1\,\a+t_2\,
\a',V_\phi\rangle\Big)\nn\\ && \qquad\qquad\qquad \qquad \times~
\e^{-\ii \Tr(C_0\, \phi) -\frac{g'}{2}\,\Tr(\phi^2)} \nn\\[4pt]
&=& \frac1{\vol(G)}\, \left(\frac{\pi}{\ii}\right)^{d_1}\,
 \frac{\ii^{d_0}}{(2\pi\, N^2)^{d_0}}\,
\int_{\ms}\,\Big[\,\frac{\ddd s}{2\pi}\,\Big]~ \prod_{k > l}\,
(s_k-s_l)^{|n_k-n_l|+1}~
\frac{\pfaff(A)}{\sqrt{\det\big(\tilde M(s)\big)}} \nn\\
&& \qquad\qquad\qquad \qquad\times~
\frac 1{N^{d-d_0-d_1}}\, \int_{\cO_{\rm max}} \,\Big(\,
\bigwedge_{j=1}^{d-d_0-d_1}\,\lambda^j\Big)~ \wedge ~\Big(\,
\bigwedge_{i=1}^{2d_0}\,\b^i \Big)~
\e^{-\ii \Tr(C_0\,s) -\frac{g'}{2}\, \Tr (s^2)}\nn\\[4pt]
&=&\frac1{\vol(G)}\,\frac{\ii^{d_0-d_1}}{(2\pi)^{d_0-d_1}}\,
\prod_{k=1}^n\, \sqrt{n_k}\, \int_{\R^n}\,\Big[\,\frac{\ddd s}{2\pi}\,
\Big]~\Delta(s)^2~
\e^{-\ii \Tr(C_0\, s) -\frac{g'}{2}\, \Tr (s^2)}\nn\\
&& \qquad\qquad\qquad \qquad\times~
\frac 1{N^{d+d_0-d_1}}\, \int_{\cO_{\rm max}} \,\Big(\,
\bigwedge_{j=1}^{d-d_0-d_1}\,\lambda^j\Big)~ \wedge ~\Big(\,
\bigwedge_{i=1}^{2d_0}\,\b^i \Big)
\label{Zmax-s1}\eea
where we have transformed the integration over $\phi =
s=\diag(s_1~\one_{n_1},\dots,s_n~\one_{n_n}) \in \ms$ to an integral
over $s=(s_1,\dots,s_n)\in\R^n$. We can carry out the integral over
the moduli space $\cO_{\rm max}$ by observing again
\be
\frac 1{N^{d+d_0-d_1}}\, \int_{\cO_{\rm max}}\,\Big(\,
\bigwedge_{j=1}^{d-d_0-d_1}\,\lambda^j\Big)~ \wedge ~\Big(\,
\bigwedge_{i=1}^{2d_0}\,\b^i \Big)
= \int_{G/S}\,\bigwedge_{j=1}^{d+d_0-d_1}\,\eta^j
= \frac{\vol(G)}{\vol(S)} \ ,
\label{Omaxint}\ee
where $\cJ^*(\lambda^i) = \eta^i$ are left-invariant one-forms on the
gauge group $G$. Note that \eq{Omaxint} includes the integral over $E_0$,
and $\dim_\R (\mg \ominus \ms) = d + d_0 - d_1$. 
We also have $\vol(S) = \prod_k\, 2\pi\, \sqrt{n_k}$ in our
metric on $\ms$, since $S=\prod_k\,U(1)\otimes\one_{n_k}$, 
and $C_0(n_i) = \frac {N}{2n_i}~\one_{n_i}$.
Using furthermore $d_0- d_1 = n^2-n$ which is an even integer,
we may then bring (\ref{Zmax-s1}) into the form
\bea
Z_{\rm max} &=& \frac{\ii^{n^2-n}}{(2\pi)^{n^2+n}}\,
\int_{\R^n}\, [\ddd s]~ \Delta(s)^{2}~
\e^{-\ii \Tr(C_0\,s) -\frac{g'}{2} \,\Tr (s^2)} \nn\\[4pt]
&=& \frac{\ii^{n^2-n}}{(2\pi)^{n^2+n}}\,
\int_{\R^n}\, [\ddd s]~ \Delta(s)^{2}~
\e^{-\frac\ii2\,  N\,\sum_i\, s_i  -\frac{g}{4}\, \sum_i
  \frac{n_i}{N}\, s_i^2 } \label{Zmax-s2} \\[4pt]
&=& \frac{\ii^{n^2-n}}{(2\pi)^{n^2+n}}\, \frac{N^{n/2}}
{\prod\limits_{k=1}^n\,\sqrt{n_k}}\,
\int_{\R^n}\, [\ddd\tilde s\,]~ \prod_{k > l} \,
\Big(\,\sqrt{\mbox{$\frac{N}{n_k}$}}\, \tilde s_k-\sqrt{
\mbox{$\frac{N}{n_l}$}}\,\tilde s_l\Big)^{2}~
\e^{-\frac\ii2\, \sum_i\, \sqrt{\frac{N^3}{n_i}}\,\tilde s_i 
-\frac{g}{4}\, \sum_i\, \tilde s_i^{\,2} } \nn
\eea
where $\tilde s_i := \sqrt{n_i/N} \,s_i$. Completing the square of the
gaussian function of $\tilde s_i$ in \eq{Zmax-s2} identifies the
Boltzmann weight of the action \eq{YM-action-prime} on the
non-degenerate solution in \eq{action-eval}. In the large $N$ limit,
we substitute \eq{nidomclass} with $\tilde s_i
\approx\big(1+\frac{m_i}{2N}\big)\, s_i$. Neglecting terms of order
$\frac 1N$ then reduces \eq{Zmax-s2} to
\beq
Z_{\rm max}\approx \pm\, \frac 1{(2\pi)^{n^2+n}}\, 
\int_{\R^n}\, [\ddd s]~ \Delta(s)^{2}~
\e^{- \frac\ii2\,N \,\sum_i\, s_i}~
\e^{\frac\ii4\, \sum_i\, m_i\, s_i  -\frac{g}{4}\, \sum_i\, s_i^2 } \
,
\label{Zmax-s3}\eeq
and an application of the integral identity (\ref{Lemma}) leads to our
final result
\beq
Z_{\rm max}\approx \pm\, \frac 1{(2\pi)^{n^2+n}}~
\e^{-\frac{n\,N^2- m\,N}{4 g}}\,
\int_{\R^n}\, [\ddd s]~\Delta(s)^{2}~
\e^{\frac\ii4\, \sum_i\, m_i\, s_i  -\frac{g}{4}\, \sum_i\, s_i^2 } \
.
\label{Zmaxfinal}\eeq
The exponential prefactor in this formula exhibits the shift of the
vacuum action, corresponding to the modification of the trace
constraint \eq{UnTrconstr} to \eq{tracemodify}, by the Chern class
${\sf c}_1=m=\sum_i\,m_i$. The remaining contributions coincide with
the classical result~\cite{Minahan:1993tp} for the contribution to the
$U(n)$ sphere partition function from the Yang-Mills instanton on
$S^2$ specified by the configuration of magnetic monopole charges
$m_1,\dots,m_n\in\Z$. In particular, using the standard manipulation
of~\cite{Minahan:1993tp} one can change integration variables in
\eq{Zmaxfinal} to identify the anticipated Boltzmann weight of the
action \eq{action-eval-2}.

\bigskip

\section{Abelianization\label{Abelianization}}

In the following sections we will describe an alternative technique of
evaluating
the partition function of $U(n)$ Yang-Mills theory on the fuzzy sphere
$S_N^2$, within the framework of our symplectic model. This method can
be regarded as a finite-dimensional version of the technique of
abelianization for ordinary Yang-Mills theory in two
dimensions~\cite{Blau:1995rs}, which can be used to derive the
strong-coupling expansion of the gauge theory and agrees with the
nonabelian localization. The advantage of this formalism is that it
captures {\it all} classical contributions to the partition function
in a single go and for any $N$, in contrast to nonabelian localization
which requires analysis of each type of critical point individually
and only yields tractable expressions in the large $N$ classical
limit. Its downfall is that it leads to somewhat cumbersome
expressions for the partition function which arise from a rather
different sort of localization. This is analogous to the case of gauge
theory on the two-dimensional noncommutative torus whose
strong-coupling expansion involves the addition of infinitely many
higher Casimir operators to the usual Migdal
formula~\cite{Paniak:2003gn}, or its matrix model regularization which
is given by a complicated combinatorial formula~\cite{szaborev1}. This
complexity makes it difficult to explicitly extract the contributions
from fuzzy sphere instantons, and we will examine this problem more
thoroughly in the next section. Here we shall derive in detail our
alternative abelianized formula for the partition function
(\ref{Z-1}), representing yet another new solution for quantum gauge
theory on the fuzzy sphere.

Let us start from the partition function in the form \eq{Z-2}. The
crucial observation is that the function $f:\mg\to\R$ defined by the
symplectic integral
\be
f(\phi):=\frac1{{\rm vol}(G)}\,\int_{\cO(\Xi)}\,
\exp\left(\omega-\ii\Tr(C_0\,\phi)-\mbox{$\frac{g'}2$}\,
\Tr\big(\phi^2\big)\right)
\label{fphidef}\ee
is gauge invariant. Analogously to what we did in
Section~\ref{LocVacSurface}, we may therefore apply the Weyl
integration formula \eq{Weylint} which reduces its integral over the
gauge algebra $\mg=\mun(n\,N)$ to an integral over the Lie algebra
$\mun(1)^{n\,N}$ of the maximal torus $T=U(1)^{n\,N}$ of
$G=U(n\,N)$. This rewriting of the $\phi$-integral in (\ref{Z-2}) is
called {\it diagonalization} or {\it abelianization}, and it can be
thought of as the eigenvalue representation of the gauge theory
regarded as a matrix model. In this way we may bring the partition
function into the form
\be
Z=\frac1{(n\,N)!}\,\int_{\R^{n\,N}}\,
\Big[\,\frac{\ddd p}{2\pi}\,\Big]~
\eee^{-\frac{g'}4\,\Tr(p^2)}~\Delta(p)^2~Z_\cO(p) \ ,
\label{partfndiag}\ee
where
\be
Z_\cO(p)=\int_{\cO(\Xi)}\,\exp\Big(\omega-\mbox{$\frac\ii2$}\,
\Tr(p\,C)\Big)
\label{calZdef}\ee
is the Fourier transform of the orbit $\cO(\Xi)$ and we have
identified $(n\,N)$-vectors with diagonal matrices $p={\rm
  diag}(p_1,\dots,p_{n\,N})\otimes\sigma^0$.

Localization can then be applied to the symplectic integral \eq{calZdef}
in three different ways, by:
\begin{enumerate}
\item Considering $p \in \mun(\cN\,)$ and observe that $Z_\cO(p)$ can 
be considered as being
  invariant under $p \to U^{-1}\, p\, U$ for $U \in U(\cN\,)$. One can then
 evaluate the integral over the orbit space $\cO(\Xi)$ directly
  using the Itzykson-Zuber formula \eq{IZformula} for the unitary group
  $U(\cN\,)$. This is essentially the calculation that was carried out
  in~\cite{matrixsphere}, which is adapted to the present formulation 
   in Section~\ref{IZ-loc}.
 It amounts to an abelian localization of
  the original orbit integral via the Duistermaat-Heckman theorem.
\item Considering $p \in \mun(nN) \otimes \sigma^0$ and apply abelian
  localization to the maximal torus $T$ of the gauge group $G = U(n\,N)$.
  This will be elaborated in detail in Section~\ref{Abel-loc}, 
taking advantage of a suitable polar decomposition of the orbit space.
 This in turn will involve a localization onto the radial $U(N_+)\times
  U(N_-)$-foliation,  accompanied by a fluctuation
  integral over the moduli space of symplectic leaves.
\item Adding a localization form $Q\a$ as in Section~\ref{NonabLoc},
  and applying nonabelian localization techniques to write the
  partition function as a sum over local contributions from Yang-Mills
  critical points. 
\end{enumerate}
Technique~3 here was of course dealt with at length in
Section~\ref{NonabLoc}, and will be compared 
in some detail to the other two approaches below.
 Comparison with
technique~1 first is interesting in its own right as a comparison
between the matrix model approach of~\cite{matrixsphere} to gauge
theory on $S_N^2$ and the results of the present paper. It is also a
useful warm-up to the abelianization approach of technique~2 which
shares some of its qualitative features. We will find that the
abelianization technique through the polar decomposition of the
configuration space exploits the radial coordinates in a rather
explicit way to describe the local geometry of Yang-Mills critical
surfaces, and it may also find useful applications in related
considerations.

\bigskip

\section{Itzykson-Zuber localization on the configuration space}
\label{IZ-loc}

The integral \eq{calZdef} can be evaluated immediately using the Itzykson-Zuber
formula~\cite{Itzykson:1979fi}, which we briefly recall.
If $X,Y$ are $m\times
m$ hermitian matrices with nondegenerate eigenvalues $x_i,y_i\in\R$,
$i=1,\dots,m$, then one has
\beq
\int_{U(m)}\,[\ddd U]~\exp\Big(\mbox{$\frac{\ii N}s$}\,\Tr\big(
X\,U\,Y\,U^\dag\big)\Big)=c_N(m,s)~\frac{\det\limits_{1\leq i,j
\leq m}\,\big(\e^{\frac{\ii N}s\,x_i\,y_j}\big)}
{\Delta(x)~\Delta(y)}
\label{IZformula}\eeq
where for $m\in\N$ and $s\in\C$ we have defined
\beq
c_N(m,s):={\rm vol}\big(U(m)\big)\,
(\ii N/s)^{-m\,(m-1)/2}~\prod_{k=1}^{m-1}\,k! \ .
\label{cmsdef}\eeq
Applied to the present situation for $U(\cN\,)$, this yields
\bea
Z_\cO(p) &=&\frac1{\vol\big(U(N_+)\big)~\vol\big(U(N_-)\big)}~
\int_{U(\cN\,)}\,[\ddd U]~ \exp\Big(-\mbox{$\frac\ii2$}\,
\Tr\big(U^{-1}\, \Xi\, U\,\Phi\big)\Big) \nn\\[4pt]
&=&c_1'(\cN,2)~\frac{\det\limits_{1\leq i,j\leq\cN}\,\big(
\e^{-\frac\ii2 \,\Xi_i\,\Phi_j}\big)}{\D(\Xi) ~\D(\Phi)}
\label{U-int}
\eea
where $\Phi = \diag(p_1,...,p_{nN}) \otimes \sigma^0$ and
$c_1'(\cN,2):=c_1(\cN,2)/\vol(U(N_+))~\vol(U(N_-))$. This
formula can be understood as an abelian localization with respect
to the action of the maximal torus group $U(1)^{\cN}$ on the flag
manifold $U(\cN\,)/U(1)^{\cN}$~\cite{szaboloc}. The corresponding
fixed points are the solutions of the equation
\beq
[C,\Phi] =0 \ ,
\label{IZsaddle}\eeq
which are the saddle-points of the Itzykson-Zuber integral, and the
expansion of the determinant in \eq{U-int} into a sum over
permutations $\pi\in\mS_\cN$ gives the sum over critical points in the
localization formula. This is completely analogous to the abelianized
localization of Section~\ref{Abel-loc}. However, the expression
\eq{U-int} is formal as it stands because both sets of eigenvalues $\Xi_i$ and
$\Phi_i$ are degenerate, and correspondingly the critical surfaces are
in fact nontrivial spaces. 
Therefore \eq{U-int} has to be defined
using an appropriate limiting procedure which removes the degeneracy.

The partition function \eq{Z-2} is then given by
\be
Z=\frac{~ c_1'(\cN,2)}{(n\,N)!}\,\int_{\R^{n\,N}}\,
\Big[\,\frac{\ddd p}{2\pi}\,\Big]~
\eee^{-\frac{g'}4\,\Tr(p^2)}~\Delta(p)^2
~\frac{\det\limits_{1\leq i,j\leq\cN}\,\big(
\e^{-\frac\ii2 \,\Xi_i\,\Phi_j}\big)}{\D(\Xi) ~\D(\Phi)}\ ,
\label{Z-full-IZ}
\ee
where the
set of eigenvalues $\Phi_i$ of $\Phi$ consists of two copies of
$(p_1,\dots,p_{nN})$ and is therefore highly degenerate.  
While this explicit formula in terms of an $nN$-dimensional integral 
is very appealing, the
ratio of degenerate determinants in \eq{Z-full-IZ} makes it
difficult to evaluate explicitly~\cite{matrixsphere}, and its
combinatorial expansion is even more intricate than that of
Section~\ref{UnAbelian}.  Thus far only an asymptotic
analysis (of a slightly modified integral) has been made possible
in~\cite{matrixsphere}. The reason for this complexity is the fact
that, without the addition of a suitable localization form $Q\a$ to
the path integral \eq{Z-2}, the localization is onto the solutions of
the equation \eq{IZsaddle} in $\cO$ which are not related to the
critical surfaces of the Yang-Mills action in any simple way.
This will be explored in more detail below.

\bigskip

\section{Abelian localization and radial coordinates\label{Abel-loc}}

We now return to  the symplectic orbit integral \eq{calZdef}, and 
observe that it
fulfills the conditions of the Duistermaat-Heckman theorem, or
equivalently the abelian version of the localization theorem of
Section~\ref{LocPrinc}. Therefore, we have mapped the original
nonabelian localization problem to the simpler problem of {\it
  abelian} localization. Indeed, $\langle\mu_T(C),p\rangle=\Tr(p\,C)$
is just the restriction of the moment map
$\mu:\cO(\Xi)\to\mun(\cN\,)^\vee$ to the maximal torus $T$ of the
gauge group $G$. The torus action on the orbit space $\cO(\Xi)$ is the
restriction of the adjoint $G$-action given by
\be
C~\longmapsto~P\,C\,P^{-1}
\ee
for $C=C_\mu\otimes\sigma^\mu=U\,\Xi\,U^{-1}\in\cO(\Xi)$, $U\in
U(\cN\,)$ and $P\in T$. To compute the corresponding localization
formula we need the fixed points of this $T$-action. 
They are given by those $C\in\cO(\Xi)$ which commute with the
$T$-action generated by the element $p\in\mun(1)^{n\,N}$, so that
\be
[C,p]=0 \ .
\label{Cp0}\ee

This equation will be studied in detail in Section~\ref{AbLoc}.
It is solved by those $U\in U(\cN\,)$ for which
$U^{-1}\,P\,U$ lies in the stabilizer subgroup $U(n\,N_+)\times
U(n\,N_-)\subset U(\cN\,)$ of the element $\Xi$ (with $N_\pm:=N\pm1$
as before). The saddle points
$U$ are generically also labelled by permutation matrices $\Sigma\in
U(\cN\,)$ representing elements $\pi\in\mS_{n\,N}$. On the
configuration space $\cO$, the saddle point equation \eq{Cp0} means
that $C$ commutes with the characteristic projectors of $p$, i.e. $C$
has the same block decomposition as $p$.

The Fourier transform (\ref{calZdef}) will thus generically localize
onto a subspace of $U(n\,N_+)\times U(n\,N_-)$ in $\cO$. It may be
evaluated with the help of the degenerate version of the
Duistermaat-Heckman theorem~\cite{szaboloc}, which expresses it in
terms of an integral over the critical submanifold $U(n\,N_+)\times
U(n\,N_-)$ with the quantum fluctuation determinants determined by the
$T$-equivariant Euler class of the normal bundle to the
stabilizer~\cite{BGVbook}. While this can be worked out in principle,
it is rather cumbersome to do in practise. Instead we will proceed in
a more direct fashion by exploiting some further geometrical properties of the
configuration space $\cO$, which in the next section will be related
to the local symplectic geometry near each Yang-Mills critical point
as analysed at length in Section~\ref{ClassSols}. This explicit
calculation will justify the abelianized localization {\it a priori},
with the quantum fluctuation determinants given by integrals over
symplectic leaves of a foliation of the configuration space
parametrized by abelian subspaces of the tangent spaces to $\cO$. The
symplectic integral (\ref{calZdef}) could also be analysed using
Fourier transform techniques along with the Guillemin-Lerman-Sternberg
theorem~\cite{GLS1}, as in~\cite{JK1,Paradan1,JKKW1}, but this leads
to much more complicated combinatorial expressions than the ones we
derive.

\subsection{Polar decomposition of the configuration
  space\label{RadialCoords}}

The key step in the evaluation of (\ref{calZdef}) is the introduction
of {\it radial coordinates} on the orbit space
(see~\cite{helgason1,casmag1,szabo1} for details). Let us go back to
the Cartan decomposition \eq{ucN-decomp} at a given point
$C\in\cO$. Let $\mt$ be a maximal abelian subalgebra in the tangent
space $T_C\cO\cong\ker(\cJ^2+\one_\cN)$. Then the radial coordinates
on the orbit space $\cO$ are given by
\be
U=V\,R\,V^{-1}=V\,R~{\sf j}\big(V^{-1}\big)
\ee
where $V\in U(n\,N_+)\times U(n\,N_-)$, modulo elements of the
centralizer of $\mt$, and $R\in\exp(\mt)$ up to the adjoint action of
the Weyl group of the {\it restricted} root system of the irreducible
symmetric space~$\cO$. By definition, they satisfy the respective
commutation and anticommutation relations
\beq
V\,\Xi = \Xi\, V \qquad \mbox{and} \qquad R \,\Xi = \Xi\, R^{-1} \ .
\label{commanticommrels}\eeq
The corresponding covariant coordinate $C\in\cO(\Xi)$ is then given by
\bea 
C &=& U\,\Xi\,U^{-1} \nn\\[4pt] &=& V\, R \, \Xi\, R^{-1}\, V^{-1}
\nonumber\\[4pt]
&=&\mbox{$\frac12$}\,V\,\left(R^2\,\Xi + \Xi\, R^{-2}\right)\,V^{-1}
~=~\mbox{$\frac12$}\,\left(V\, R^2\, V^{-1}\,\Xi  + \Xi\, V\, R^{-2}\,
V^{-1}\right) \ .
\label{orbit-coordinates}
\eea

The jacobian for the change of invariant integration measure on $\cO$
can be computed by standard techniques with the result
\be
\ddd C=r(n,N)~[\ddd V]~\prod_{i=1}^{\dim(\mt)}\,\ddd r_i~
\prod_{\alpha>0}\,\bigl|\sin(\alpha,\log R)\bigr|^{m_\alpha}
\label{changemeas}\ee
where\footnote{The normalization constant $r(n,N)$ is determined by
  the requirement $\int_\cO\,\ddd C={\rm vol}(\cO)$.}
\beq
r(n,N)=\frac{\vol\big(U(\cN\,)\big)}{\vol\big(U(n\,N_+)\big)^2~
\vol\big(U(n\,N_-)\big)^2}\,\frac{2^{n^2\,(N^2-1)/2}}
{2^{n\,(N-2n\,N-3)/2}} \ .
\label{rnNdef}\eeq
The radial coordinates $r_i\in[0,\frac\pi2]$ are the eigenvalues of
$U$, while $V$ are the angular coordinates with $[\ddd V]$ denoting the
standard invariant Haar measure. The second product runs over positive
roots of the restricted root lattice on $\cO$, and $m_\alpha$ is the
multiplicity of the root $\alpha$ in the Cartan decomposition
\eq{ucN-decomp}. The pairing is defined by choosing an orthonormal
basis $\vec e_i$ in weight space and identifying a root vector
$\alpha$ with the dual element $\alpha^\vee=\sum_i\alpha_i\,\vec
e_i$. Then $(\alpha,\log R)=\sum_i\alpha_i\,r_i$. This polar
decomposition defines a foliation of the configuration space $\cO$ by
conjugacy classes under the adjoint action of the stabilizer
subgroup. The radial symplectic leaves ${\cal L}(R)$ of this foliation
are parametrized by the abelian Lie group $\exp(\mt)$.

Let us make this decomposition more explicit using the known data for
the symmetric space \eq{nonaborbit}~\cite{casmag1}. The
restricted root lattice is given by the root system
$BC_{n\,N_-}=B_{n\,N_-}\cup C_{n\,N_-}$ which has positive weights $\vec e_i\pm
\vec e_j$, $2\vec e_i$ and $\vec e_i$ with $i,j=1,\dots,n\,N_-$,
$i<j$. The corresponding multiplicities are $m_{\vec e_i\pm\vec
  e_j}=2$, $m_{2\vec e_i}=1$ and $m_{\vec e_i}=4n$. The gauge
invariant volume form on $\cO$ thereby becomes
\be
\ddd C=r(n,N)~[\ddd V]~\prod_{i=1}^{n\,N_-}\,\ddd r_i~\sin2r_i\,\sin^{4n}r_i~
\prod_{i<j}\,\sin^2(r_i-r_j)\,\sin^2(r_i+r_j) \ .
\ee
Using the trigonometric identities
\be
\sin(r_i-r_j)\,\sin(r_i+r_j)=\mbox{$\frac12$}\,(
\cos2r_j-\cos2r_i) \qquad \mbox{and}
\qquad \sin^2r_i=\mbox{$\frac12$}\,(1-\cos2r_i) \ ,
\ee
and defining $\lambda_i:=\cos2r_i\in[-1,1]$, we may bring the measure to
the form
\be
\ddd C=\frac{r(n,N)}{2^{n^2\,(N^2-1)}}~[\ddd V]~\Delta(\lambda)^2~
\prod_{i=1}^{n\,N_-}\,\ddd\lambda_i~(1-\lambda_i)^{2n} \ .
\ee

A convenient choice for the radial coordinates is provided by setting
\beq
\rho:={\rm diag}(r_1,\dots,r_{n\,N_-})
\label{rhodefr}\eeq
and defining
\be
R={\rm diag}\bigl(\sigma^0\otimes\one_n\,,\,
\exp(\ii\sigma^1\otimes\rho)\bigr)={\rm diag}\big(
\sigma^0\otimes\one_n\,,\,\sigma^0\otimes\cos(\rho)+\ii
\sigma^1\otimes\sin(\rho)\big) \ .
\label{Rconv}\ee
We also choose a basis in which
\beq
\Xi=\mbox{$\frac N2$}~{\rm diag}\big(\one_{n\,N_+}\,,\,-\one_{n\,N_-}
\big)=\mbox{$\frac N2$}~{\rm diag}\big(\sigma^0\otimes\one_n\,,\,
\sigma^3\otimes\one_{n\,N_-}\big)
\label{Xichoice}\eeq
and $V\in U(n\,N_+)\times U(n\,N_-)$ is given by
\be
V={\rm diag}(V_+,V_-) \ ,
\label{Vconv}\ee
with $V_\pm\in U(n\,N_\pm)$ and $[\ddd V]=[\ddd V_+]~[\ddd V_-]$. The
relations \eq{commanticommrels} are then automatically satisfied.

\subsection{Evaluation of the abelianized partition function: $U(1)$
  gauge theory\label{U1Abelian}}

We will now explicitly evaluate the Fourier transform
\eq{calZdef}, beginning with the abelian case $n=1$. Using
\eq{orbit-coordinates} and \eq{rhodefr}--\eq{Vconv}, it is
straightforward to work out the abelian moment map in (\ref{calZdef})
with the result
\bea
\big\langle\mu_T(C)\,,\,p\big\rangle
&=&\Tr\big(p\,U\,\Xi\,U^{-1}\big)\nonumber\\[4pt]
&=&\mbox{$\frac12$}\,\Tr\bigl(p\,\Xi\,V\,(R^2+R^{-2})\,V^{-1}
\bigr) \nonumber\\[4pt]
&=&\Tr\Bigl(p\,\Xi\,V~{\rm diag}\bigl(
\sigma^0\,,\,\cos(2\sigma^1\otimes\rho)\bigr)\,V^{-1}\Bigr)
\nonumber\\[4pt]
&=&\mbox{$\frac N2$}\,\Tr\bigl({\rm diag}(p_1\,\sigma^0,p_2,\dots,
p_{N})\,V_+~{\rm diag}(\sigma^0,\lambda_1,\dots,\lambda_{N_-})\,
V_+^{-1}\bigr)
\nonumber\\ && -\,\mbox{$\frac N2$}\,\Tr\bigl({\rm diag}(p_2,
\dots,p_{N})\,V_-~{\rm diag}(\lambda_1,\dots,\lambda_{N_-})
\,V_-^{-1}\bigr)
\label{calZactionexpl}
\eea
where we have used an inconsequential redefinition of the unitary
matrix $V_+$ by multiplication with an appropriate permutation
matrix. Upon substitution into (\ref{calZdef}), we see that the two
angular integrals decouple from each other.

The integral over $V_-\in U(N_-)$ is now easily evaluated with the
help of \eq{IZformula} with the result
\be
\frac{c_N(N_-,4)}{\Delta(p_2,\dots,p_{N})\,\Delta(\lambda)}~
\sum_{\pi_-\in\mS_{N_-}}\,{\rm sgn}(\pi_-)~
\prod_{i=1}^{N_-}\,\eee^{\frac{\ii N}4\,
p_{i+1}\,\lambda_{\pi_-(i)}} \ .
\label{VminusIZ}\ee
The integral over $V_+\in U(N_+)$ is more delicate since the
Itzykson-Zuber formula will involve a ratio of degenerate
determinants. Since both numerator and denominator of \eq{IZformula}
are completely antisymmetric functions of the eigenvalues $x_i$ and
$y_i$ independently, the limit where some eigenvalues coalesce gives a
well-defined analytic function in $(x_i,y_i)$ because all poles are
cancelled by zeroes in the determinant. We will regularize the
$V_+$-integral by replacing the first $p_1$ entry in the last line of
(\ref{calZactionexpl}) with an auxilliary momentum variable
$p_0\in\R$, the second entry of $1$ with an auxilliary radial variable
$\lambda_0\in[-1,1]$, and then afterwards take the limits $p_0\to
p_1$, $\lambda_0\to1$. Defining $\lambda_N:=1$, the Itzykson-Zuber
formula \eq{IZformula} applied to the regularized $V_+$-integral
yields
\be
\frac{c_N(N_+,-4)}{\Delta(p_0,p_1,\dots,p_N)\,\Delta(\lambda_0,
\lambda_1,\dots,\lambda_N)}~\sum_{\pi_+\in\mS_{N_+}}\,
{\rm sgn}(\pi_+)~\eee^{-\frac{\ii N}4\,p_0\,\lambda_{\pi_+(N)}}~
\prod_{i=0}^{N_-}\,\eee^{-\frac{\ii N}4\,p_{i+1}\,\lambda_{\pi_+(i)}}
\ .
\label{VplusIZ}\ee
Taking the limit $p_0\to p_1$ first  using
l'H\^opital's rule gives
\bea
&&\frac{\frac{\ii N}4\,c_N(N_+,-4)}{p_1~\prod\limits_{i=2}^N\,(p_1-p_i)~
\Delta(p)~\Delta(\lambda_0,\lambda_1,\dots,\lambda_N)}\nonumber\\ &&
\qquad\qquad\times\,
\sum_{\pi_+\in\mS_{N_+}}\,{\rm sgn}(\pi_+)\,\lambda_{\pi_+(N)}~
\eee^{-\frac{\ii N}4\,p_1\,\lambda_{\pi_+(N)}}~
\prod_{i=0}^{N_-}\,\eee^{-\frac{\ii N}4\,p_{i+1}\,\lambda_{\pi_+(i)}}
\ .
\label{VplusIZp}\eea
Finally, taking the limit $\lambda_0\to1$ again
using l'H\^opital's rule yields
\bea
&&\frac{-\frac{\ii N}4\,c_N(N_+,-4)}{p_1~\prod\limits_{i=2}^N\,(p_1-p_i)~
\Delta(p)~\prod\limits_{i=1}^{N_-}\,(1-\lambda_i)^2~\Delta(\lambda)}
\\ && \qquad\qquad\times\,
\left|\begin{matrix}\left(1-\frac{\ii N}4\right)~\eee^{-\frac{\ii N}4\,p_1}&
\lambda_1~\eee^{-\frac{\ii N}4\,p_1\,\lambda_1}&\dots&
\lambda_{N_-}~\eee^{-\frac{\ii N}4\,p_1\,\lambda_{N_-}}&
\eee^{-\frac{\ii N}4\,p_1}\cr-\frac{\ii N}4\,p_1~\eee^{-\frac{\ii
    N}4\,p_1}&\eee^{-\frac{\ii N}4\,p_1\,\lambda_1}&\dots&
\eee^{-\frac{\ii N}4\,p_1\,\lambda_{N_-}}&\eee^{-\frac{\ii N}4\,p_1}\cr
\vdots&\vdots& &\vdots&\vdots\cr-\frac{\ii N}4\,p_N~\eee^{-\frac{\ii
    N}4\,p_N}&\eee^{-\frac{\ii N}4\,p_N\,\lambda_1}&\dots&
\eee^{-\frac{\ii N}4\,p_N\,\lambda_{N_-}}&\eee^{-\frac{\ii N}4\,p_N}
\end{matrix}\right| \ . \nonumber
\eea

Substituting the above into (\ref{calZdef}) gives us the expression
\bea
Z_\cO(p)&=&-~\frac{4~\vol(\cO)}{p_1~\Delta(p)^2}\,
\frac{N!\,(N-1)!\,\prod\limits_{k=1}^{N-2}\,(k!)^2}
{\left(\,\sqrt8\,N\right)^{N^2-N}}~\prod_{l=1}^{N_-}\,
\int_{-1}^1\,\ddd\lambda_l~\sum_{\pi_-\in\mS_{N_-}}\,{\rm sgn}(\pi_-)~
\prod_{i=1}^{N_-}\,\eee^{\frac{\ii N}4\,
p_{i+1}\,\lambda_{\pi_-(i)}}\nonumber\\ &&
\times~\left|\begin{matrix}\left(1-\frac{\ii N}4\right)~
\eee^{-\frac{\ii N}4\,p_1}&
\lambda_1~\eee^{-\frac{\ii N}4\,p_1\,\lambda_1}&\dots&
\lambda_{N_-}~\eee^{-\frac{\ii N}4\,p_1\,\lambda_{N_-}}&
\eee^{-\frac{\ii N}4\,p_1}\cr-\frac{\ii N}4\,p_1~\eee^{-\frac{\ii
    N}4\,p_1}&\eee^{-\frac{\ii N}4\,p_1\,\lambda_1}&\dots&
\eee^{-\frac{\ii N}4\,p_1\,\lambda_{N_-}}&\eee^{-\frac{\ii N}4\,p_1}\cr
\vdots&\vdots& &\vdots&\vdots\cr-\frac{\ii N}4\,p_N~\eee^{-\frac{\ii
    N}4\,p_N}&\eee^{-\frac{\ii N}4\,p_N\,\lambda_1}&\dots&
\eee^{-\frac{\ii N}4\,p_N\,\lambda_{N_-}}&\eee^{-\frac{\ii N}4\,p_N}
\end{matrix}\right| \ .
\label{calZalmost}\eea
We will now write the product of determinants in (\ref{calZalmost}) as
a single sum over the Weyl group $\mS_N$ of the original gauge
symmetry group $U(N)$. For this, we embed $\mS_{N_-}$ in the Weyl
group $\mS_N$ as the subgroup of permutations $\pi_-$ of
$\{1,\dots,N_-,N\}$ with $\pi_-(N)=N$. We perform a Laplace expansion
of the second determinant in (\ref{calZalmost}) into minors along the
first row to write
\bea
&&\left|\begin{matrix}\left(1-\frac{\ii N}4\right)~
\eee^{-\frac{\ii N}4\,p_1}&
\lambda_1~\eee^{-\frac{\ii N}4\,p_1\,\lambda_1}&\dots&
\lambda_{N_-}~\eee^{-\frac{\ii N}4\,p_1\,\lambda_{N_-}}&
\eee^{-\frac{\ii N}4\,p_1}\cr-\frac{\ii N}4\,p_1~\eee^{-\frac{\ii
    N}4\,p_1}&\eee^{-\frac{\ii N}4\,p_1\,\lambda_1}&\dots&
\eee^{-\frac{\ii N}4\,p_1\,\lambda_{N_-}}&\eee^{-\frac{\ii N}4\,p_1}\cr
\vdots&\vdots& &\vdots&\vdots\cr-\frac{\ii N}4\,p_N~\eee^{-\frac{\ii
    N}4\,p_N}&\eee^{-\frac{\ii N}4\,p_N\,\lambda_1}&\dots&
\eee^{-\frac{\ii N}4\,p_N\,\lambda_{N_-}}&\eee^{-\frac{\ii N}4\,p_N}
\end{matrix}\right|\nonumber\\ && \qquad\qquad~=~
\sum_{\pi_+\in\mS_N}\,{\rm sgn}(\pi_+)\,
\Biggl[\left(1-\mbox{$\frac{\ii N}4$}\,p_1\right)~
\eee^{-\frac{\ii N}4\,p_1}~\prod_{i=1}^N\,\eee^{-\frac{\ii N}4\,
\lambda_i\,p_{\pi_+(i)}}\Biggr.\nonumber\\ && \qquad \qquad \qquad
\Biggl.-\,\frac{\ii N}4\,\sum_{i=1}^N\,\lambda_i~
\eee^{-\frac{\ii N}4\,\lambda_i\,p_1}\,p_{\pi_+(i)}~
\eee^{-\frac{\ii N}4\,p_{\pi_+(i)}}~
\prod_{\stackrel{\scriptstyle k=1}{\scriptstyle k\neq i}}^N\,
\eee^{-\frac{\ii N}4\,\lambda_k\,p_{\pi_+(k)}}\Biggr] \ .
\label{2nddetexpand}\eea
When inserted into the expression (\ref{calZalmost}), we can use the
invariance of the radial integration measure and domain under
permutations of the $\lambda_i$'s to reduce the double sum over the
Weyl groups to a {\it single} sum over the relative permutation
$\pi:=\pi_+\,\pi_-^{-1}\in\mS_N$ with $\pi(N)=\pi_+(N)$. The sum over
$\pi_+$ can be replaced by a sum over $\pi$, while the remaining sum
over $\pi_-$ simply produces the order $N!$ of the Weyl group of
$U(N)$.

In this way we may bring the Fourier transform of the orbit into the
form
\bea
Z_\cO(p)&=&-~\frac{4~\vol(\cO)}{p_1~\Delta(p)^2}~
\frac{\prod\limits_{k=1}^{N}\,(k!)^2}
{(N-1)!\,\left(\,\sqrt8\,N\right)^{N^2-N}}\nn\\ && \times~\sum_{\pi
\in\mS_N}\,{\rm sgn}(\pi)\,\left[\left(1-\mbox{$\frac{\ii N}4$}\,
p_1\right)~\eee^{-\frac{\ii N}4\,(p_1+p_{\pi(N)})}~
\prod_{i=1}^{N_-}\,\int_{-1}^1\,\ddd\lambda_i~
\eee^{-\frac{\ii N}4\,\lambda_i\,(p_{\pi(i)}-p_{i+1})}\right.
\nonumber\\ && \qquad\qquad 
-\,\frac{\ii N}4~\sum_{j=1}^{N_-}\,p_{\pi(j)}~
\eee^{-\frac{\ii N}4\,(p_{\pi(j)}+p_{\pi(N)})}
\nn\\ && \qquad\qquad\qquad\qquad\quad
\times~\int_{-1}^1\,\ddd\lambda_j~\lambda_j~\eee^{-\frac{\ii N}4\,
\lambda_j\,(p_1-p_{j+1})}~\prod_{\stackrel{\scriptstyle i=1}
{\scriptstyle i\neq j}}^{N_-}\,\int_{-1}^1\,\ddd\lambda_i~
\eee^{-\frac{\ii N}4\,\lambda_i\,(p_{\pi(i)}-p_{i+1})}
\nonumber\\ && \qquad\qquad
\left.-\,\mbox{$\frac{\ii N}4$}\,p_{\pi(N)}~
\eee^{-\frac{\ii N}4\,(p_{\pi(N)}+p_1)}~
\prod_{i=1}^{N_-}\,\int_{-1}^1\,\ddd\lambda_i~\eee^{-
\frac{\ii N}4\,\lambda_i\,(p_{\pi(i)}-p_{i+1})}\right] \ .
\label{calZb4int}\eea
Finally, the radial integrations can be expressed in terms of the
spectral sine-kernel of the unitary ensemble of random matrix theory
and its derivative given by
\be
\sK(x):=\frac{\sin x}x=\frac12\,\int_{-1}^1\,\ddd\lambda~
\eee^{-\ii\lambda\,x} \quad \mbox{and} \quad \sK'(x)=\frac1x\,
\left(\cos x-\frac{\sin x}x\right)=-\frac\ii2\,\int_{-1}^1\,
\ddd\lambda~\lambda~\eee^{-\ii\lambda\,x} \ .
\label{sinekernel}\ee
Then the abelianized partition function \eq{partfndiag} is
written as an exact expansion in gaussian momentum transforms given by
\bea
Z&=&-~\frac{8~\vol(\cO)\,N!\,(N-1)!\,\prod\limits_{k=1}^{N-2}\,
(k!)^2}{2^{N^2}\,(2\pi)^N\,\left(\,\sqrt2\,N\right)^{N^2-N}}~
\sum_{\pi\in\mS_N}\,{\rm sgn}(\pi)~\int_{\R^N}\,
[\ddd p]~\frac{\e^{-\frac g{4N}\,\sum_i\,p_i^2}}{p_1}\nn\\ && \times~
\left[\left(1-\mbox{$\frac{\ii N}4$}\,p_1\right)~
\eee^{-\frac{\ii N}4\,(p_1+p_{\pi(N)})}~\prod_{i=1}^{N_-}\,
\sK\left(\mbox{$\frac N4$}\,(p_{\pi(i)}-p_{i+1})\right)\right.
\nonumber \\ && \qquad\qquad +\,\frac N4~\sum_{j=1}^{N_-}\,p_{\pi(j)}~
\eee^{-\frac{\ii N}4\,(p_{\pi(j)}+p_{\pi(N)})}~
\sK'\left(\mbox{$\frac N4$}\,(p_1-p_{j+1})\right)~
\prod_{\stackrel{\scriptstyle i=1}{\scriptstyle i\neq j}}^{N_-}\,
\sK\left(\mbox{$\frac N4$}\,(p_{\pi(i)}-p_{i+1})\right)\nonumber\\ &&
\qquad\qquad \left.-\,\mbox{$\frac{\ii N}4$}\,p_{\pi(N)}~
\eee^{-\frac{\ii N}4\,(p_1+p_{\pi(N)})}~\prod_{i=1}^{N_-}\,
\sK\left(\mbox{$\frac N4$}\,(p_{\pi(i)}-p_{i+1})\right)\right] \ .
\label{calZfinalp}\eea

For low values of $N$, the momentum integrals in this formula can be
computed in terms of transcendental error functions, which are the
typical contributions in nonabelian localization~\cite{witten1} and
reflect the occurence of non-gaussian quantum fluctuation integrals.
Note that there is a single momentum $p_1$ singled out in the formula
\eq{calZfinalp}. In the $U(n)$ case of Section~\ref{UnAbelian} below
there will be $n$ momenta singled out which is where the sum over sets
of $n$ integers required by the nonabelian localization formula in the
large $N$ limit will come from. At $N\to\infty$, the spectral kernels
$\sK\big(\frac{N}4\, (p_{\pi(i)} -
p_{i+1})\big)\approx\frac{4\pi}{N}\,\delta(p_{\pi(i)} -
p_{i+1})$ provide the  necessary groupings of variables into
partitions of $N$ arising from the sum over the residual gauge
symmetry group $\mS_N$. The conjugacy class of a given permutation
$\pi\in\mS_N$ is characterized entirely by its cycle decomposition,
which contains $n_k\geq0$ cycles of length $k$ for $k=1,\dots,N$ with
$N=\sum_k\,k\,n_k$ and ${\rm
  sgn}(\pi)=(-1)^{\sum_k\,(k-1)\,n_k}$. However, the saddle-point
  partitions here do {\it not} correspond to the cycles themselves,
  but rather to the {\it numbers} $N_{n_1,\dots,n_N}$ of cycles
  $(n_1,\dots,n_N)$. For instance, the vacuum state now corresponds to
  the instanton configuration with $N$ fluxons, i.e. only trivial
  representations due to the abelianization, with moduli space
  \eq{calM0} as described in Section~\ref{CritPoints}. The higher
  critical points consist of an even number of irreducible
  representations which are suppressed roughly as
  $\e^{-N^3/2g\,n_i}$. This indicates that the radial coordinates on
  the configuration space $\cO$ are not so nicely adapted to the local
  symplectic geometry of the Yang-Mills critical surfaces. We will
  return to these issues in the next section.

\subsection{Evaluation of the abelianized partition function: $U(n)$
  gauge theory\label{UnAbelian}}

The nonabelian case $n>1$ becomes very complicated due to the
increasing complexity of the combinatorics involved in regulating the
Itzykson-Zuber integral (\ref{IZformula}) over $V_+\in U(n\,N_+)$. We
will therefore only briefly sketch the essential features, defering
the explicit evaluation in favour of a more formal, regulated
combinatorial expansion. Consider the radial coordinates $\lambda_i$,
$i = 1,\dots,n\,N_-$ on $\cO$ and add $2n$ new real variables
$1+\varepsilon_i$. We assemble them into the ordered set defined by
\be
\big(\,\obar\lambda_1,\dots,\obar\lambda_{n\,N_+}\big):=
\big(1+ \varepsilon_1,\dots,1+
\varepsilon_{2n},\lambda_{1},\dots, \lambda_{n\,N_-}\big) \ .
\label{lambdabar}\ee
Similarly, we double the first $n$ entries of the momentum vector
$p=(p_1,\dots,p_{n\,N})$ and gather them into the ordered set defined
by
\be
\big(\,\obar p_1,\dots,\obar p_{n\,N_+}\big):=
\big(p_1+\kappa,\dots, p_{n}+\kappa,p_1,
\dots,p_{n},p_{n+1},\dots, p_{n\,N}\big) \ .
\label{pbar}
\ee
At the end we will take the limits $\varepsilon_i, \kappa \to 0$.

The evaluation of the Fourier transform \eq{calZdef} now proceeds
exactly as in Section~\ref{U1Abelian} above. To organize the
combinatorics, we use the identity
\bea
&& \lim_{\varepsilon_i \to 0} \,
\frac{\det\limits_{1\leq i,j\leq n\,N_+}\,
\big(\eee^{-\frac{\ii N}4\,\obar p_{i}\,\obar \lambda_j}\big)}
{\Delta(\varepsilon)}\label{expansion-lambdabar}
\\ && \qquad\qquad ~=~\frac{\vol\big(U(2n)
\big)}{c_N(2n,-4)}~\sum_{\cQ \subset \{\,\obar p_i\}} \,
{\rm sgn}\big(\cQ\hookrightarrow\{\,\obar p_i\}\big)~
\eee^{-\frac{\ii N}4\,\sum_{i}\, q_i}~ \Delta(q)~
\det_{1\leq i,j\leq n\,N_-}\,
\big(\eee^{-\frac{\ii N}4\,\hat p_{i}\,\lambda_j}\big) \nn
\eea
where $\{\hat p_1,\dots,\hat p_{n\,N_-}\} = \{\,\obar p_1,\dots,\obar
p_{n\,N_+}\} \setminus\cQ $ with $\cQ =\{q_1,\dots,q_{2n}\}$ a subset
of $\{\,\obar p_1,\dots,\obar p_{n\,N_+}\}$ which is ordered according
to \eq{pbar}, and the sign is determined by the parity of the
embedding. The identity \eq{expansion-lambdabar} can be derived by
performing a Laplace expansion of the determinant on the left-hand
side into the $2n$ rows containing the variables $1 + \varepsilon_i$,
and using the limit formula
\be
\lim_{\varepsilon_i \to 0} \,
\frac{\det\limits_{1\leq i,j\leq 2n}\,
\big(\eee^{-\frac{\ii N}4\,q_i\,\varepsilon_j}\big)}
{\Delta(\varepsilon)} 
= \frac{\vol\big(U(2n)\big)}{c_N(2n,-4)}~\Delta(q) 
\ee
which follows from the Itzykson-Zuber formula \eq{IZformula}. The
Vandermonde determinants can also be factorized as
\be
\Delta(\,\obar \lambda\,) = \Delta(\lambda)~ \Delta(\varepsilon)~
\prod_{i=1}^{n\,N_-}\, (1-\lambda_i)^{2n}
\ee
up to higher order terms in $\varepsilon_i\to0$, along with
\be
\Delta(\,\obar p\,)~\Delta( p_{n+1},\dots, p_{n\,N}) 
= \kappa^n~ \Delta(p)^2~ \Delta(p_1 ,\dots, p_n)^2
\ee
in the limit $\kappa\to0$.

In this way the partition function \eq{partfndiag} can be expanded as
\bea
Z &=& \zeta_{n,N}~\lim_{\kappa\to0}\,\frac1{\kappa^n}~
\sum_{\cQ \subset \{\,\obar p_i\}} \,
{\rm sgn}\big(\cQ\hookrightarrow\{\,\obar p_i\}\big)~
\int_{\R^{n\,N}}\,[\ddd p]~\frac{\eee^{-\frac{g}{4N}\,\sum_i\,p_i^2}}
{\Delta(p_1 ,\dots, p_n)^2 }~
\eee^{-\frac{\ii N}4\,\sum_i\, q_i}~\Delta(q)\nn\\ &&\times~
\prod_{l=1}^{n\,N_-}\,
\int_{-1}^1\, \ddd\lambda_l~\det_{1\leq i,j\leq n\,N_-}\,
\big(\eee^{\frac{\ii N}4\, p_{i+n}\,\lambda_j}\big)~
\det_{1\leq i,j\leq n\,N_-}\,
\big(\eee^{-\frac{\ii N}4\,\hat p_{i}\,\lambda_j}\big)
\label{Znonab-1}\eea
where
\beq
\zeta_{n,N}:=\frac{\vol(\cO)}{(n\,N)!\,(2\pi)^{n\,N}}~
\frac{(\ii N)^{2n^2+n\,N\,(1-n\,N_+)}}{2^{n\,N_-\,(2-n\,N_+)}}~
\prod_{k=1}^{n\,N_--1}\,(k!)^2~\prod_{m=1}^{2n}\,
\frac{(m+n\,N_--1)!}{m!} \ .
\label{zetanNdef}\eeq
We now expand the two determinants in \eq{Znonab-1} into a double sum
over the Weyl group $\mS_{n\,N_-}$, and use permutation symmetry of the
radial integration to rewrite it as a sum over a single relative
permutation exactly as in Section~\ref{U1Abelian} above. Using
\eq{sinekernel} we arrive finally at the exact combinatorial expansion
\bea
Z &=& 2^{n\,N_-}\,(n\,N_-)!~\zeta_{n,N}~\lim_{\kappa\to0}\,
\frac1{\kappa^n}~ \sum_{\cQ \subset \{\,\obar p_i\}} \,
{\rm sgn} \big(\cQ\hookrightarrow\{\,\obar p_i\}\big)~
\sum_{\pi\in\mS_{n\,N_-}}\,{\rm sgn}(\pi)\nn\\ &&\times~
\int_{\R^{n\,N}}\,[\ddd p]~\frac{\eee^{-\frac{g}{4N}\,\sum_i\,p_i^2}}
{\Delta(p_1 ,\dots, p_n)^2 }~
\eee^{-\frac{\ii N}4\,\sum_i\, q_i}~ \Delta(q)~\prod_{i =1}^{n\,N_-}\, 
\sK\big(\mbox{$\frac{N}4$}\, (\hat p_{\pi(i)} - p_{i+n})\big) \ .
\label{Znonabfinal}\eea

The combinatorics of the large $N$ limit of the partition function
\eq{Znonabfinal} can be described as follows. The sine-kernels
$\sK\big(\frac{N}4\, (\hat p_{\pi(i)} - p_{i+n})\big) \approx\frac{4\pi}{N}\,
\delta(\hat p_{\pi(i)} - p_{i+n})$ define a link from $\hat p_{\pi(i)}$
to $p_{i+n}$. Following these, we obtain a set of open or closed links
determined by $\pi\in\mS_{n\,N_-}$. The open links must start at
$\{p_1+\kappa,\dots,p_n+\kappa, p_1,\dots,p_n\}$ (since those are not
contained in the $p_{i+n}$) and end at $\{q_1,\dots,q_{2n}\}$ (since
those are not contained in the $\hat p_i$). The closed links
correspond to cycles in the conjugacy class of the permutation
$\pi$. In particular, there are no factors $\eee^{-\frac{\ii
    N}4\,p_i}$, $i=1,\dots,n$ or $\Delta(p_1 ,\dots, p_n)^2$ for the
internal variables, and hence we can explicitly evaluate the internal
integrals. The difficulty lies in evaluating the sum over all possible
distinct cycles for the internal variables in a closed form. 

\subsubsection*{\it Comparison with the constrained matrix model}

In~\cite{matrixsphere}, quantum gauge theory on the fuzzy sphere
$S_N^2$ was formulated as a multi-matrix model with action
\be
S_{\rm mm} =\mbox{$\frac 1{N\,g}$}\, \Tr\big(C^2 -
\mbox{$\frac{N^2}4$}~\one_\cN\big)^2
\ee
and the constraint $C_0 = \frac 12~\one_N$. It was shown that
this matrix model also reproduces Yang-Mills theory on $S^2$ in the
large $N$  limit. This differs from the formulation of the present
paper essentially by replacing the pair (action , constraint) given by
$\big((C^2 - \frac{N^2}4~\one_\cN)^2\,,\, (C_0-\frac 12~\one_N)\big)$
with the permuted pair  $\big((C_0-\frac 12~\one_N)^2\,,\, (C^2 -
\frac{N^2}4~\one_\cN)\big)$. This can be understood by imposing the
respective constraints using gaussian terms in the actions, as then
the tangential degrees of freedom are essentially the same in both
cases.  The symplectic formulation of the present paper has not only
the advantage of applying the equivariant localization principle to
systematically construct the instanton expansion of gauge theory on
the fuzzy sphere, but it also somewhat simplifies the evaluation of
the matrix integral. It also enables one in principle to keep control
of the $\frac1N$ corrections to Yang-Mills theory on $S^2$, and the
approximate delta-functions at $N\to\infty$ responsible for the
groupings of variables are more transparent along the lines explained
in Sections~\ref{U1Abelian} and~\ref{UnAbelian}.

\bigskip

\section{Yang-Mills critical surfaces in abelianized
  localization\label{AbLoc}}

In this final section we will elucidate the relationship between the
nonabelian and abelianized localization approaches to the exact
instanton expansion of Yang-Mills theory on the fuzzy sphere $S_N^2$.
As discussed above, the critical surfaces for 
abelian localization are determined by 
the saddle-point equation \eq{IZsaddle}, \eq{Cp0}
\be
[C,\Phi] =0 \ 
\ee
for  $\Phi = \phi\otimes\sigma^0$ with $\phi\in\mun(N)$, which can be
assumed to be diagonal by using a gauge transformation. 
Its distinct eigenvalues $\Phi_\nu$ are arranged
into degenerate blocks as
\be
\Phi = \bigoplus_{\nu=1}^k \, \Phi_\nu~ \one_{n_\nu}\otimes \sigma^0
\label{phi-block}
\ee
with $\sum_\nu\,n_\nu=N$. Then $[C,\Phi] =0$ implies that the
covariant coordinate
\beq
C = U^{-1}\, \Xi\, U = \bigoplus_{\nu=1}^k\, C_\nu
\label{C-block-decomp}\eeq
has the same block decomposition as $\Phi$. Thus it can be
diagonalized as
\be
C_\nu = V^{-1}_\nu \,\Xi_\nu \,V_\nu
\label{C-alpha-blocks}
\ee
where $V_\nu$ is a $2n_\nu\times2n_\nu$ unitary matrix on the block
defined by $\one_{n_\nu}\otimes \sigma^0$ in \eq{phi-block}, and
$\Xi_\nu$ has eigenvalues $\pm\, \frac N2$. Then comparing
\eq{C-block-decomp} and \eq{C-alpha-blocks} implies
\be
\Big(\,\bigoplus_{\nu=1}^k\, V_\nu\Big)\, U^{-1} \,\Xi\, U \,
\Big(\,\bigoplus_{\nu=1}^k\, V^{-1}_\nu\Big)
= \bigoplus_{\nu=1}^k \,\Xi_\nu 
= \Sigma^{-1}\,\Xi\,\Sigma
\ee
for some permutation matrix $\Sigma\in U(\cN\,)$ representing an
element $\pi\in \mS_\cN/\mS_{N_+}\times\mS_{N_-}$, since both $\Xi$ and
$\bigoplus_\nu\, \Xi_\nu$ are diagonal $\cN\times\cN$ matrices with
the same set of degenerate eigenvalues. It follows that
\be
U\,\Big(\,\bigoplus_{\nu=1}^k\, V^{-1}_\nu\Big)\, \Sigma^{-1} \in
U(N_+)\times U(N_-) \ ,
\ee
and therefore $U\in U(\cN\,)$ is equal to
$\Sigma\,\big(\bigoplus_{\nu}\,V_\nu\big)$ times an element of the
stablizer subgroup $U(N_+) \times U(N_-)\subset U(\cN\,)$ of the
element $\Xi$.

We conclude that the gauge equivalence classes of solutions of the
saddle point equation $[C,\Phi] =0$ in the configuration space $\cO$
are described by the following data:
\begin{itemize}
\item A quotient permutation $\pi
  \in\mS_\cN/\mS_{N_+}\times\mS_{N_-}$;
\item A unitary matrix in the stabilizer group $U(N_+) \times U(N_-)$;
  and
\item A unitary block transformation $\bigoplus_\nu\,V_\nu$ adapted to
  the block decomposition \eq{phi-block} of $\Phi$.
\end{itemize}
It is evident that these critical surfaces are much larger than the
critical surfaces of the original Yang-Mills action, and they are not
even in any one-to-one correspondence with the Yang-Mills saddle
points. Any such block configuration is degenerate for the action in
\eq{Z-2}, and contains some Yang-Mills blocks of
Section~\ref{ExplYMDecomp} (with the irreducible low-energy critical
surface $\cC_{(N,1)}$ and possibly fluxons or other purely
noncommutative solutions). The reason is the absence of any
localization form $Q\a$, without which there is no way to separate the
desired Yang-Mills blocks of Section~\ref{ExplYMDecomp} from these
abelianized critical surfaces.

\subsection{Itzykson-Zuber localization on the symplectic leaves}

We now consider the foliation of the orbit $\cO(\Xi) \cong
U(2N)/R$  by conjugacy classes under the adjoint
action of the stabilizer group $R = U(N_+) \times U(N_-)$. The
corresponding symplectic leaves $\cL(\la)$ are parametrized by
the radial coordinates $\lambda_i\in[-1,1]$, $i=1,\dots,N_-$. For a
given leaf $\cL(\la)$, the integral  $\int_R\, [\ddd
V]~\e^{-\frac\ii2\,\langle\mu_T(C),p\rangle}$ is obtained by using the
Itzykson-Zuber formula for the unitary groups $U(N_+)$ and $U(N_-)$,
as we did in Sections~\ref{U1Abelian} and~\ref{UnAbelian}. As in
Section~\ref{IZ-loc} above, the Itzykson-Zuber formula
can itself be regarded as a consequence of abelian localization, and
the expansions of the resulting determinants in
Section~\ref{U1Abelian} is precisely the sum over the saddle-points on
each leaf $\cL(\la)$.

Let us identify these saddle-points explicitly. Choosing $\Xi$ as in
\eq{Xichoice}, the critical points of the moment map
\eq{calZactionexpl} with respect to arbitrary variations of
$(V_+,V_-)\in R$ are given by the solutions of the equations
\bea
\big[{\rm diag}(p_2,\dots, p_N)\,,\,V_-~{\rm
  diag}(\lambda_1,\dots,\lambda_{N_-})\,V_-^{-1}\big] &=& 0 \ ,
\nn\\[4pt]
\big[{\rm  diag}(p_1\,\sigma^0,p_2,\dots, p_N)\,,\,V_+
~{\rm  diag}(\sigma^0,\lambda_1,\dots,\lambda_{N_-})\,V_+^{-1}
\big] &=& 0 \ .
\label{saddle-Vpm}
\eea
As in Section~\ref{UnAbelian}, we consider for convenience the
extended sets of radial coordinates \eq{lambdabar} and momentum
variables \eq{pbar} for $n=1$. Then the first equation in
\eq{saddle-Vpm} means that the matrix $V_-~{\rm
  diag}(\lambda_1,\dots,\lambda_{N_-})\,V_-^{-1}$ commutes with the
spectral projectors of $(p_2,\dots,p_N)$, i.e. it has the same
block decomposition, and similarly the second equation in
\eq{saddle-Vpm} implies that the matrix $V_+~{\rm diag}(\,\obar
\lambda_1,\dots,\obar\lambda_{N_+})\,V_+^{-1}$ commutes with the
spectral projectors of $\obar p$.

Using unitary transformations on each of these blocks, the matrix
$V_-~{\rm diag}(\lambda_1,\dots,\lambda_{N_-})\,V_-^{-1}$ can then be
diagonalized with the same eigenvalues $\lambda_i$. It follows that
\bea
&& \Big(\,\bigoplus_{\nu=1}^k\, U_\nu\Big)\, V_- ~
{\rm diag}(\lambda_1,\dots,\lambda_{N_-})\, V_-^{-1}\,
\Big(\,\bigoplus_{\nu=1}^k\,
U_\nu^{-1}\Big) \nn\\ && \qquad\qquad\qquad
~=~ \diag (\lambda_{\pi_-(1)},\dots,\lambda_{\pi_-(N_-)})~=~
\Sigma_-~\diag (\lambda_1,\dots,\lambda_{N_-})\, \Sigma_-^{-1}
\eea
for some $U_\nu\in SU(n_\nu)$, where $n_\nu$ labels the degenerate
blocks of $(p_2,\dots,p_N)$ with $\sum_\nu\,n_\nu=N_-$ and
$\Sigma_- \in SU(N_-)$ is a permutation matrix corresponding to an
element $\pi_-\in\mS_{N_-}$. If $\lambda_i$ are {\em nondegenerate},
this implies that $\big(\bigoplus_\nu\, U_{\nu}\big)\, V_- =
\Sigma_-$ and hence
\be
V_- = \Big(\,\bigoplus_{\nu=1}^k\,U_\nu^{-1}\Big) \,\Sigma_- \ .
\ee
If some $\lambda_i$ are degenerate, it only follows that
$\Sigma_-^{-1}\,\big(\bigoplus_\nu\,U_\nu\big)\, V_-$ commutes with
the spectral projectors of $\lambda$, so that
$\Sigma_-^{-1}\,\big(\bigoplus_\nu\,U_\nu\big)\,
V_-=\bigoplus_\nu\,\tilde U_\nu$ for some $\tilde U_\nu\in
SU(n_\nu)$. It follows that the angular saddle-point $V_-\in U(N_-)$
is given by
\be
V_- = \Big(\,\bigoplus_{\nu=1}^k\,U_\nu^{-1}\Big)\,\Sigma_-\,
\Big(\,\bigoplus_{\nu=1}^k\, \tilde U_\nu\Big) \ .
\label{V-minus}
\ee
Similar statements hold for the angular saddle-point $V_+\in U(N_+)$,
with the additional feature that the first two entries of $\obar p$
and $\obar \lambda$ are degenerate by definition.

In each case, the value of the action \eq{calZactionexpl} is given by
\be
\big\langle\mu_T(C)\,,\,p\big\rangle
=\frac N4\,\sum_{i=1}^{N_+}\,\obar p_{i}\,\obar
\lambda_{\pi_+(i)} - \frac N4\,\sum_{i=1}^{N_-}\,
p_{i+1}\,\lambda_{\pi_-(i)} \ .
\label{action-saddle}
\ee
Therefore, each saddle-point is characterized by two permutation
matrices $\Sigma_\pm$ corresponding to $\pi_\pm\in\mS_{N_\pm}$, which
may or may not generate non-trivial  fibers on the homogeneous spaces
of the group $\prod_\nu\, U(n_\nu)$ depending on the degeneracies of
$p$ and $\la$. The integral over these $V_\pm$ orbits can then be
evaluated using the Itzykson-Zuber formula leading to \eq{VminusIZ}
and \eq{VplusIZ}, which gives precisely the sum over the saddle
points. The regularization required in \eq{VplusIZ} reflects the fact
that the critical surfaces are no longer isolated points, due to the
degeneracies of $\obar \la_i$ and $\obar p_i$.

The main point of this analysis is that these critical surfaces are
again not in any one-to-one correspondence with those of the original
Yang-Mills action. In fact, the abelian critical surfaces above
contain as subspaces those of the Itzykson-Zuber localization on
$\cO(\Xi)$ discussed in Section~\ref{IZ-loc} above, 
which are not only stationary on the symplectic leaves $\cL(\la)$ but
also with respect to variations of the radial coordinates
$\la_i$. However, even the critical surfaces for the Itzykson-Zuber
localization on the configuration space $\cO(\Xi)$ are not simply
related to those of the Yang-Mills action. In particular, the
variational problem for the action (\ref{action-saddle}) does not
determine the $\la_i$. A given radial saddle-point $\pi_\pm$ can thus
correspond to various types of Yang-Mills solutions by appropriately
choosing some $\la_i$, as we show explicitly in Section~\ref{YMRadial}
below. This arbitrariness in the radial coordinates $\la_i$ is lifted
by the addition of the localization one-form $\alpha$ of
Section~\ref{NonabLoc}, which serves to single out the Yang-Mills
saddle points from the new critical points. Nevertheless, it is
instructive to work out the radial coordinates of some Yang-Mills
saddle-points to illustrate the powerful workings of the polar
decomposition.

\subsection{Radial coordinates for Yang-Mills critical
  surfaces\label{YMRadial}}

We will now work out the radial coordinates for the solutions of the
Yang-Mills equations $[C_0, C_i]=0$, which will identify precisely the
appropriate localization values of $\la_i$ for each critical surface of
Section~\ref{CritPoints}. Given (\ref{Xichoice}) we now consider the
fuzzy sphere coordinates $\Sigma_0\,\Xi\,\Sigma_0^{-1}$ and
correspondingly modify the radial coordinates \eq{Rconv} to
\beq
R=\Sigma_0 \,\left(\begin{array}{ccc} \sigma^0 & 0 \\
     0 & \exp(\ii\sigma^1\otimes\rho)
   \end{array}\right) \,\Sigma_0^{-1} \ ,
\label{radialcoords-modified}
\eeq
where $\Sigma_0\in U(\cN\,)$ is a permutation matrix representing the
cyclic permutation
\beq
\pi_{(N_+)}=(1\,2\,\cdots\,N_+) \ .
\label{piNplusperm}\eeq
As we will see, the modification by $\Sigma_0$, although irrelevant
from the point of view of the path integral, will greatly simplify the
explicit parametrization.

Using this parametrization and \eq{Vconv}, we can
write the covariant coordinates \eq{orbit-coordinates} in the explicit
form
\bea
C &=& \frac N2\,V\,\Sigma_0\left(\begin{array}{ccc}
\sigma^0 & 0 \\ 0 & \sigma^3\otimes\cos(2\rho)+
   \sigma^2\otimes\sin(2\rho) \end{array}\right)\,
\Sigma_0^{-1}\,V^{-1} \nn\\[4pt]
&=& \frac N2\,  \left(\begin{array}{ccc}
  V_+ \,\left(\begin{array}{ccc} 1 &   & \\ 
            & \cos(2\rho) & \\
            &    & 1 \end{array}\right) \, V_+^{-1} &
 -\ii V_+\, \left(\begin{array}{c} 0 \\
      \sin(2\rho) \\ 0 \end{array}\right)\, V_-^{-1} \\
\ii V_- \,\left(\begin{array}{ccc} 0~, &  \sin(2\rho)~, &0 
       \end{array}\right)\,V_+^{-1} &
 - V_- \,\cos(2\rho)\, V_-^{-1}\end{array}\right)
\label{C-radial-explicit}
\eea
where we have applied the commutation relation $\ii[\sigma^1,\sigma^3]
= 2 \sigma^2$. The role of the cyclic permutation matrix $\Sigma_0$ is
to move the unit entries of $\sigma^0$ symmetrically around the matrix
$\cos(2\rho)$. We note for later use that if the
unitary matrices $V_\pm\in U(N_\pm)$ are block-diagonal, then so is
$C$. We will now use this parametrization to illustrate the use of the
radial coordinates by working out \eq{C-radial-explicit} explicitly
for various classical gauge field configurations of
Section~\ref{CritPoints}.

\subsubsection*{\it The vacuum solution}

The generators of the irreducible $N$-dimensional representation of
the $\ms\mun(2)$ Lie algebra \eq{FS-l} are given explicitly by
\beq
(\xi_3)_{ij}=-\delta_{ij}~\mbox{$\frac{N+1-2i}2$} \qquad \mbox{and}
\qquad (\xi_+)_{ij}=\delta_{i+1,j}~\sqrt{(N-i)\,i}
\label{Ndimrep}\eeq
where $i,j=1,\dots,N$ and $\xi_{\pm}=\xi_1\pm\ii\xi_2$ with
  $\xi_-=\xi_+^\dag$. The vacuum solution \eq{Xi-collective} in the
  abelian case $n=1$ thus has the explicit form 
\bea
C &=& \left(\begin{array}{cc} \frac 12~\one_N + \xi_3 & \xi_+ \nn\\
   \xi_- & \frac 12~\one_N - \xi_3 \end{array}\right)\\[4pt]
&=&  \left(\begin{array}{cc}\frac 12~\diag(-N+2,\dots,N-2,N) & \xi_+ \\
  \xi_- & \frac 12~\diag(N,\dots,-N+4,-N+2) \end{array}\right) 
\label{C-vac-block}
\eea
using the splitting into equal blocks of size $N$. This should be
identified with \eq{C-radial-explicit}, which splits into blocks of
sizes $N_\pm$. Noting the explicit form of $\xi_\pm$ in \eq{Ndimrep} as
raising and lowering operators, it follows that one can consistently
take both $V_+~{\rm diag}(\la_1,\dots,\la_{N_-},1,1)\,V_+^{-1} $ and
$V_-~{\rm diag}(\la_1,\dots,\la_{N_-})\,V_-^{-1} $ to be diagonal
matrices.

We can then consistently match the eigenvalues as
\beq
N \,(\la_1,\dots,\la_{N_-},1,1)=(-N+2,\dots, N-2,N,N) \ ,
\label{eigenvacmatch}\eeq
which gives
\beq
\la_i=-\mbox{$\frac{N-2i}N$}\qquad \mbox{for} \quad i=1,\dots,N_-
\label{la-vacuum}
\eeq
and provides the eigenvalues of the radial matrix $R$ for the vacuum
critical surface $\cC_{(N,1)}$. Note that the eigenvalue $\frac N2$
from the second diagonal block $\frac 12~\one_N-\xi_3$ of $C$ in
\eq{C-vac-block} is contained in the matrix $\frac N2\,V_+~{\rm
  diag}(\la_1,\dots,\la_{N_-},1,1)\,V_+^{-1}$. It follows that $V_-=
\Sigma_-$ is a permutation matrix in $U(N_-)$, while 
$V_+ = \Sigma_+\,U_2$ is a permutation matrix up to a possible
conjugation with a unitary matrix $U_2\in SU(2)\subset U(N_+)$ acting
on the two marked indices labelling the unit entries. We can absorb
$\Sigma_-$ by a redefinition of the $\la_i$, and hence take
\beq
V_- = \one_{N_-}
\label{Vminuswithout}\eeq
without loss of generality. It is also enough to consider the case
$U_2=\one_{N_+}$. Comparing \eq{C-radial-explicit} with
\eq{C-vac-block}, it follows that 
\be
V_+ = \Sigma_+
\label{cycle-vacuum}
\ee
is a permutation matrix representing the irreducible cycle
\eq{piNplusperm} of length $N_+$. Furthermore, one has
\be
\sin(2\rho_i) = \sqrt{1-\la_i^2} 
= \sqrt{\mbox{$\frac {4i}N - \frac{4i^2}{N^2}$}}
= \mbox{$\frac 2N$}\, \sqrt{i\, (N - i)} = \mbox{$\frac 2N$}\,
(\xi_+)_{i,i+1}
\label{sin2rho}
\ee
for $i = 1,\dots, N_-$, which is indeed the correct representation of 
$\xi_\pm$ in \eq{Ndimrep}, embedded in the correct off-diagonal way 
in \eq{C-radial-explicit} due to the block decomposition into sizes
$N_\pm$.

Let us point out one interesting feature of the covariant coordinate
\eq{C-vac-block}. The two diagonal entries of $\frac N2$ in the center
of the matrix constitute a trivial $2\times 2$ unit matrix $\sigma^0$
which completely decouples from the rest of $C$. This block can be
traced to the $\sigma^0$ in the upper-left corner of the first line in
\eq{C-radial-explicit}, whose position is determined by the
permutation matrix $\Sigma_0$, or equivalently to the auxilliary
radial coordinates $\obar\la_i =1+\varepsilon_i$, $i=1,2$. In fact,
any explicit entry of $\pm\, \frac N2$ in $C$ necessarily decouples
from the rest of $C$, for otherwise $C$ would have eigenvalues of
modulus larger than $\frac N2$. This means, in particular, that we can
permute these two entries using a suitable permutation matrix
$V_+=\Sigma_+$ without any effect on $C$ (but it will have an effect
on the momenta $p_i$ if they are included). This observation will be
useful below. This construction clearly generalizes to give the blocks
$C(n_a)$ of size $2n_a$ of the critical surfaces
$\cC_{(n_1,s_1),\dots,(n_k,s_k)}$ corresponding to irreducible $SU(2)$
representations of dimensions $n_a< N$. The most extreme case $n_a=1$
consists of the one-dimensional representation with $C_0(n_a=1)=\frac
N2$ and $C_i(n_a=1)=0$, whereby
\beq
C(n_a=1)=\mbox{$\frac N2$}\,\sigma^0
\label{Csigma0expl}\eeq
and hence only the explicit $\sigma^0$ block survives.

\subsubsection*{\it Nonabelian generalization}

For $n \geq 2$, the vacuum critical surface $\cC_{(N,1), ..., (N,1)}$
is associated with the solution \eq{vacsolnonab} which is a direct sum
of $n$ irreducible $SU(2)$ representations of dimension $N$. This can
clearly be obtained by repeating the above construction $n$ times. In
particular, $V_+ = (\Sigma_+)^{\oplus n}$ is a product of $n$ ``marked cycles'' as
above. Notice, however, that the {\em same} saddle point is obtained
if one acts with an additional permutation of the $2n$ auxilliary
radial coordinates $\obar \la_i = 1$, $i=1,\dots,2n$ (recall that the
explicit entries $\pm\, \frac N2$ of $C$ are always isolated). In
doing this, the decomposition of $V_+$ into irreducible cycles gets
modified. It can nonetheless be made into one irreducible cycle with
$2n$ marked points which come in groups of two at equal distance, for
example. This demonstrates that the mapping between the Yang-Mills
saddle-points and those of the abelianization approach in
Section~\ref{Abel-loc} is complicated. In particular, it is not
injective. Again, this construction generalizes to blocks of the
critical surfaces $\cC_{(n_1,s_1),\dots, (n_k,s_k)}$ corresponding to irreducible
$SU(2)$ representations of various dimensionalities.

\subsubsection*{\it Fluxons}

Fix an integer $1\leq n\leq N$ and consider the block gauge
field configuration of size $2n$ given by
\bea
C &=& \frac N2\,V\,\bigl(\sigma^3\otimes\cos(2\rho) 
+ \sigma^2\otimes\sin(2\rho)\bigr)\,V^{-1} \nn\\[4pt]
&=&\frac N2\,  \left(\begin{array}{ccc}
  V_+ \, \cos(2\rho) \,  V_+^{-1} &
 -\ii V_+ \,  \sin(2\rho)\,  V_-^{-1} \\
\ii V_-\,  \sin(2\rho)\,  V_+^{-1} &
 - V_- \,\cos(2\rho)\, V_-^{-1}\end{array}\right) \ ,
\label{C-cycle}
\eea
which is almost the same as (\ref{C-radial-explicit}) above but
without the $\sigma^0$ block. We choose
\beq
\la_i=-\mbox{$\frac{n-2i}n$} \qquad \mbox{for} \quad i=1,\dots,n-1 \ ,
\label{lainchoose}\eeq
along with
\beq
V_+=\Sigma_{(n)} \qquad \mbox{and} \qquad V_- = \one_{n-1}
\label{cycle-coords}
\eeq
where $\Sigma_{(n)}\in U(n+1)$ is a cyclic permutation matrix
representing $\pi_{(n)}:=(1\,2\,\cdots\,n)$. Then we get explicitly
\beq
C=\frac N{2n}\,\left(\begin{array}{cc} \,\diag(-n+2,\dots,
    n-2) & \tilde\xi_+ \\
  \tilde\xi_- & \diag(n-2,\dots, -n+2) \end{array}\right) 
\label{C-cycle-explicit}
\eeq
where $\tilde \xi_\pm$ are cyclic operators (rather than
raising/lowering operators as before).

In this case $C_0 = 0$, and hence this solution is part of the
orbifold singularities for $n$ coincident fluxons in the moduli space
\eq{calM0} of Section~\ref{CritPoints}, rather than an irreducible
representation of the isometry group $SU(2)$. This construction is
further used below. In particular, the special case $n=1$ gives a
single fluxon $C(n=1) = \frac N2\,\sigma^3$. Then there exists a
unitary transformation $U\in SU(2)$ such that
\beq
U\,C(n=1)\,U^{-1}=\mbox{$\frac N2$}\,U\,\sigma^3\,U^{-1}=c_i\,\sigma^i
\ ,
\eeq
which gives the position $c_i$ of the fluxon on the sphere $S^2$.

\subsubsection*{\it Multi-block solutions}

Let us modify the previous radial solution by setting
$\la_1=\pm\,1$ and taking $\la_{i+1}$ to be given by \eq{lainchoose},
while keeping the angular variables \eq{cycle-coords} in $U(n+2)$ and
$U(n)$ the same. Then the block covariant coordinates \eq{C-cycle} of
size $2(n+1)$ are given explicitly as
\beq
C=\frac N{2n}\,\left(\begin{array}{cc} \,\diag(-n+2, \dots,
    n-2,\pm\, n) & \xi_+ \\
  \xi_- & \diag(\mp\, n,n-2,\dots, -n+2) \end{array}\right) \ ,
\label{C-cycle-1-explicit}
\eeq
which is almost the same as the vacuum configuration \eq{C-vac-block}
for an $n$-dimensional irreducible representation except that there
are two explicit diagonal entries $\frac N2,-\frac N2$ instead of
$\frac N2,\frac N2$. In particular, $C_0$ is no longer constant and 
 hence the gauge fields \eq{C-cycle-1-explicit} are not solutions of
 the Yang-Mills equations of motion. This can be cured by the addition
 of extra irreducible representations as follows.

One can now construct solutions of the Yang-Mills equations with
several blocks and arbitrary parameters, i.e. the generic critical
surfaces $\cC_{(n_1,s_1),\dots,(n_k,s_k)}$, by joining an even number
of copies of \eq{C-cycle-1-explicit} in a suitable way. Fix another
integer $m\geq1$ such that $n+m\leq N$, and consider again the block
covariant coordinate \eq{C-cycle} of size $2(n+m)$ with
\beq
\la_1=1 \ , \quad \la_{i}=
-\mbox{$\frac{n-2(i-1)}n$}\quad\mbox{for}\quad i=2,\dots,n
\quad\mbox{and}\quad \la_{j+n-1}=-\mbox{$\frac{m-2(j-1)}m$}\quad
\mbox{for}\quad j=1,\dots,m \ .
\label{lamultichoice}\eeq
The angular degrees of freedom are given by
\beq
V_+=\Sigma_{(n+m)} \qquad \mbox{and} \qquad V_-=\one_{n+m-1}
\label{2irreps-coords}\eeq
in $U(n+m\pm1)$, corresponding to the cyclic permutation $\pi_{(n+m)}$
decomposed as
\beq
\pi_{(n+m)}=(\pi_{(n)})_{1,\dots,n}\circ(\pi_{(m)})_{n+1,\dots,n+m}
\circ(1~n{+}1)
\label{pinmdecomp}
\eeq
where the subscripts indicate the indices that the permutations act
on. The role of the transposition $(1~n{+}1)$ is to first interchange
the explicit $1$ and $-1$ in \eq{lamultichoice} for the upper block in
\eq{C-cycle}, which then takes the form of two copies of the matrix
\eq{C-cycle-1-explicit} but with the correct explicit diagonal entries
$\pm\, \frac N2$. Since $V_+ =\Sigma_{(n+m)}$ corresponds to an irreducible
cycle, $C$ is a direct sum of two irreducible representations with
opposite sign and hence lives on the critical surface block
$\cC_{(n,1),(m,-1)}$ with vanishing overall trace. This construction
clearly generalizes to an arbitrary number of irreducible
representations of the $SU(2)$ isometry group.

\subsection{Action of the gauge group\label{sec:gaugegroup-embed}}

Finally, let us describe how the gauge symmetry acts on the radially
foliated solutions. Recall that the gauge group $G \cong SU(n\,N)$ is
embedded in the symmetry group of the orbit space $\cO$ as $\phi  =
\phi_0 \otimes\sigma^0$ in the Lie algebra of $G \subset
SU(2n\,N)$. This embedding is well adapted to the modification of the
radial coordinates in \eq{radialcoords-modified} by the permutation
matrix $\Sigma_0$. Indeed, there is an embedding of the ``diagonal''
subgroup $U(n\,N_-) \subset U(n\,N_+) \times U(n\,N_-)$ into $G$ given
by taking $V_-$ into $\diag(\one_n,V_-) \otimes \sigma^0$ as
\be
V_- ~\longmapsto ~ 
\left(\begin{array}{cc}\left(\begin{array}{ccc}\one_n &&\\
                                        & V_-&\\
                                     && \one_n  \end{array}\right) & 0\\ 
                                0 & V_- \end{array}\right) \ .
\ee
This shows explicitly that a large  part of the gauge group
is part of the stabilizer group $R=U(n\,N_+) \times U(n\,N_-)$ which
defines the foliation  of the radial coordinates.

Furthermore, there is an additional symmetry $SU(n)\subset U(n\,N_+)$
embedded into $G$ by taking $U$ into $\diag( U,\one_{n\,N_-}) \otimes
\sigma^0$ as
\be
 U ~\longmapsto~
\left(\begin{array}{cc}\left(\begin{array}{ccc}U &&\\
                                        & \one_{n\,N_-}&\\
                                     && U  \end{array}\right) & 0\\ 
                                0 & \one_{n\,N_-} \end{array}\right) \
                            .
\label{extra-sun}
\ee
This extra $SU(n)$ symmetry acts on the marked momenta
$p_{1},\dots, p_n$ of Section~\ref{UnAbelian}, and together with the
degenerate Itzykson-Zuber localization it is thus responsible for the
emergence of the nonabelian gauge symmetry in the commutative
limit. The remainder $SU(n\,N) / SU(n\,N_-) \times SU(n)$ of the gauge
group mixes the symplectic leaves, so that the radial foliation is not
$G$-equivariant.

\bigskip

\section*{Acknowledgments}

\noindent
We thank C.-S.~Chu, B.~Dolan, H.~Grosse, X.~Martin and D.~O'Connor for helpful
discussions. The work of H.S. was supported in part by the FWF Project
P16779-N02 and in part by the FWF Project P18657. 
The work of R.J.S. was supported in part by the EU-RTN Network
Grant MRTN-CT-2004-005104.

\end{document}